\documentclass[prd,notitlepage,showpacs,nofootinbib,superscriptaddress]{revtex4-2}

\usepackage[utf8]{inputenc} 

\usepackage{amsmath}
\usepackage{amssymb}
\usepackage{float}
\usepackage{comment}
\usepackage{slashed}
\usepackage[normalem]{ulem}
\usepackage{bbold}
\usepackage{cancel}
\usepackage{graphicx}
\usepackage{booktabs}
\usepackage{multirow}
\usepackage{rotating}
\usepackage{caption}
\usepackage{subcaption}
\usepackage{tabulary}

\DeclareMathOperator{\diag}{diag}

\usepackage[usenames,dvipsnames]{color}

\newcommand {\ignore}[1]{}

\usepackage[colorinlistoftodos]{todonotes}
\usepackage[colorlinks=true,linkcolor=darkblue,citecolor=darkblue,urlcolor=darkblue, pdfborder={0 0 0}]{hyperref}
\usepackage[normalem]{ulem}

\usepackage{caption}
\usepackage{subcaption}
\definecolor{linkcolor}{rgb}{0,0,0.8}

\definecolor{darkgreen}{rgb}{0,0.5,0}
\definecolor{darkred}{rgb}{0.6,0,0}
\definecolor{brown}{rgb}{0.59, 0.29, 0.0}
\definecolor{mightnightblue}{RGB}{25,25,112}
\definecolor{darkblue}{rgb}{0,0,0.8}

\newcommand {\darkblue} {\color{darkblue}}

\newcommand{\M}{\mathbf{M}}
\newcommand{\U}{\mathbf{U}}

\newcommand{\BR}{{\rm BR}}
\newcommand{\GeV}{{\rm GeV}}
\newcommand{\TeV}{{\rm TeV}}

\def\EW{$\mathrm{SU(2)}_L \otimes \mathrm{U(1)}_Y$}

\def\U{\mathbf{U}}

\newcolumntype{K}[1]{>{\centering\arraybackslash}m{#1}}
\def\gsim{\raise0.3ex\hbox{$\;>$\kern-0.75em\raise-1.1ex\hbox{$\sim\;$}}}
\def\lsim{\raise0.3ex\hbox{$\;<$\kern-0.75em\raise-1.1ex\hbox{$\sim\;$}}}

\newcommand{\AddrCFTP}{%
Departamento de F\'{\i}sica and CFTP, Instituto Superior T\'ecnico, Universidade de Lisboa, Av. Rovisco Pais 1, 1049-001 Lisboa, Portugal}

\newcommand{\AddrIEAP}{%
Institute of Experimental and Applied Physics, \\ Czech Technical University in Prague,
Prague 160 00, Czech Republic}

\usepackage[force]{feynmp-auto}

\usepackage{colortbl}

\begin{document}

\title{\darkblue Fermionic dark linear seesaw with an anomalous Majoron}

\author{\textbf{A. Batra}}\email{aditya.batra@tecnico.ulisboa.pt}
\affiliation{\AddrCFTP}

\author{\textbf{H.~B. C\^amara}}\email{henrique.camara@cvut.cz}
\affiliation{\AddrIEAP}

\author{\textbf{F.~R. Joaquim}}\email{filipe.joaquim@tecnico.ulisboa.pt }
\affiliation{\AddrCFTP}

\begin{abstract}
\begin{center}
{ \center  \bf ABSTRACT}\\    
\end{center}
We propose a minimal fermionic realisation of the \emph{dark} linear seesaw with spontaneous breaking of lepton number. The small lepton-number-violating parameter responsible for linear-seesaw neutrino masses is generated radiatively by a \emph{dark} sector consisting of a singlet and a doublet fermion, together with scalar singlets. The global U(1)$_L$ symmetry is broken to a residual $\mathcal{Z}_2$ under which the \emph{dark} sector is odd. This breaking yields a Majoron $J$ and makes the lightest \emph{dark} particle stable. The charged \emph{dark} fermion acquires its mass solely from U(1)$_L$ breaking and thus induces an anomalous Majoron coupling to photons. Current bounds on this coupling exclude a massless Majoron for TeV-scale U(1)$_L$ breaking. We add a linear term that breaks lepton number softly and gives the Majoron a GeV-scale mass. Such a Majoron decays into two photons and can be probed at beam-dump experiments. We study the dark matter phenomenology, focusing on the case where the stable particle is a fermionic weakly interacting massive particle. We analyse how the framework can be tested through charged-lepton flavour-violating decays such as $\mu \to e\gamma$, which receive contributions from both the \emph{dark} and the seesaw sectors. Moreover, both sectors enhance the Majoron couplings to charged leptons, so that the Majoron-emitting decays $\tau \to \ell J$ ($\ell = e,\mu$) typically exceed $\tau \to \ell\gamma$ by orders of magnitude and can probe parameter space beyond the reach of $\mu \to e\gamma$.
\end{abstract}

\maketitle
\noindent

\section{Introduction}
\label{sec:intro}

The discovery of neutrino flavour oscillations~\cite{Kajita:2016cak,McDonald:2016ixn} and the overwhelming evidence that about $27\%$ of the energy budget of the Universe is in the form of dark matter~(DM)~\cite{Planck:2018vyg} (for a review on DM, see Ref.~\cite{Cirelli:2024ssz}) provide two of the most compelling motivations for physics beyond the Standard Model~(SM). While these two puzzles may well be unrelated, it is tempting to regard them as different manifestations of a common new-physics sector. Regarding neutrino masses, the required new states may act at tree level, the prime example being the type-I seesaw mechanism~\cite{Minkowski:1977sc,Gell-Mann:1979vob,Yanagida:1979as,Schechter:1980gr,Glashow:1979nm,Mohapatra:1979ia}, or solely at the quantum level, as in radiative models~\cite{Zee:1980ai,Ma:1998dn} (for a review, see Ref.~\cite{Cai:2017jrq}). In its canonical form, the type-I seesaw requires either heavy right-handed neutrinos far above the electroweak~(EW) scale or tiny Yukawa couplings in order to accommodate neutrino data, leaving little hope for direct tests. In contrast, its low-scale variants, such as the inverse~\cite{Mohapatra:1986bd} and linear~\cite{Akhmedov:1995ip,Akhmedov:1995vm,Malinsky:2005bi} seesaws, evade this limitation by linking the smallness of neutrino masses to parameters controlling lepton number violation~(LNV), which are naturally small in the 't~Hooft sense~\cite{tHooft:1979rat}. Such schemes therefore allow for TeV-scale heavy neutrinos with sizeable Yukawa couplings, opening up the prospect of testing the seesaw paradigm at colliders and high-intensity facilities, most notably through charged-lepton flavour violation~(cLFV)~\cite{Alonso:2012ji,Abada:2015oba,Abada:2018nio,Hernandez-Tome:2019lkb,Camara:2020efq}. Linear seesaw models, in particular, exhibit a rich phenomenology, encompassing heavy-neutrino production, LNV collider signals and gravitational-wave signatures~\cite{Batra:2022arl,Batra:2023ssq,Batra:2023mds,Mandal:2026wbz}. Radiative models, in turn, offer an economical bridge to the DM puzzle: the new \emph{dark}-sector particles mediating neutrino masses can be odd under a stabilising symmetry as simple as a $\mathcal{Z}_2$, rendering the lightest of them a viable weakly interacting massive particle~(WIMP) DM candidate. This is the essence of the scotogenic mechanism~\cite{Tao:1996vb,Ma:2006km}, in which DM acts as the seed of neutrino masses (for a review, see Ref.~\cite{Avila:2025qsc}).

The appealing features of both approaches can be combined by making a \emph{dark} sector the seed of LNV in a low-scale seesaw. This is the idea behind our proposal of the \emph{dark} linear seesaw~\cite{Batra:2023bqj}, in which the LNV parameter responsible for linear-seesaw neutrino masses vanishes at tree level and is induced at one loop by particles odd under a \emph{dark} $\mathcal{Z}_2$ parity symmetry (see also Ref.~\cite{CarcamoHernandez:2023atk} for a related construction, and Refs.~\cite{Wang:2015saa,Das:2017ski,CarcamoHernandez:2021tlv,CarcamoHernandez:2024edi,Abada:2025ozu}). The same strategy has been employed in the context of the inverse seesaw~\cite{Ma:2009gu,Bazzocchi:2009kc,Baldes:2013eva,Mandal:2019oth,Abada:2021yot,Abada:2025ozu}. In the minimal \emph{dark} linear seesaw of Ref.~\cite{Batra:2023bqj}, the \emph{dark} sector comprises a fermion singlet together with doublet and singlet scalars, while lepton number is broken explicitly, though softly, by a single cubic scalar interaction that preserves the $\mathcal{Z}_2$ symmetry imposed to stabilise DM. In that work, we explored the predictions of the model for scalar WIMP DM and showed that cLFV signals arise from the interplay of sizeable low-scale seesaw and \emph{dark}-sector contributions, lying within the projected MEG~II sensitivities for $\mu \to e \gamma$~\cite{MEGII:2018kmf}. Pursuing a fermionic realisation of this setup is therefore particularly well motivated, as singlet-doublet fermion DM is one of the simplest and most extensively studied WIMP candidates~\cite{Cohen:2011ec,Cheung:2013dua,Calibbi:2015nha,Yaguna:2015mva,Banerjee:2016hsk,Paul:2024prs}, including within radiative neutrino-mass frameworks~\cite{Restrepo:2015ura,Bhattacharya:2018fus,Konar:2020wvl,Paul:2026snc}. More recently, the flavour and DM phenomenology of a scotogenic model featuring singlet and doublet fermions and scalars in its \emph{dark} sector has also been investigated in detail~\cite{Esch:2018ccs,Sarazin:2021nwo,Alvarez:2023dzz,Darricau:2025vcs,Darricau:2026iwg}.

Another compelling avenue is the spontaneous breaking of lepton number. If the Lagrangian respects an exact global U(1)$_L$ symmetry that is spontaneously broken by the vacuum expectation value~(VEV) of a scalar singlet, the physical spectrum contains one of the most emblematic Goldstone bosons~(GBs), the Majoron $J$~\cite{Chikashige:1980ui,Schechter:1981cv}. Moreover, the breaking of U(1)$_L$ can leave behind a residual $\mathcal{Z}_2$ symmetry, so that DM stability emerges naturally as a remnant of lepton number~\cite{Ma:2015xla}. Far from being a phenomenological spectator, the Majoron acquires one-loop couplings to charged leptons~\cite{Escribano:2020wua,Herrero-Brocal:2023czw} which, in low-scale seesaws or scotogenic scenarios, can be strongly enhanced compared to the canonical dynamical type-I seesaw~\cite{Heeck:2019guh,Escribano:2021uhf,CentellesChulia:2024uzv,CentellesChulia:2025eck}. Consequently, Majoron-emitting cLFV decays such as $\mu \to e J$, alongside astrophysical and cosmological Majoron signatures, provide an additional probe of LNV. Majoron phenomenology has likewise been explored in dynamical scotogenic scenarios~\cite{Bonilla:2019ipe,Mandal:2019oth,Escribano:2021ymx,DeRomeri:2022cem,Portillo-Sanchez:2023kbz,Chun:2023vbh,Herrero-Brocal:2023czw,Bonilla:2023egs,Ardila-Tafurth:2025qdf,Datta:2026aks,CentellesChulia:2026ats,Hieu:2026pfh}. An explicit breaking of U(1)$_L$, for instance via a soft term in the scalar potential, endows the Majoron with a mass. If, in addition, U(1)$_L$ possesses an electromagnetic~(EM) anomaly, the Majoron couples to photons like an axion, allowing axion detection experiments to probe Majoron DM~\cite{Liang:2024vnd}. In minimal models, however, the Majoron is the GB of the anomaly-free U(1)$_{B-L}$ rather than of U(1)$_L$, and a coupling to photons does not by itself signal an anomaly~\cite{Herrero-Brocal:2026nmc}.

Inspired by these ideas, we propose a minimal fermionic realisation of the \emph{dark} linear seesaw in which, unlike in our previous work~\cite{Batra:2023bqj}, lepton number is broken spontaneously. We extend the SM by a single pair of fermion singlets $\nu_R$ and $S_R$, a complex scalar singlet $\sigma$, and a \emph{dark} sector consisting of a fermion doublet and singlet together with two complex scalar singlets. While U(1)$_L$ forbids bare fermion mass terms -- so that all heavy-fermion masses originate from the VEV of $\sigma$ -- the same VEV turns a lepton-number-conserving cubic scalar interaction into a bilinear mixing between the two \emph{dark} scalar singlets, which is essential for neutrino mass generation. Spontaneous breaking of U(1)$_L$ leaves a residual $\mathcal{Z}_2$ symmetry under which only the \emph{dark}-sector fields are odd, thereby stabilising the lightest \emph{dark} particle as either a fermionic or a scalar WIMP DM candidate. The physical spectrum also contains a Majoron $J$, identified as the phase of $\sigma$. Because the charged \emph{dark} fermion acquires its mass solely from the VEV of $\sigma$, U(1)$_L$ exhibits an EM anomaly that, in contrast to minimal models~\cite{Herrero-Brocal:2026nmc}, cannot be rotated away by redefining lepton number. The resulting anomalous Majoron coupling to photons is fixed exclusively by the VEV of $\sigma$, and for a TeV-scale VEV current bounds exclude a massless Majoron. We therefore introduce a term linear in $\sigma$ that breaks lepton number softly, preserves the residual $\mathcal{Z}_2$ symmetry, and gives the Majoron a GeV-scale mass. 

The remainder of this paper is organised as follows. In Sec.~\ref{sec:model}, we present the particle content, symmetries and neutrino mass generation mechanism of our model. In Sec.~\ref{sec:pheno}, we investigate its phenomenological implications for the Majoron, cLFV and DM: we examine the anomalous Majoron coupling to photons in Sec.~\ref{sec:anomalous}, review the relevant collider constraints in Sec.~\ref{sec:collider}, analyse the cLFV predictions in Sec.~\ref{sec:cLFV}, and study the WIMP DM phenomenology in Sec.~\ref{sec:DM}. Our concluding remarks are given in Sec.~\ref{sec:concl}. Technical details on the scalar sector, \emph{dark}-fermion masses and mixing, \emph{dark}-loop functions and the generalised Casas-Ibarra parameterisation are collected in the appendices.

\section{Model}
\label{sec:model}

%
\begin{table}[t!]
\renewcommand*{\arraystretch}{1.3}
	\centering
	\begin{tabular}{| c | c | c | c |}
		\hline 
&Fields&\EW &  U$(1)_L \rightarrow \mathcal{Z}_2$  \\
		\hline \hline
		\multirow{6}{*}{Fermions} 
&$\ell_L$&($\mathbf{2}, {-1/2}$) & $1 \rightarrow +$  \\
&$e_R$&($\mathbf{1}, {-1}$) & $1 \rightarrow +$  \\
&$\nu_R$&($\mathbf{1}, {0}$) & $1 \rightarrow +$ \\
&$S_R$&($\mathbf{1}, {0}$) & $-2 \rightarrow +$ \\
\cline{2-4}
&$D_{L,R}$&($\mathbf{2}, {-1/2}$) & $5/2, 7/2 \rightarrow -$ \\
&$F_{L,R}$&($\mathbf{1}, {0}$) & $3/2 , 5/2 \rightarrow -$ \\
		\hline \hline
\multirow{3}{*}{Scalars}  &$\Phi$&($\mathbf{2}, {1/2}$) & $0 \rightarrow +$ \\
&$\sigma$&($\mathbf{1}, {0}$) & $1 \rightarrow +$ \\
\cline{2-4}
&$\chi_{1},\chi_2$&($\mathbf{1},0$) & $-5/2, 7/2 \rightarrow -$  \\		
\hline
	\end{tabular}
	\caption{Field content and transformation properties under \EW~and U$(1)_L \rightarrow \mathcal{Z}_2$.}
	\label{tab:model} 
\end{table}
The minimal linear seesaw is implemented by adding to the SM two sterile singlet fermions $\nu_R$ and $S_R$. LNV stems from the $\textbf{m}_S\overline{\nu_L}S_R$ term, where $\nu_L$ is the SM neutrino field from the lepton doublet $\ell_L$. The resulting light neutrino masses are then proportional to $\mathbf{m}_S$ which can be naturally small in the t'Hooft sense~\cite{tHooft:1979rat}. Consequently, in contrast with the canonical type-I seesaw, which requires heavy mediators, the linear seesaw allows for low-scale neutrino mass generation. In this work, we consider, as in Ref.~\cite{Batra:2023bqj}, that $\mathbf{m}_S$ is generated radiatively by a \emph{dark} sector, containing viable DM candidates. Our novel model is based on the field content and symmetries shown in Table~\ref{tab:model}. It is clear, due to the global lepton number symmetry U$(1)_L$, that $\mathbf{m}_S=0$ at tree-level. To generate it at one-loop, we introduce a vector-like fermion doublet and singlet, $D_{L,R}$ and $F_{L,R}$, respectively, as well as two scalar singlets~$\chi_{1,2}$. These fields constitute the \emph{dark}-sector which is odd under the residual $\mathcal{Z}_2$ discrete  symmetry arising from the spontaneous breaking of U$(1)_L$ by the scalar $\sigma$. The imaginary component of $\sigma$, is the so-called Majoron $J \equiv \sqrt{2} \text{Im}(\sigma)$, the GB associated with spontaneous lepton number breaking. The most general Yukawa Lagrangian allowed by the SM gauge and U$(1)_L$ symmetries is given by~\footnote{Without loss of generality, we consider the charged leptons in their physical basis, focusing in what follows on the neutrino sector.}
\begin{align}
    -\mathcal{L}_{\text{Yuk.}} & = \mathbf{Y}_e~\overline{\ell_L} \Phi e_R + \mathbf{Y}_\nu~\overline{\ell_L} \tilde{\Phi} \nu_R + \mathbf{Y}_\chi~\overline{\ell_L} \chi_1 D_R 
     + y_S~\overline{F_L} \chi_2 S_R + y_\chi~\overline{\nu_R^c} \chi_2^\ast F_R \nonumber \\
    &+ y_R~\overline{\nu_R^c} \sigma S_R + y_D~\overline{D_L} \sigma^\ast D_R + y_{DF}~\overline{D_L} \tilde{\Phi} F_R + y_F~\overline{F_L} \sigma^\ast F_R + \text{H.c.} \; ,
    \label{eq:LYuk}
\end{align}
where ``H.c." stands for Hermitian conjugate and the Yukawa couplings $\mathbf{Y}_e$, $\mathbf{Y}_{\nu,\chi}$ and $y_{S,\chi,R,D,DF,F}$ are $3 \times 3$ matrices, $3 \times 1$ vectors and numbers, respectively. Furthermore, the Higgs doublet is defined as $\Phi = (\phi^+ , \phi^0)^T$, with $\tilde{\Phi}= i \tau_2 \Phi^\ast$, $\tau_2$ being the complex Pauli matrix.  The relevant scalar potential term for neutrino mass generation is
    \begin{align}
    V \supset \kappa \ \chi_1 \chi_2 \sigma^\ast + \text{H.c.} \; ,
    \label{eq:Vpotential}
    \end{align}
where $\kappa$ can be naturally small in the t'Hooft sense~\cite{tHooft:1979rat}, since in the limit $\kappa \rightarrow 0$ the scalar potential exhibits an enlarged symmetry being entirely U(1) symmetric. Note that, in our previous work~\cite{Batra:2023bqj}, lepton number was explicitly broken by a soft cubic term in the scalar potential, which triggered neutrino mass generation. Here, the cubic interaction in Eq.~\eqref{eq:Vpotential} conserves lepton number. Once $\sigma$ acquires its VEV, this interaction generates the bilinear $\kappa v_\sigma\chi_1\chi_2/\sqrt{2}+\mathrm{H.c.}$, which violates lepton number. Hence, the spontaneous breaking of lepton number mixes the inert scalars and closes the \emph{dark} loop. The full scalar sector is presented in Appendix~\ref{sec:scalarsector}. The Higgs doublet and scalar $\sigma$ acquire non-zero VEVs, where $\left< \phi^0 \right> = v_\Phi/\sqrt{2} \simeq 174$ GeV and $\left< \sigma \right> = v_\sigma/\sqrt{2}$, breaks the EW and lepton number symmetry, respectively. Since the residual $\mathcal{Z}_2$ symmetry needs to remain unbroken to stabilize the DM candidate (see Sec.~\ref{sec:DM}), the \emph{dark} singlets $\chi_{1,2}$ are inert, i.e. $\left< \chi_{1,2} \right> = 0$.

\begin{figure}[t!]
        \centering
        \includegraphics[scale=0.55]{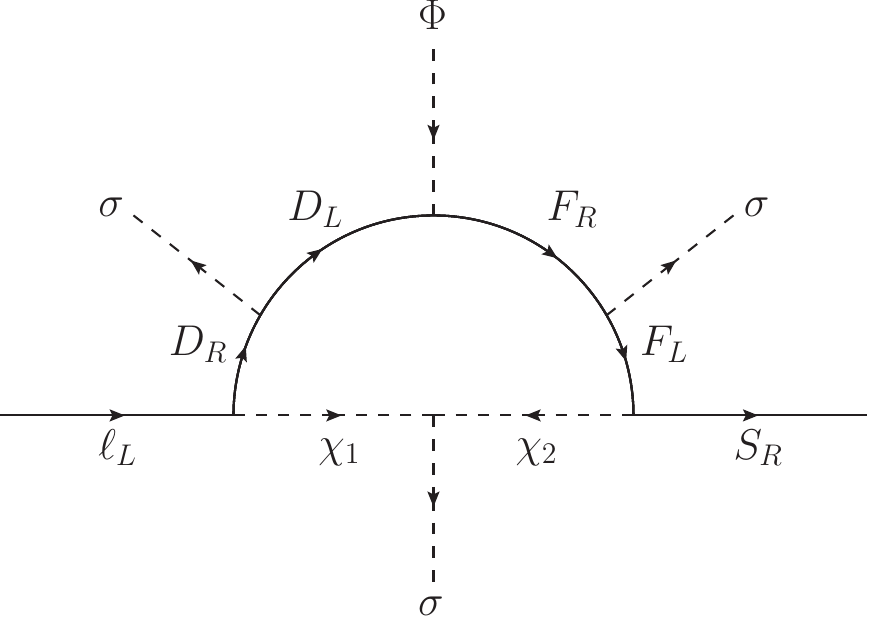} \hspace{+0.3cm} \includegraphics[scale=0.55]{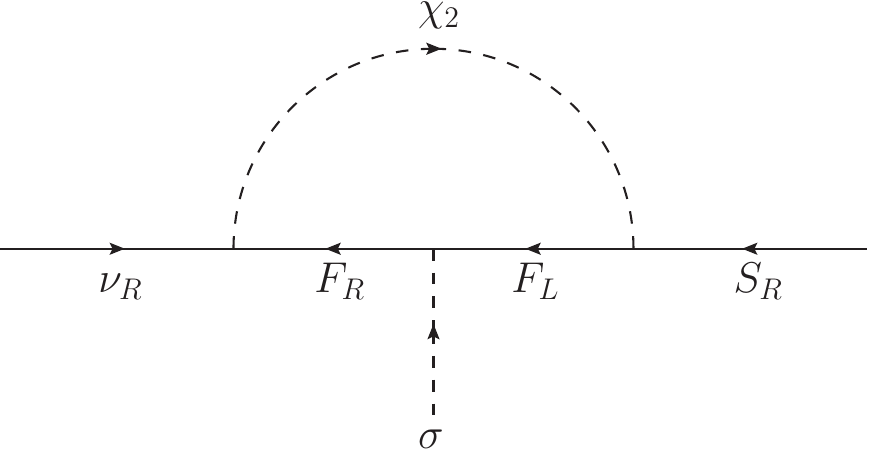}
        \caption{Lowest order one-loop diagrams for neutrino mass generation: fermionic \emph{dark}-seeded linear seesaw (left) and additional correction to $\overline{\nu_R^c} S_R$ (right) (see text for details).}
    \label{fig:neutrino}
    \end{figure}

After EW and lepton number symmetry breaking, the full tree-level neutrino mass matrix $\mathcal{M}_\nu^0$ defined in the $(\nu_L, \nu_R^c,S_R^c)$ basis is given by,
\begin{align}
\mathcal{M}_\nu^0 = \begin{pmatrix}
 0 & \mathbf{M}_\nu & 0    \\
 \mathbf{M}_\nu^{T} & 0 & M_R^0    \\
 0 & M_R^0 & 0    \\
\end{pmatrix} \; , \; \mathbf{M}_\nu = \frac{v_\Phi \mathbf{Y}_\nu}{\sqrt{2}} \; , \; M_R^0 = \frac{v_\sigma y_R}{\sqrt{2}} \; ,
\label{eq:treemassmatrix}
\end{align}
where it is clear that active neutrinos are massless at tree-level. As shown in Fig.~\ref{fig:neutrino}, neutrino mass generation arises at one-loop level, mediated by the $\mathcal{Z}_2$-odd \emph{dark}-sector fields $D,F$ and $\chi_{1,2}$. The left diagram is responsible for generating a non-zero LNV parameter $\mathbf{m}_S$ triggering the linear seesaw mechanism. In addition, the right diagram is a one-loop correction $\Delta M_R$ to $M_R^0$. Taking this into account the neutrino mass matrix can be now written as
\begin{align}
\mathcal{M}_\nu = \mathcal{M}_\nu^0 + \Delta \mathcal{M}_\nu = \begin{pmatrix}
 0 & \mathbf{M}_\nu & \mathbf{m}_S    \\
 \mathbf{M}_\nu^T & 0 & M_R    \\
 \mathbf{m}_S^T & M_R & 0    \\
\end{pmatrix} \; , \; M_R = M_R^0 + \Delta M_R \; ,
\label{eq:loopmassmatrix}
\end{align}
where $\mathbf{m}_S$ is a $3\times 1$ vector. The one-loop mass terms computed in the \emph{dark} fermion and scalar mass basis are:
\begin{align}
 \mathbf{m}_S  = \mathcal{F}_S\left(m_{\psi^0_a} , m_{\zeta_k}\right)~m_{\psi^0_2} \; y_S~\mathbf{Y}_\chi \; , \;
 \Delta M_R = \mathcal{F}_R\left(m_{\psi^0_a} , m_{\zeta_k}\right)~m_{\psi^0_2} \; y_S~y_{\chi} \; ,
\label{eq:mSandDeltaMR}
\end{align}
where the complete expressions for the dimensionless loop factors $\mathcal{F}_{S,R}\left(m_{\psi^0_a},m_{\zeta_k}\right)$ are given in Appendix~\ref{sec:darkloop}. These functions depend on the \emph{dark} scalar and \emph{dark}-neutral fermion masses, denoted $m_{\zeta_k}$ ($k=1,\cdots,4$) and $m_{\psi^0_a}$ ($a=1,2$), respectively. Furthermore, $\mathcal{F}_{S,R}$ also depend on the mixing between $\chi_1$ and $\chi_2$, as well as between the neutral component of the fermion doublet $D_{L,R}$ and $F_{L,R}$. Details regarding the scalar and \emph{dark}-fermion mass spectrum and mixing are provided in Appendix~\ref{sec:scalarsector} and~\ref{sec:darkfermion}, respectively.

In the linear seesaw approximation $\mathbf{m}_S \ll \mathbf{M}_\nu \ll M_R$, the full neutrino mass matrix $\mathcal{M}_\nu $, can be diagonalised following Ref.~\cite{Schechter:1981cv}. The resulting effective light neutrino mass matrix is given by,
\begin{align}
\mathbf{M}_{\text{eff}} \simeq - \frac{\mathbf{M}_\nu~\mathbf{m}_S^T + \mathbf{m}_S~\mathbf{M}_\nu^T}{M_R} \;.
\label{eq:Mnu}
\end{align}
Since we consider the minimal model containing a single $\nu_R-S_R$ pair, the above $\mathbf{M}_{\text{eff}}$ matrix has rank $2$, resulting in a massless light neutrino at one-loop level. Evidently, considering two $\nu_R-S_R$ pairs would yield three massive light neutrinos. The seesaw diagonalisation procedure is controlled by a small parameter $\boldsymbol{\epsilon}$, which in the limit $\mathbf{m}_S \rightarrow 0$, is obtained as~\cite{Schechter:1981cv}
\begin{align}
\boldsymbol{\epsilon} \simeq - \frac{\mathbf{M}_\nu}{M_R} \; \Rightarrow \; \mathbf{K}=\begin{pmatrix}
 \mathbf{K}_{\text{light}}  &  \mathbf{K}_{\text{heavy}} 
 \end{pmatrix} \; , \; \mathbf{K}_{\text{light}} =\left(1-\frac{1}{2} \boldsymbol{\epsilon}^\ast \boldsymbol{\epsilon}^T\right) \mathbf{U}_{\nu} \; , \; \mathbf{K}_{\text{heavy}} =\left(\frac{\boldsymbol{\epsilon}^\ast}{\sqrt{2}}  \,\,\,\,\, \frac{i \boldsymbol{\epsilon}^\ast}{\sqrt{2}} \right) \; ,
 \label{eq:K} 
\end{align}
where $\mathbf{K}$ is the $3 \times 5$ matrix appearing in charged-current interactions which induces cLFV as discussed in Sec.~\ref{sec:cLFV}. $\mathbf{K}_{\text{light}}$ is the $3 \times 3$ matrix that diagonalises $\mathbf{M}_{\text{eff}}$, with $\boldsymbol{\epsilon}^\ast \boldsymbol{\epsilon}^T$ encoding deviations from unitarity of the lepton mixing matrix $\mathbf{U}_{\nu}$. $\mathbf{K}_{\text{heavy}}$ is the $3 \times 2$ heavy-light mixing matrix. Neglecting those deviations, $\mathbf{M}_{\text{eff}}$ can be diagonalised through a unitary rotation $\nu \rightarrow \mathbf{U}_{\nu}\, \nu$, as follows
\begin{equation}
\mathbf{U}_{\nu}^{T}\, \mathbf{M}_{\text{eff}} \, \mathbf{U}_{\nu}  = \mathbf{d}_{\nu} = \text{diag}\left(m_1, m_2, m_3\right) \; , \; m_4 \simeq m_5 \simeq M_R \; ,
\label{eq:Unudef}
\end{equation}
where $m_{1,2,3}$ are the real and positive light neutrino masses. The heavy neutrinos stemming from $\nu_R$ and $S_R$ form a pseudo-Dirac pair. Their masses $m_{4,5}$ are approximately equal to $M_R$, apart from terms proportional to the small parameters $\mathbf{m}_S$ and $\boldsymbol{\epsilon}$. As mentioned before, the linear seesaw can be accommodated as a low-scale neutrino mass generation mechanism. Namely, the smallness of the LNV $\mathbf{m}_S$ term is provided by $\kappa$ and loop suppression, implying that we can have $\mathbf{m}_S \sim \mathcal{O}(1\,{\rm eV})$, leading to light neutrino masses $\sim \mathcal{O}(0.1\,{\rm eV})$ for heavy neutrino masses $M_R \sim \mathcal{O}(1\,{\rm TeV})$. As discussed in Sec.~\ref{sec:cLFV}, this has important implications with regard to to cLFV. The above unitary lepton mixing matrix $\mathbf{U}_{\nu}$ can be, as usual, parameterised by three mixing angles $\theta_{12}$, $\theta_{23}$, and $\theta_{13}$, and two CP-violating phases: a Dirac phase $\delta$ and, for a massless neutrino, a single Majorana phase $\alpha$~\cite{ParticleDataGroup:2026mpi}. Besides the mixing angles and $\delta$, neutrino oscillation experiments constrain the two neutrino mass squared differences $\Delta m_{21}^2= m_2^2 - m_1^2$ and $\Delta m_{31}^2= m_3^2 - m_1^2$, with two possible orderings for neutrino masses: normal and inverted ordering (NO and IO, respectively). In our minimal case the lightest neutrino is massless: $m_1=0$ for NO and $m_3=0$ for IO. Our model is testable at future experiments looking for the neutrinoless double beta decay process which could elucidade whether neutrinos are Majorana or Dirac~\cite{Schechter:1981bd}. Interestingly, in the case of a massless neutrino, the effective Majorana mass parameter has a lower bound for NO neutrino masses~\cite{Barreiros:2018bju}. Furthermore, the IO prediction falls within the sensitivity of future experiments such as SNO+ II~\cite{SNO:2015wyx}, LEGEND~\cite{LEGEND:2017cdu} and nEXO~\cite{nEXO:2017nam}.

\section{Phenomenology}
\label{sec:pheno}

In this section, we study the phenomenology of our model, focusing on the Majoron, cLFV and DM. We perform a numerical scan with fermion and scalar input parameters as shown in Table~\ref{tab:Scan} [see Eqs.~\eqref{eq:LYuk} and~\eqref{eq:Vpotentialfull}, respectively].

As a benchmark, we take $v_\sigma/\sqrt{2}=5$~TeV and the heavy-scalar mass $m_H=2$~TeV [see Eq.~\eqref{eq:nondarkscalarmasses}]. The SM Higgs mass is $m_h=125.13$~GeV~\cite{ParticleDataGroup:2026mpi}. The $\Phi-\sigma$ scalar mixing of Eq.~\eqref{eq:mixingnondark} is controlled by the angle $\theta$ which we vary between the interval $|\sin\theta| \in [10^{-3},0.1]$ [see discussion in Sec.~\ref{sec:collider} and the bound of Eq.~\eqref{eq:invisible_Majoron_only}].  The \emph{dark}-scalar self-couplings are fixed to $\lambda_{\chi_1}=\lambda_{\chi_2}=\lambda_{\chi_1\chi_2}=0.5$, since they do not affect our analysis, while the remaining quartic couplings are varied over a wide range. As mentioned before, neutrino mass suppression stems from the loop-generated $\mathbf{m}_S$ [see Eq.~\eqref{eq:mSandDeltaMR}], which is controlled by $y_S$ and the naturally small parameter $\kappa$. For simplicity, we set $y_S=1$ and vary $\kappa\in[10^{-6},10^{-3}]$~GeV.

Since we are interested in a low-scale seesaw, we choose the \emph{dark}-scalar mass parameters $m^2_{\chi_{1,2}}$ such that the \emph{dark} scalars lie at the GeV-TeV scale, and $M_R^0\in[0.5,5]$~TeV, i.e.\ $y_R\in[0.1,1]$, which ensures $\mathbf{m}_S\ll \mathbf{M}_\nu \ll M_R^0$ [see Eq.~\eqref{eq:treemassmatrix}]. The one-loop correction $\Delta M_R$ to $M_R^0$ [see Eq.~\eqref{eq:mSandDeltaMR}] is small, and we simplify the scan by setting $y_\chi=0$, hence $M_R\equiv M_R^0$. For the \emph{dark} fermions we take $m_{\psi^\pm}\in[100,5000]$~GeV, i.e.\ $y_D\in[0.02,1]$, satisfying the bound in Eq.~\eqref{eq:mpsipm_LEPbound}. We vary the Yukawa coupling $y_F$ within $[10^{-3},1]$, so that the resulting \emph{dark}-fermion masses [see Eqs.~\eqref{eq:neutraldarkfermionmass} and~\eqref{eq:chargedfemionmass}] also lie at the GeV-TeV scale. The $y_{DF}$ parameter that controls \emph{dark}-neutral fermion singlet-doublet mixing, as shown in Eq.~\eqref{eq:neutraldarkfermionmixing}, is taken within a wider range $y_{DF} \in [10^{-5},1]$, being relevant for the DM analysis of Sec.~\ref{sec:DM}.

Compatibility with neutrino oscillation data is ensured by reconstructing the Yukawa vectors $\mathbf{Y}_{\nu}$ and $\mathbf{Y}_{\chi}$ through the generalised Casas-Ibarra parametrisation of Appendix~\ref{sec:CasasIbarra}, where we vary $\text{Re}(z)\in[0,2\pi]$, $\text{Im}(z)\in[-3,3]$, and the Majorana phase $\alpha\in[0,2\pi]$. Assuming the currently favoured NO, we vary the oscillation parameters (the two mass-squared differences, the three mixing angles and the Dirac CP phase) within their $3\sigma$ ranges from the most recent global fit of Ref.~\cite{Esteban:2026phq} (see also Refs.~\cite{deSalas:2020pgw,Esteban:2024eli,Capozzi:2025wyn,Capozzi:2025ovi}). Finally, we impose perturbativity on the reconstructed Yukawa couplings, requiring all entries of $\mathbf{Y}_{\nu}$ and $\mathbf{Y}_{\chi}$ to be at most unity in absolute value. Over the scanned parameter ranges, the Yukawa norms span $10^{-5}\lesssim|\mathbf{Y}_{\nu,\chi}|\leq 1$.
\begin{table}[!t]
\renewcommand*{\arraystretch}{1.3}
\centering
\begin{tabular}{|c|c|c|}   
\hline
Sector & Parameters & Scan range \\
\hline
\multirow{4}{*}{Fermion} & $M_R^0$ & $\left[500, 5000\right]$ GeV \\
& $m_{\psi^\pm}$ & $\left[100 , 5000\right]$ GeV \\
& $y_F$ & $\left[10^{-3} , 1\right]$ \\
& $y_{DF}$ & $\left[10^{-5} , 1\right]$ \\
\hline
\multirow{4}{*}{Scalar} & $|\sin \theta|$ &  $[10^{-3}, 0.1]$ \\
& $m_{\chi_1}^2 , m_{\chi_2}^2$ &  $[10^2 , 10^8]$ (GeV$^2$) \\
& $\kappa$ &  $[10^{-6}, 10^{-3}]$ (GeV)  \\
& $|\lambda_{\Phi \chi_1}|, |\lambda_{\Phi \chi_2}|, |\lambda_{\sigma \chi_1}|, |\lambda_{\sigma \chi_2}|$ & $[10^{-5} , 1]$ \\
\hline
\end{tabular}
\caption{Input parameters of our model and ranges used in our numerical scan (see text for details).}
\label{tab:Scan}
\end{table}
%

\subsection{Anomalous Majoron coupling to photons}
\label{sec:anomalous}

In our framework the masses of the new fermions are generated dynamically by the VEV of $\sigma$, which breaks U(1)$_L$ spontaneously. To this end, the \emph{dark} doublet $D$ and singlet $F$, although vector-like under the SM gauge group, are chiral under U(1)$_L$ (see Table~\ref{tab:model}). As a consequence, U(1)$_L$ has an EM anomaly, with the effective Majoron-to-photon interaction generically written as~\cite{Liang:2024vnd}
\begin{align}
\mathcal{L}_{J\gamma\gamma}=\frac{g_{J\gamma\gamma}}{4}\,J\,F_{\mu\nu}\widetilde{F}^{\mu\nu} \; ,
\label{eq:LJgg}
\end{align}
where $F_{\mu\nu}$ is the photon field-strength tensor and $\widetilde{F}^{\mu\nu}=\epsilon^{\mu\nu\rho\sigma}F_{\rho\sigma}/2$ its dual. The coupling is given by
\begin{align}
|g_{J\gamma\gamma}|=\frac{\alpha_e\,|\mathcal{A}_\gamma|}{\pi\,v_\sigma} \; , \;
\mathcal{A}_\gamma=\sum_f\left[L(f_L)-L(f_R)\right]Q_f^2 \; ,
\label{eq:gJgg}
\end{align}
with $\alpha_e=e^2/4\pi$, $Q_f$ the electric charge of the fermion $f$, and lepton numbers normalised to $L(\sigma)=1$ as in Table~\ref{tab:model}. Since the left- and right-handed SM charged leptons carry the same lepton number, only the charged \emph{dark}-fermion $\psi^\pm$ contributes, and $\mathcal{A}_\gamma=L(D_L)-L(D_R)=-1$. Note that, baryon number and hypercharge have no EM anomaly, so $\mathcal{A}_\gamma$ cannot be removed by any redefinition $L\to L+aB+bY$, and the coupling of Eq.~\eqref{eq:gJgg} is physical. $D$ also couples $J$ to the EW gauge bosons, with mixed anomaly coefficients $C_{WW}=C_{BB}=-1/2$ for SU(2)$_L$ and U(1)$_Y$, such that $\mathcal{A}_\gamma=C_{WW}+C_{BB}$~\footnote{The SM leptons also contribute to the SU(2)$_L$ and U(1)$_Y$ anomalies of U(1)$_L$. These contributions equal those of baryon number and are removed by $L\to L-B$~\cite{Herrero-Brocal:2026nmc}.}. In particular, the triangle loops of $D$ induce the effective interaction
\begin{align}
\mathcal{L}_{JWW}=\frac{g^2}{16\pi^2 v_\sigma}\,C_{WW}\,J\,W^a_{\mu\nu}\widetilde{W}^{a\,\mu\nu}\supset\frac{g^2}{4\pi^2 v_\sigma}\,C_{WW}\,J\,\epsilon^{\mu\nu\rho\sigma}(\partial_\mu W^+_\nu)(\partial_\rho W^-_\sigma) \; ,
\label{eq:LJWW}
\end{align}
where $g$ is the SU(2)$_L$ gauge coupling. A further $W$-boson loop turns $JW^+W^-$ into flavour-changing $b\to sJ$ transitions, a two-loop effect in the full theory~\cite{Izaguirre:2016dfi,Bauer:2017ris,Bauer:2021mvw}. For our benchmark $m_J=1~\GeV$ [see upcoming discussion below Eq.~\eqref{eq:mJ}] the kaon decays $K\to\pi J$ are kinematically closed, while $B\to K^{(*)}J$ are open. For $v_\sigma/\sqrt{2}=5~\TeV$ and $C_{WW}=-1/2$ the predicted branching ratios~(BRs) are of $\mathcal{O}(10^{-9})$~\cite{Izaguirre:2016dfi,Bauer:2021mvw}, several orders of magnitude below current BaBar~\cite{BaBar:2013npw} and Belle~II~\cite{Belle-II:2023esi} bounds.

\begin{figure}[t!]
    \centering
    \includegraphics[width=\textwidth]{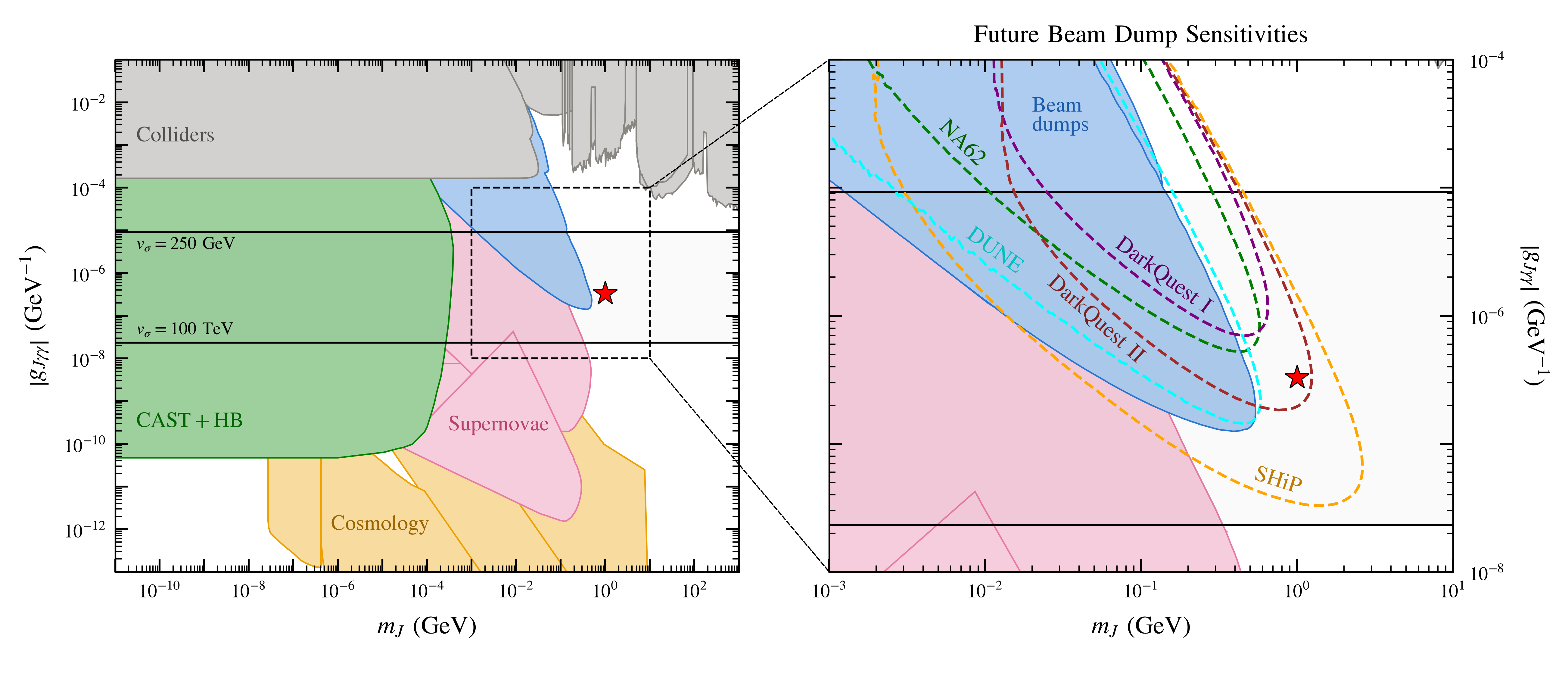}
    \caption{Majoron-to-photon coupling $|g_{J\gamma\gamma}|$ versus the Majoron mass $m_J$. The band between the two black lines is the prediction of Eq.~\eqref{eq:gJgg}, for $250~\GeV\leq v_\sigma\leq100~\TeV$, and the red star marks our benchmark $v_\sigma/\sqrt{2}=5$~TeV and $m_J=1$~GeV. The left panel shows the current bounds from CAST and HB stars (green), supernovae (SNe) (pink), beam dumps (blue), cosmology (orange) and colliders (grey), as compiled in Ref.~\cite{AxionLimits}. Bounds that assume the pseudoscalar to be the DM are not shown. The right panel zooms into the region marked by the dashed rectangle and shows, besides the current beam-dump bounds, as dashed lines the projected sensitivities for NA62 (green), DarkQuest Phase~I (purple) and II (brown), DUNE (cyan) and SHiP (orange), obtained with ALPINIST~\cite{Jerhot:2022chi}.}
    \label{fig:Anomalous_Majoron_photon}
\end{figure}

As mentioned before, the \emph{dark} linear seesaw model, is a low-scale scenario where the masses from the heavy neutrinos and \emph{dark}-sector particles are naturally taken at the TeV. Thus, leading to a low scale for the spontaneous breaking of U(1)$_L$ set by the $\sigma$ VEV. However, for our benchmark $v_\sigma/\sqrt{2}=5$~TeV, Eq.~\eqref{eq:gJgg} gives a Majoron-to-photon coupling of $|g_{J\gamma\gamma}|\simeq3.3\times10^{-7}~\GeV^{-1}$. A massless Majoron with such a coupling is excluded as shown in Fig.~\ref{fig:Anomalous_Majoron_photon}. Namely, the CAST helioscope sets $|g_{J\gamma\gamma}|<5.8\times10^{-11}~\GeV^{-1}$~\cite{CAST:2024eil}, observations of horizontal-branch~(HB) stars require $|g_{J\gamma\gamma}|<4.7\times10^{-11}~\GeV^{-1}$~\cite{Ayala:2014pea,Dolan:2022kul}, and SN~1987A implies $|g_{J\gamma\gamma}|\lesssim5\times10^{-12}~\GeV^{-1}$~\cite{Payez:2014xsa,Hoof:2022xbe}. Evading the HB bound alone would require $v_\sigma\gtrsim 10^{8}$~GeV, far above the TeV scale. Hence, we consider the possibility that the Majoron is massive. We assume that U(1)$_L$ is also broken explicitly, though softly, by the following linear term~\cite{Barger:2008jx,Gonderinger:2012rd,Chiang:2017nmu,Alanne:2020jwx,Bodas:2020yho,Coito:2021fgo,Kim:2024tnn,Chiang:2026ayw}
\begin{align}
V_{\text{soft}} = -\frac{\mu_J^3}{\sqrt{2}}\left(\sigma+\sigma^\ast\right) \;\Rightarrow\; m_J^2=\frac{\mu_J^3}{v_\sigma} \; ,
\label{eq:mJ}
\end{align}
with the Majoron mass $m_J$ being calculable from the single parameter $\mu_J>0$. Since lepton number is restored for $\mu_J\to0$, $\mu_J$ is naturally small in the 't~Hooft sense~\cite{tHooft:1979rat}. The full scalar sector, including the effect of the above soft breaking, is presented in Appendix~\ref{sec:scalarsector}. From now on, we take the Majoron mass to be $m_J=1$ GeV, which for $v_\sigma/\sqrt{2}=5$~TeV corresponds to $\mu_J \simeq 19.2$~GeV. As shown in this section, this benchmark evades all current bounds, lies within reach of upcoming beam-dump experiments, and, as studied in Sec.~\ref{sec:cLFV}, allows for cLFV decays with Majoron emission.

We stress that the linear soft-breaking term is not only the simplest source of the Majoron mass but also essentially the only one without drawbacks. It preserves the residual $\mathcal{Z}_2$, under which the \emph{dark} fields are odd, and leaves the one-loop neutrino masses of Fig.~\ref{fig:neutrino} unaffected. Quantum gravity is also expected to break all global symmetries through Planck-suppressed operators~\cite{Giddings:1987cg,Kallosh:1995hi,Banks:2010zn,Kamionkowski:1992mf,Holman:1992us,Barr:1992qq}, whose effect on the Majoron was studied in Refs.~\cite{Akhmedov:1992hi,Rothstein:1992rh}. At our benchmark, however, the leading dimension-5 operator $\propto \sigma^5/\overline{M}_{\rm Pl}$, with the reduced Planck mass $\overline{M}_{\rm Pl}=2.44\times10^{18}$~GeV, only gives $m_J\simeq1$~MeV for an $\mathcal{O}(1)$ coefficient, inside the region excluded in Fig.~\ref{fig:Anomalous_Majoron_photon}. Soft-breaking terms with higher powers of $\sigma$ inevitably lead to the cosmological domain-wall~(DW) problem~\cite{Zeldovich:1974uw,Sikivie:1982qv,Vilenkin:1984ib}. For example, a quadratic term $\sigma^2+\sigma^{\ast2}$, preserves a $\mathcal{Z}_4$ subgroup of U(1)$_L$ under which $\sigma\to-\sigma$. Its spontaneous breaking by the $\sigma$ VEV then leaves two degenerate vacua separated by DWs. Bare $\Delta L=1$ masses for the \emph{dark} fermions or for $\overline{\nu_R^c}S_R$ also break U(1)$_L$ softly while preserving the residual $\mathcal{Z}_2$. However, fermion loops then regenerate the linear term of Eq.~\eqref{eq:mJ}~\cite{Khan:2026eqx}, and all Majoron couplings, including $g_{J\gamma\gamma}$, are rescaled by the fraction of the corresponding fermion mass that still stems from $v_\sigma$. Finally, the remaining soft terms, $\chi_1\chi_2$, $\overline{\nu_R^c}\nu_R$ and $\overline{S_R^c}S_R$, reintroduce explicit LNV into the neutrino masses, which then no longer arise solely from the \emph{dark} linear seesaw mechanism.

Since the Majoron couples to photons, it will decay mainly into two photons. The decay width is given by
\begin{align}
\Gamma(J\to\gamma\gamma)=\frac{g_{J\gamma\gamma}^2\,m_J^3}{64\pi} \; ,
\label{eq:GammaJ}
\end{align}
with the corresponding lifetime being
\begin{align}
\tau_J\simeq1.2\times10^{-9}~{\rm s}~\left(\frac{1~\GeV}{m_J}\right)^3\left(\frac{v_\sigma/\sqrt{2}}{5~\TeV}\right)^2 \; ,
\label{eq:tauJ}
\end{align}
which is clearly orders of magnitude shorter than the age of the Universe $t_0\simeq4.4\times10^{17}$~s~\cite{Planck:2018vyg}. Therefore, in our scenario the Majoron cannot be DM. Instead, the DM candidate corresponds to the lightest neutral \emph{dark} particle, a fermion or scalar WIMP stabilised by the residual $\mathcal{Z}_2$, as studied in Sec.~\ref{sec:DM}. Furthermore, the Majoron also decays long before Big-Bang nucleosynthesis~(BBN). In the early Universe, decays and inverse decays $J\leftrightarrow\gamma\gamma$ keep it in thermal equilibrium with the SM plasma, since at $T\simeq m_J$ they are much faster than the Hubble rate $H=1.66\sqrt{g_\ast}\,T^2/M_{\rm Pl}$:
\begin{align}
\left.\frac{\Gamma(J\to\gamma\gamma)}{H}\right|_{T=m_J}=\frac{g_{J\gamma\gamma}^2\,m_J\,M_P}{64\pi\times1.66\sqrt{g_\ast}}\simeq5\times10^{2}\left(\frac{m_J}{1~\GeV}\right)\left(\frac{5~\TeV}{v_\sigma/\sqrt{2}}\right)^2 \; ,
\label{eq:GammaH}
\end{align}
where $M_{\rm Pl}=1.22\times10^{19}$~GeV is the Planck mass and $g_\ast\simeq66$ is the number of relativistic degrees of freedom at the temperature $T=1$~GeV~\cite{Saikawa:2018rcs}. The Majoron abundance therefore follows its equilibrium value, which is Boltzmann suppressed by $e^{-m_J/T}$ for $T<m_J$, and no relic population survives. At neutrino decoupling, $T\simeq2$~MeV, this suppression reaches $10^{-217}$. The decay photons thus thermalise while neutrinos are still in equilibrium, giving $\Delta N_{\rm eff}=0$, and no Majorons are left when BBN begins at $t\sim1$~s~\cite{Fields:2019pfx}. BBN bounds on pseudoscalars heavier than $0.1$~GeV only apply to $\tau\gtrsim0.1$~s~\cite{Depta:2020wmr}. For the same reason, our Majoron evades X-ray line searches~\cite{Foster:2021ngm,Roach:2022lgo,Calore:2022pks} and the diffuse X-ray background and the extragalactic background light~\cite{Cadamuro:2011fd}.

In Fig.~\ref{fig:Anomalous_Majoron_photon} we show the Majoron-to-photon coupling $|g_{J\gamma\gamma}|$ as a function of the Majoron mass $m_J$. The shaded-band between the two black lines is the prediction of Eq.~\eqref{eq:gJgg}, for $250~\GeV\leq v_\sigma\leq100~\TeV$, lying within the range $2 \times 10^{-8}~\text{GeV}^{-1}\lesssim |g_{J \gamma \gamma}| \lesssim10^{-5}~\text{GeV}^{-1}$. The red star marks our benchmark, $v_\sigma/\sqrt{2}=5$~TeV and $m_J=1$~GeV, where $|g_{J\gamma\gamma}|\simeq3.3\times10^{-7}~\GeV^{-1}$. The left panel shows the current bounds on pseudoscalars coupled to photons, as compiled in Ref.~\cite{AxionLimits}, without the bounds that assume the pseudoscalar to be the DM. Within the band, the lightest Majorons up to $m_J \lesssim 0.1$ MeV, are excluded by CAST~\cite{CAST:2024eil} and HB stars~\cite{Ayala:2014pea,Dolan:2022kul} (green). Heavier Majorons $0.1~\text{MeV}~\lesssim m_J \lesssim 0.5$ GeV are excluded by SNe (pink)~\cite{Jaeckel:2017tud,Lucente:2020whw,Caputo:2022mah,Hoof:2022xbe,Muller:2023vjm}, for couplings below $|g_{J \gamma \gamma}| \lesssim~2 \times 10^{-7}~\text{GeV}^{-1}$, and by beam dumps (blue)~\cite{CHARM:1985anb,Riordan:1987aw,Bjorken:1988as,Blumlein:1990ay,Dolan:2017osp,NA64:2020qwq}, for couplings above. Cosmological bounds (orange)~\cite{Cadamuro:2011fd,Depta:2020wmr}, are only relevant for couplings below the band $|g_{J \gamma\gamma}| \lesssim 10^{-10}$~GeV$^{-1}$ and masses $m_J\lesssim20$~MeV. Collider searches (grey)~\cite{Jaeckel:2015jla,Mariotti:2017vtv,Knapen:2016moh,CMS:2018erd,Aloni:2019ruo,ATLAS:2020hii,Belle-II:2020jti,BESIII:2022rzz,BESIII:2024hdv}, lie above the band, reaching $|g_{J\gamma\gamma}|\gtrsim10^{-4}~\GeV^{-1}$ for $0.1~\GeV\lesssim m_J\lesssim10~\GeV$. The right panel zooms into the region $10^{-3}~\GeV\leq m_J\leq10~\GeV$, marked by the dashed rectangle in the left panel. With the lifetime of Eq.~\eqref{eq:tauJ}, a Majoron of momentum $p_J$ travels on average a distance
\begin{align}
L_J=\frac{p_J}{m_J}\,c\,\tau_J\simeq0.37~{\rm m}\left(\frac{p_J}{1~\GeV}\right)\left(\frac{1~\GeV}{m_J}\right)^4\left(\frac{v_\sigma/\sqrt{2}}{5~\TeV}\right)^2 \; ,
\label{eq:LJ}
\end{align}
before decaying. For the momenta of tens of GeV reached at proton beam dumps, $L_J$ is of order $10$~m at our benchmark. Majorons in this mass range are therefore prime targets for beam-dump experiments. Produced in the target through their photon coupling, e.g. via the Primakoff process~\cite{Dobrich:2015jyk}, they cross the shielding and decay into photons in a detector a few to hundreds of meters downstream. The dashed-coloured lines show the sensitivities projected with ALPINIST~\cite{Jerhot:2022chi}, for NA62 (green), DarkQuest Phase~I (purple) and II (brown), DUNE (cyan) and SHiP (orange). DarkQuest~\cite{Berlin:2018pwi,Blinov:2021say} can probe our benchmark in Phase~II with $10^{20}$ protons on target, and SHiP~\cite{Alekhin:2015byh,Climescu:2026omx}, approved in 2024 for the ECN3 facility of the CERN SPS, probe it with $6\times10^{20}$ protons on target.

\subsection{Collider constraints}
\label{sec:collider}

The portal coupling $\lambda_{\Phi\sigma}$ in the scalar potential of Eq.~\eqref{eq:Vpotentialfull} in Appendix~\ref{sec:scalarsector} mixes the CP-even neutral component of the SM Higgs doublet $\phi_{\text{R}}$, with that of the singlet $\sigma_{\text{R}}$. The mixing angle $\theta$ is defined in Eq.~\eqref{eq:mixingnondark}, and the mass eigenstates are
\begin{align}
h = \cos\theta\,\phi_{\text{R}} - \sin\theta\,\sigma_{\text{R}} \;, \quad
H = \sin\theta\,\phi_{\text{R}} + \cos\theta\,\sigma_{\text{R}} \;,
\label{eq:hH_mixing}
\end{align}
where $h$ is the SM-like state with mass $m_h=125.13$~GeV~\cite{ParticleDataGroup:2026mpi}, and $H$ is the heavy neutral scalar, for which we take $m_H=2$~TeV. The $\Phi - \sigma$ mixing rescales all couplings of $h$ to SM fermions and gauge bosons by $\cos\theta$. It also induces an $hJJ$ coupling proportional to $\sin\theta$, which opens the invisible decay $h\to JJ$. Both effects are probed at colliders, and we begin our phenomenological analysis by reviewing the relevant constraints on the parameter space of the model:
\begin{itemize}

\item \textbf{Higgs invisible decay:} In dynamical low-scale seesaw scenarios, the SM Higgs boson can have sizeable invisible decays into Majorons, since the scale of spontaneous LNV is typically at the TeV. In our model, the $\sigma$ VEV sets this scale, which as a benchmark we take $v_\sigma/\sqrt{2} = 5$ TeV. For a Majoron whose mass $m_J$ arises from the linear soft term of Eq.~\eqref{eq:mJ} (see Appendix~\ref{sec:scalarsector}), the decay width for $h \rightarrow JJ$ is~\cite{Abe:2021nih}
\begin{align}
\Gamma(h\to JJ)=\frac{1}{32\pi} \frac{m_h^3} {v_\sigma^2} \sin^2\theta \left(1-\frac{m_J^2}{m_h^2}\right)^{2}\left(1-\frac{4m_J^2}{m_h^2}\right)^{1/2} \; ,
\label{eq:inv_Majoron}
\end{align}
which reduces to the well-known result for a massless Majoron~\cite{Joshipura:1992hp,Joshipura:1992ua} and depends on the $\theta$ angle and $v_\sigma$. Furthermore, if kinematically allowed $h$ can decay into the \emph{dark}-neutral fermions $\psi_a^0$ ($a=1,2$) and scalars $\zeta_k$ ($k=1,\cdots,4$). In our model the total invisible width of $h$ is
\begin{align}
\Gamma^{\rm inv}(h)=\Gamma(h\to JJ) +\sum_{a,b}\Theta\big(m_h-m_{\psi^0_a} - m_{\psi^0_b}\big)\,\Gamma(h\to\psi^0_a\overline{\psi^0_b})+\sum_{k\leq j}\Theta\big(m_h-m_{\zeta_k}-m_{\zeta_j}\big)\,\Gamma(h\to\zeta_k\zeta_j)\;,
\label{eq:inv_Higgs_total}
\end{align}
with its BR being
\begin{align}
\BR^{\rm inv}(h)=\frac{\Gamma^{\rm inv}(h)}{\cos^2\theta\,\Gamma^{\text{SM}}(h)+\Gamma^{\rm inv}(h)}\;,
\label{eq:inv_BR_Higgs}
\end{align}
where $\Gamma^{\text{SM}}(h)=4.07$ MeV~\cite{ParticleDataGroup:2026mpi}. Large Hadron Collider~(LHC) Higgs data sets a current 95\%~CL upper limit of~\cite{ParticleDataGroup:2026mpi}
\begin{align}
\BR^{\rm inv}(h)<0.107\; .
\label{eq:invisible_bound}
\end{align}
If only the Majoron channel is open, and since $m_J\ll m_h$, we obtain the following bound
\begin{align}
\frac{v_\sigma}{|\tan\theta|}\;\gtrsim\;6.3~\TeV
\Rightarrow \theta\ll1 :
|\sin\theta|\;\lesssim\;0.23\,\left(\frac{v_\sigma/\sqrt{2}}{1~\TeV}\right) \;.
\label{eq:invisible_Majoron_only}
\end{align}
Hence, by taking $|\sin \theta| \in [10^{-3},0.1]$ and $v_\sigma/\sqrt{2} = 5$ TeV, we safely satisfy this constraint. Albeit, if kinematically allowed, the Higgs invisible decay will still constrain the coupling between the Higgs boson and the \emph{dark}-neutral fermion and scalars.

\item \textbf{Higgs signal strengths:} Due to $\Phi-\sigma$ mixing, as shown in Eq.~\eqref{eq:hH_mixing}, all tree-level couplings of $h$ to SM fermions and gauge bosons are rescaled by $\cos \theta$. Hence, the production cross sections of $h$ will be modified as follows
\begin{align}
\sigma(pp\to h) = \cos^2\theta\;\sigma^{\text{SM}}(pp\to h)\;,
\label{eq:xsec}
\end{align}
where $\sigma^{\text{SM}}(pp\to h)$ denotes the SM cross sections for the Higgs boson. Consequently, this will modify the Higgs signal strengths, defined as
\begin{align}
\mu_{XX}=\frac{\sigma(pp\to h)\,\BR(h\to XX)}
{\sigma^{\text{SM}}(pp\to h)\,\BR^{\text{SM}}(h\to XX)}\;,
\label{eq:muf-def}
\end{align}
that quantifies the deviations from the SM predictions and $XX$ denotes the final states. Due to the invisible Higgs decays of
Eq.~\eqref{eq:inv_Higgs_total}, the BRs becomes
\begin{align}
\BR(h\to XX)=
\frac{\cos^2\theta\,\Gamma^{\text{SM}}_{XX}(h)}{\cos^2\theta\,\Gamma^{\text{SM}}(h)+\Gamma^{\rm inv}(h)}\;,
\label{eq:BR_visible}
\end{align}
so that, using Eq.~\eqref{eq:inv_BR_Higgs}, we obtain
\begin{align}
\mu_{XX}=\cos^2\theta\,\left[1-\BR^{\rm inv}(h)\right]\;.
\label{eq:muXX}
\end{align}
The LHC constrains the final states $XX=\gamma \gamma,ZZ,WW,\tau\tau,b\bar{b}$, with the averages being $\mu_{\gamma \gamma}=1.10\pm0.06$, $\mu_{ZZ}=1.02\pm0.08$, $\mu_{WW}=1.00\pm0.08$, $\mu_{\tau\tau}=0.91\pm0.09$ and $\mu_{b\bar{b}}=0.94\pm0.11$~\cite{ParticleDataGroup:2026mpi}, dominated by the full Run-2 ATLAS and CMS data~\cite{ATLAS:2022vkf,CMS:2022dwd}. Combining all final states gives the world average $\mu=1.03\pm0.04$~\cite{ParticleDataGroup:2026mpi}, consistent with the latest ATLAS and CMS global fits, $\mu=1.023^{+0.056}_{-0.053}$~\cite{ATLAS:2025qxq} and $\mu=1.014^{+0.055}_{-0.053}$~\cite{CMS:2026zxu}, respectively. We therefore require
\begin{align}
\mu\;=\;\cos^2\theta\,\left[1-\BR^{\rm inv}(h)\right]\;\geq\;0.95\;,
\label{eq:mu_bound}
\end{align}
i.e. two standard deviations below the world average. For a negligible invisible width this gives $|\sin\theta|\lesssim0.22$. Any additional invisible decay channel into the \emph{dark}-neutral fermions and/or scalars tightens the bound. In particular, Eq.~\eqref{eq:mu_bound} alone requires $\BR^{\rm inv}(h)\leq0.05$, which is stronger than the direct limit $\BR^{\rm inv}(h)<0.107$ of Eq.~\eqref{eq:invisible_bound}. In our scan we take $|\sin\theta|\in[10^{-3},0.1]$, so that Eq.~\eqref{eq:mu_bound} holds whenever $\BR^{\rm inv}(h)\lesssim0.04$.

\item \textbf{Higgs to diphoton decay:} Note that, the Higgs decay into photons, i.e. $XX = \gamma \gamma$, is a radiative process which, besides the usual $W$-boson and fermions loops, in our model receives an additional contribution mediated by the \emph{dark}-charged fermion $\psi^{\pm}$. In fact, the coupling of the SM-like Higgs and the \emph{dark}-charged fermion is, 
\begin{equation}
g_{h\psi\psi}= \frac{m_{\psi^\pm}}{v_\Phi} \frac{v_\Phi}{v_\sigma} \sin\theta \; ,
\end{equation}
with the $\psi^\pm$ mass given in Eq.~\eqref{eq:chargedfemionmass}. At one-loop the decay width for $h \rightarrow \gamma \gamma$ is obtained as~\cite{Djouadi:2005gj,Posch:2010hx,Fontes:2014xva,Aguilar-Saavedra:2023tql}
\begin{align}
\Gamma (h \rightarrow \gamma \gamma) = \frac{G_F \alpha^2 m_h^3}{128 \sqrt{2} \pi^3} ~\left|\cos\theta A_{\text{SM}} + \sin\theta~\frac{v_\Phi}{v_\sigma}~A_{\psi^\pm} \right|^2 \; ,
\end{align}
with the form factors being
\begin{align}
A_{\text{SM}} &= \frac{\left[ 3 \tau_W + 2 \tau_W^2 + 3 \left(2 \tau_W -1\right) g(\tau_W) \right]}{\tau_W^{2}} - \sum_f N_c^f Q_f^2 \frac{2 \left[\tau_f + \left(\tau_f - 1 \right) g(\tau_f) \right]}{\tau_f^{2}} \; , \nonumber \\
A_{\psi^\pm} &= \frac{2 \left[\tau_{\psi^\pm} + \left(\tau_{\psi^\pm} - 1 \right) g(\tau_{\psi^\pm}) \right]}{\tau_{\psi^\pm}^{2}} \; ,
\end{align}
where $\tau_X=m_h^2/4 m_X^2$, $N_c^f$ and $Q_f$ are respectively the color factor and electric charge of the SM fermions $f=e,d,u$. The loop function is~\cite{Djouadi:2005gj,Posch:2010hx,Fontes:2014xva,Aguilar-Saavedra:2023tql}
\begin{eqnarray}
  g(\tau) =\left\{\begin{array}{ll} \displaystyle
    \arcsin^2\sqrt{\tau} & \ \text{for} \ \tau\leq 1 \\
    \displaystyle -\frac{1}{4}\left[ \log\frac{1+\sqrt{1-\tau^{-1}}}
    {1-\sqrt{1-\tau^{-1}}}-i\pi \right]^2 \hspace{0.5cm} & \ \text{for} \ \tau>1
  \end{array} \right. \; .
\end{eqnarray}
From the above, we write the Higgs to diphoton signal strength as [see Eq.~\eqref{eq:muXX}]:
\begin{align}
\mu_{\gamma\gamma}=\cos^2\theta\,\left[1-\BR^{\rm inv}(h)\right]\,
\left|1+ \tan \theta~\frac{v_\Phi}{v_\sigma}~\frac{A_{\psi^\pm}}{A_{\text{SM}}} \right|^2\; ,
\label{eq:mugammagamma}
\end{align}
to be compared with the measured average~\cite{ParticleDataGroup:2026mpi}
\begin{equation}
\mu_{\gamma\gamma}=1.10\pm 0.06 \,.
\label{eq:mugammagamma_exp}
\end{equation}
The $\psi^\pm$ contribution in Eq.~\eqref{eq:mugammagamma} is controlled by the same combination $\tan\theta/v_\sigma$ that governs $h\to JJ$. The invisible-decay bound of Eq.~\eqref{eq:invisible_Majoron_only} thus implies $v_\Phi\tan\theta/v_\sigma\lesssim0.039$, so that the $\psi^\pm$ loop shifts $\mu_{\gamma\gamma}$ by at most $1.8\%$ for any $m_{\psi^\pm}\gtrsim100~\GeV$ allowed by the bound of Eq.~\eqref{eq:mpsipm_LEPbound}. Given the $6\%$ precision of Eq.~\eqref{eq:mugammagamma_exp}, the diphoton rate follows the universal rescaling of Eq.~\eqref{eq:muXX} and provides no independent handle on the \emph{dark} sector. This is in sharp contrast with inert-doublet~\cite{Belyaev:2016lok} and scotogenic models~\cite{Barreiros:2022aqu}, where the \emph{dark}-charged scalar couples to $h$ through an unsuppressed quartic coupling and $\mu_{\gamma\gamma}$ is an important constraint. Here, by contrast, $\psi^\pm$ couples to the Higgs only through $\Phi-\sigma$ mixing, since its mass is generated dynamically by the $\sigma$ VEV once lepton number is spontaneously broken. As we showed, this mixing is already constrained by invisible Higgs decays into Majorons and by the Higgs signal strengths.

\item \textbf{LEP constraints:} The masses of the \emph{dark} fermions of our model are constrained by Large Electron-Positron~(LEP) collider data. Note that, the Majoron $J$ being the phase component of the singlet $\sigma$, as well as the \emph{dark}-neutral scalar $\zeta_k$ ($k=1,\cdots, 4$) being the mass-eigenstates of the singlets $\chi_{1,2}$, are not subject to the stringent constraints set by the precise LEP-I measurements on the $Z$-boson decay width, and thus can have masses below $m_Z/2$. In fact, this is precisely what distinguishes the singlet Majoron~\cite{Chikashige:1980ui} from the triplet Majoron~\cite{Gelmini:1980re}, which is essentially ruled out by the $Z$-boson decay width measurement. The \emph{dark}-neutral fermions, on the other hand, couple to the $Z$-boson through their doublet component. Although the singlet $F_{L,R}$ has no gauge interactions, it mixes with the neutral component of the doublet $D_{L,R}$ via the Yukawa coupling $y_{DF}$ of Eq.~\eqref{eq:LYuk}, so that both mass eigenstates $\psi^0_{1,2}$ contribute to the invisible $Z$ width. A neutral \emph{dark}-fermion lighter than $m_Z/2$ is therefore allowed only if it is almost a pure singlet, with a doublet admixture of at most a few per cent, which translates into a constraint on the doublet-singlet mixing. In our analysis we conservatively require
\begin{align}
m_{\psi^0_{1,2}}\;> \frac{m_Z}{2} \simeq \;45.6~\GeV\;.
\label{eq:mpsi0_LEPbound}
\end{align}
Furthermore, LEP-II searches for pair-produced heavy charged leptons and charginos essentially exclude \emph{dark}-charged fermion masses below~\cite{L3:2001xsz}
\begin{align}
m_{\psi^\pm}\;\gtrsim\;100~\GeV\;.
\label{eq:mpsipm_LEPbound}
\end{align}

\end{itemize}

\newpage

\subsection{Charged lepton flavour violation}
\label{sec:cLFV}

Searches for lepton flavour violating processes such as $\ell_\alpha\to\ell_\beta\gamma$, $\ell_\alpha\to3\ell_\beta$ and $\mu\to e$ conversion in nuclei are important probes of physics beyond the SM. They become particularly relevant in low-scale seesaw scenarios, where the heavy-light neutrino mixing can be sizeable inducing non-negligible contributions to cLFV through the $W$-boson loops~\cite{Minkowski:1977sc,Marciano:1977wx,Cheng:1980tp,Lim:1981kv,Langacker:1988up,Ilakovac:1994kj,Alonso:2012ji,Abada:2015oba,Abada:2018nio,Hernandez-Tome:2019lkb,Camara:2020efq}. In our framework there are, in addition, one-loop contributions mediated by the \emph{dark}-sector particles $\psi^\pm$ and $\zeta_k$ ($k=1,\dots,4$), of the type encountered in singlet-doublet fermion DM models with radiative neutrino masses~\cite{Esch:2018ccs,Darricau:2025vcs,Darricau:2026iwg}. Here we focus on the radiative decays $\ell_\alpha\to\ell_\beta\gamma$, whose BRs are:
\begin{align}
\frac{\BR(\ell_{\alpha} \rightarrow \ell_{\beta} \gamma)}{\BR\left(\ell_{\alpha} \rightarrow \ell_{\beta} \nu_{\alpha} \overline{\nu_{\beta}} \right)} & = \frac{3 \alpha_{e}}{2 \pi} \left| G_{W^\pm}^{\alpha \beta}+ G_{\psi^{\pm}}^{\alpha \beta}\right|^2 \nonumber \\
&= \frac{3 \alpha_{e}}{2 \pi} \left|\sum_{i=1}^5 \mathbf{K}_{\alpha i}^{*} \mathbf{K}_{\beta i} G_{W^\pm}\left(\frac{m_{i}^2}{m_W^2}\right) - \mathbf{Y}_{\chi}^{\alpha \ast} \mathbf{Y}_{\chi}^{\beta} \sum_{k=1}^4 \frac{\left|\mathbf{V}_{1 k} + i \mathbf{V}_{3 k} \right|^2}{2} \frac{v_\Phi^2}{4 m_{\zeta_k}^2} G_{\psi^{\pm}}\left(\frac{m_{\psi^{\pm}}^2}{m_{\zeta_k}^2}\right)\right|^2 \; ,
\label{eq:cLFV}
\end{align}
where $\alpha_e=e^2/4\pi$, $\BR(\mu\to e\nu_\mu\overline{\nu_e})\simeq1.0$, $\BR(\tau\to e\nu_\tau\overline{\nu_e})\simeq0.18$ and $\BR(\tau\to\mu\nu_\tau\overline{\nu_\mu})\simeq0.17$~\cite{ParticleDataGroup:2026mpi}. The mixing matrix $\mathbf{K}$ and neutrino masses $m_{i}$ are given by Eqs.~\eqref{eq:K} and~\eqref{eq:Unudef}, respectively. The Yukawa coupling $\mathbf{Y}_\chi$ is defined in Eq.~\eqref{eq:LYuk}, the \emph{dark} scalar mixing matrix $\mathbf{V}$ and masses $m_{\zeta_k}$ are given in Eqs.~\eqref{eq:mixingdark} and~\eqref{eq:neutraldarkscalarmass}, respectively, and the \emph{dark} charged-fermion mass $m_{\psi^{\pm}}$ is given by Eq.~\eqref{eq:chargedfemionmass}. Furthermore, the above loop functions $G_{W^\pm}(x)$~\cite{Cheng:1980tp,Lim:1981kv,Langacker:1988up,Ilakovac:1994kj} and $G_{\psi^{\pm}}(x)$~\cite{Darricau:2025vcs,Darricau:2026iwg} are given by
\begin{align}
G_{W^\pm}(x) = \frac{x (1-5x-2x^2)}{4 (1-x)^3} - \frac{3 x^3}{2 (1-x)^4} \ln x \; ,  \; G_{\psi^{\pm}}(x) = \frac{2 + 3 x - 6 x^2 + x^3 + 6 x \ln x}{6(1-x)^4} \; .
\label{eq:loopfunctionmeg}
\end{align}
Table~\ref{tab:cLFV} summarises the current limits and the future sensitivities for the cLFV processes relevant to our analysis. The most stringent constraint on $\ell_\alpha\to\ell_\beta\gamma$ processes is set by the MEG~II collaboration~\cite{MEGII:2018kmf}, for $\mu\to e\gamma$. For the $\tau$ processes the current best limits are set by BaBar~\cite{BaBar:2009hkt} and Belle~\cite{Belle:2021ysv}, with projected sensitivities from Belle~II~\cite{Belle-II:2018jsg}.

\begin{table}[t]
\centering
\renewcommand{\arraystretch}{1.3}
\begin{tabular}{|c|c|c|}
\hline
Process & Current limit & Future sensitivity \\
\hline \hline
$\mu\to e\gamma$ & $<1.5\times10^{-13}$ (MEG~II)~\cite{MEGII:2025gzr} & $<6\times10^{-14}$ (MEG~II)~\cite{MEGII:2018kmf,MEGII:2025gzr} \\
$\tau\to e\gamma$ & $<3.3\times10^{-8}$ (BaBar)~\cite{BaBar:2009hkt} & $<1.2\times10^{-8}$ (Belle~II)~\cite{Belle-II:2018jsg} \\
$\tau\to\mu\gamma$ & $<4.2\times10^{-8}$ (Belle)~\cite{Belle:2021ysv} & $<5\times10^{-9}$ (Belle~II)~\cite{Belle-II:2018jsg} \\
\hline \hline
$\mu\to eJ$ & $\lesssim10^{-5}$ (TRIUMF)~\cite{Jodidio:1986mz,Hirsch:2009ee} & \begin{tabular}{@{}c@{}} $<7.3\times10^{-8}$ (Mu3e)~\cite{Calibbi:2020jvd} \\ $\lesssim 10^{-8}$ (COMET)~\cite{COMET:2018auw,Xing:2022rob} \end{tabular} \\
$\tau\to eJ$ & $<6.4\times10^{-4}$ (Belle)~\cite{Belle:2025bpu} & $<8.3\times10^{-6}$ (Belle~II)~\cite{Calibbi:2020jvd} \\
$\tau\to\mu J$ & $<3.4\times10^{-4}$ (Belle)~\cite{Belle:2025bpu} & $<2.0\times10^{-5}$ (Belle~II)~\cite{Calibbi:2020jvd} \\
\hline
\end{tabular}
\caption{Current upper limits and future sensitivities on the BRs of the cLFV processes $\ell_\alpha \rightarrow \ell_\beta \gamma$ and $\ell_\alpha \rightarrow \ell_\beta J$. All values are at 90\% CL, except those for $\tau\to\ell J$ ($\ell = e, \mu$) which are at 95\% CL. The current $\tau\to\ell J$ bounds depend on the Majoron mass $0 \leq m_J \leq 1.6$ GeV~\cite{Belle:2025bpu}, we indicate the values for our benchmark $m_J=1$~GeV.}
\label{tab:cLFV}
\end{table}
Our model is based on a global U(1)$_L$ lepton number symmetry that is spontaneously broken by $\sigma$ whose phase is the Majoron $J$ -- see Table~\ref{tab:model}. In our model the Majoron has no tree-level coupling to charged leptons. However, at one-loop level will be generated the effective $J-\ell_\alpha-\ell_\beta$ interactions, being written as~\cite{Escribano:2020wua,Herrero-Brocal:2023czw}
\begin{align}
\mathcal{L}_{J\ell_\alpha\ell_\beta}=J\,\overline{\ell_\beta}
\big(S_L^{\beta\alpha}P_L+S_R^{\beta\alpha}P_R\big)\ell_\alpha+\text{H.c.}
=J\,\overline{\ell_\beta}\big[S^{\beta\alpha}P_L+(S^{\alpha\beta})^\ast P_R\big]\ell_\alpha\;,
\label{eq:MajoronInt}
\end{align}
with $P_{L,R}=(1\mp\gamma_5)/2$ and $S^{\beta\alpha}=S_L^{\beta\alpha} +(S_R^{\alpha\beta})^\ast$. Analogously to what happens to cLFV processes such as $\ell_\alpha \rightarrow \ell_\beta \gamma$, in our model the above effective interactions will receive two distinct contributions, from the seesaw and \emph{dark} sectors:
\begin{equation}
    S=S_{\rm seesaw}+S_{\rm \emph{dark}} \; .
\end{equation}

The Majoron couples directly to the heavy-neutrinos via $y_R~\overline{\nu_R^c} \sigma S_R$ in Eq.~\eqref{eq:LYuk}, with the small LNV~$\mathbf{m}_S$, triggering linear seesaw, being radiatively generated by the \emph{dark}-sector as shown in the left diagram of Fig.~\ref{fig:neutrino}. Thus, in contrast with the minimal dynamically generated seesaws, where $J-\ell_\alpha-\ell_\beta$ interactions are suppressed by the light neutrino masses~\cite{Heeck:2019guh,Herrero-Brocal:2023czw}, the seesaw sector of our model leads to an enhanced Majoron coupling to charged leptons~\cite{CentellesChulia:2024uzv,CentellesChulia:2025eck}. This coupling is generated via $W$ and $Z$-boson loops, through the heavy-light mixing entries of $\mathbf{K}$ given in Eq.~\eqref{eq:K}, which are controlled by $\boldsymbol{\epsilon}$. Applying the general results of Ref.~\cite{Herrero-Brocal:2023czw}, in the seesaw approximation $\mathbf{m}_S \ll \mathbf{M}_\nu \ll M_R$, we obtain
\begin{align}
\mathcal{L}^{\rm seesaw}_{J\ell\ell}
=\frac{i\,J}{16\pi^2 v_\sigma v_\Phi^2}\,\frac{M_R^0}{M_R}\;
\overline{\ell}\left[\mathbf{m}_\ell \text{Tr} \left(\mathbf{M}_\nu\mathbf{M}_\nu^\dagger\right)\gamma_5
+2\,\mathbf{m}_\ell\,\mathbf{M}_\nu\mathbf{M}_\nu^\dagger\,P_L
-2\,\mathbf{M}_\nu\mathbf{M}_\nu^\dagger\,\mathbf{m}_\ell\,P_R\right]\ell\;,
\label{eq:Sseesaw}
\end{align}
with $\mathbf{m}_\ell=\mathrm{diag}(m_e,m_\mu,m_\tau)$, $M_R^0$ and $\mathbf{M}_\nu$ are given in Eq.~\eqref{eq:treemassmatrix}, and $M_R$ is given by Eqs.~\eqref{eq:loopmassmatrix} and~\eqref{eq:mSandDeltaMR}. In the notation of Eq.~\eqref{eq:MajoronInt}, and in terms of $\boldsymbol{\epsilon}$ defined in Eq.~\eqref{eq:K}, we have
\begin{align}
S^{\beta\alpha}_{L,\rm seesaw}=\frac{i\,y_R}{8\sqrt2\,\pi^2v_\Phi^2} M_R~m_{\beta}
\left(\boldsymbol{\epsilon}_\beta\boldsymbol{\epsilon}_\alpha^\ast-\frac{1}{2}\delta_{\beta\alpha}|\boldsymbol{\epsilon}|^2\right)\;,\;
S^{\beta\alpha}_{R,\rm seesaw}=\left(S^{\alpha\beta}_{L,\rm seesaw}\right)^\ast\; .
\label{eq:Sseesawcouplings}
\end{align}

The \emph{dark}-sector also contributes to $J-\ell_\alpha-\ell_\beta$ in a similar fashion as in scotogenic models with spontaneous lepton number breaking~\cite{Escribano:2021ymx,Portillo-Sanchez:2023kbz,Herrero-Brocal:2023czw,Chun:2023vbh}. In those scenarios, \emph{dark}-charged scalars and \emph{dark}-neutral fermions generate $J-\ell_\alpha-\ell_\beta$ at one-loop, leading to an enchanced coupling when compared to the standard seesaw Majoron models~\cite{Escribano:2021ymx}. In our scenario, the loop responsible for generating $J-\ell_\alpha-\ell_\beta$ involves the \emph{dark}-charged fermion $\psi^\pm$ and the \emph{dark}-neutral scalars $\zeta_k$ ($k=1,\dots,4$). In fact, the Majoron couples directly to $\psi^\pm$ via the $y_D~\overline{D_L} \sigma^\ast D_R$ term in Eq.~\eqref{eq:LYuk}. One may wonder if there is an additional diagram involving the direct coupling between $J$ and $\zeta_k$, stemming from the $\kappa \ \chi_1 \chi_2 \sigma^\ast$ term in the potential~\eqref{eq:Vpotentialfull}. However, this contributions is exactly zero since the charged-leptons only couple to $\chi_1$ and not to $\chi_2$, as it is clear from the Yukawa Lagrangian~\eqref{eq:LYuk}. We obtain the following result:
\begin{align}
S^{\beta\alpha}_{L,\rm \emph{dark}}=-\frac{i}{16\pi^{2}v_{\sigma}} m_\beta \mathbf{Y}_{\chi}^\beta \mathbf{Y}^{\alpha \ast}_\chi \sum_{k=1}^4 \frac{\left|\mathbf{V}_{1 k} + i \mathbf{V}_{3 k} \right|^2}{2} g_{\psi^{\pm}}\left(\frac{m_{\psi^{\pm}}^2}{m_{\zeta_k}^2}\right) \; , \;
S^{\beta\alpha}_{R,\rm \emph{dark}}=\left(S^{\alpha\beta}_{L,\rm \emph{dark}}\right)^\ast\; ,
\label{eq:darkMajoroncoupling}
\end{align}
where the loop function is given by~\cite{Escribano:2021ymx,Herrero-Brocal:2023czw}
\begin{align}
g_{\psi^{\pm}}(x) = - \frac{x - x^2 + x\ln x}{(1-x)^2} \; .
\label{eq:loopfunctionmeJdark}
\end{align}

The off-diagonal Majoron couplings induce the flavour-violating decays $\ell_\alpha\to\ell_\beta J$, which are kinematically open for $m_J<m_\alpha-m_\beta$. In the limit $m_\beta \ll m_\alpha$, the BR normalised as in Eq.~\eqref{eq:cLFV}, is given by~\cite{Escribano:2020wua,Herrero-Brocal:2023czw}
\begin{align}
\frac{\BR(\ell_\alpha\to\ell_\beta J)}{\BR\left(\ell_\alpha\to\ell_\beta\nu_\alpha\overline{\nu_\beta}\right)}&=\frac{6\pi^2}{G_F^2\,m_\alpha^4}\left(\left|S^{\beta\alpha}_{L}\right|^{2}+\left|S^{\beta\alpha}_{R}\right|^{2}\right)\left(1-\frac{m_J^2}{m_\alpha^2}\right)^{2} \; ,
\label{eq:Wmue2J}
\end{align}
where $G_F$ is the Fermi constant. As shown in Table~\ref{tab:cLFV} the current most constrained $\ell_\alpha \to \ell_\beta J$ process is $\mu \to e J$. The $S^{\beta\alpha}$ couplings give rise to other muon cLFV process involving the Majoron such as $\mu \rightarrow e \gamma J$~\cite{Escribano:2020wua,Hirsch:2009ee,Jho:2022snj} and $\mu \rightarrow 3 e J$~\cite{Knapen:2023zgi}. The impact of generic ultralight scalars in leptonic observables has been studied in detail in Ref.~\cite{Escribano:2020wua}. For the benchmark $m_J=1$~GeV, $\mu\to eJ$ is kinematically forbidden. The relevant processes are therefore the $\tau$ decays~\cite{Hirsch:2009ee}, $\tau\to eJ$ and $\tau\to\mu J$, with their current limits and the future sensitivities given in Table~\ref{tab:cLFV}. For $\tau\to\ell J$ ($\ell = e,\mu$) we use the Belle limits at $m_J=1$~GeV~\cite{Belle:2025bpu}, and the Belle~II projections for $50~\mathrm{ab}^{-1}$ of Ref.~\cite{Calibbi:2020jvd}.

The diagonal Majoron couplings to charged leptons can be written as
\begin{align}
\mathcal{L}_{J\ell_{\alpha}\ell_{\alpha}}=-i J \overline{\ell_{\alpha}} \; \text{Im} S^{\alpha\alpha} \; \gamma_{5} \ell_{\alpha} \; ,
\label{eq:MajoronDiagonalCoupling}
\end{align}
being purely imaginary due to the pseudoscalar nature of $J$. Ultralight bosons such as axions and axion-like particles (ALPs), which include Majorons, can be copiously produced in stellar interiors~\cite{Raffelt:1994ry,DiLuzio:2020wdo}. Their emission drains energy from stellar cores and can modify the observed evolution of red giants, HB stars, white dwarfs (WDs) and SNe. For a massless or light Majoron, WD cooling strongly constrains the coupling to electrons~\cite{Calibbi:2020jvd} and SN~1987A constrains the coupling to muons~\cite{Bollig:2020xdr,Croon:2020lrf,Caputo:2021rux}. These flavour-conserving bounds are usually stronger than the flavour-violating ones. However, these rely on thermal production of the boson and do not apply to our massive Majoron with $m_J = 1$ GeV. In WDs, red giants and HB stars the core temperature is about $1$ to $10$~keV, and the bounds shut off for masses above about twice this value~\cite{Calibbi:2020jvd}. The proto-neutron star of SN~1987A reaches $\sim 30$~MeV~\cite{Calibbi:2020jvd}. Therefore, for our benchmark, stellar cooling does not constrain the diagonal Majoron couplings to electrons and muons.

\begin{figure}[t!]
  \centering
    \includegraphics[scale=0.4]{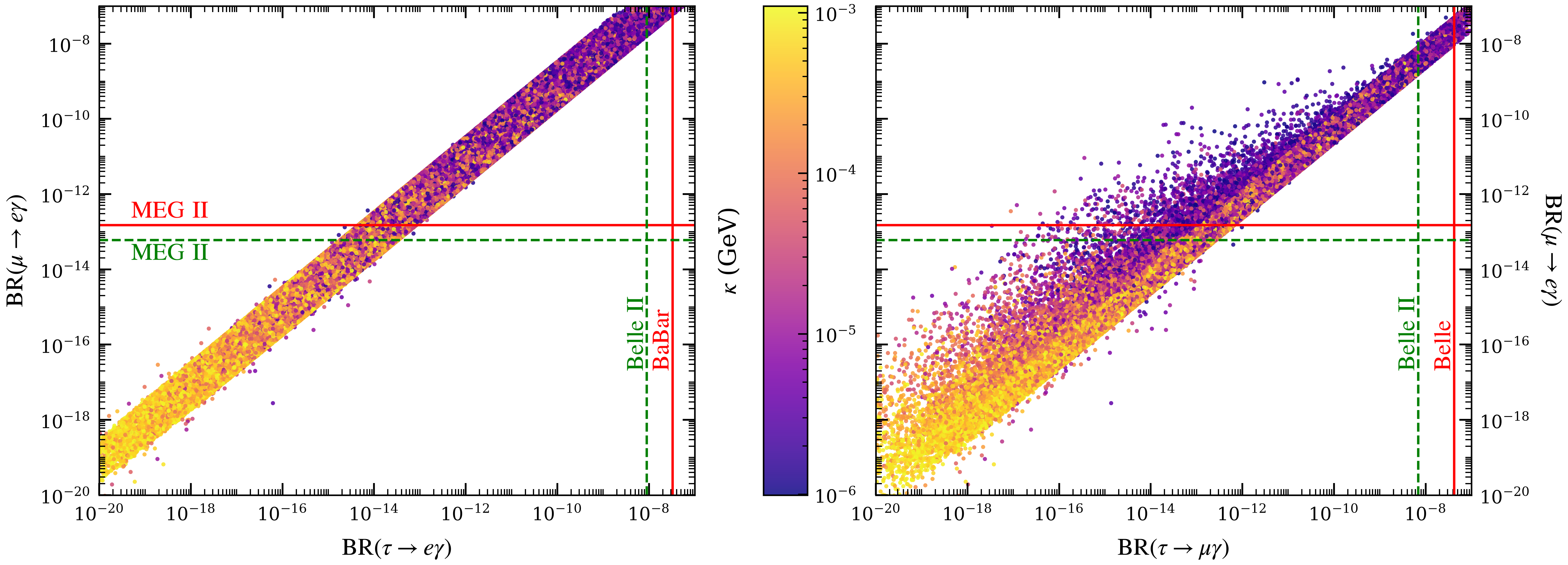} \\
    \includegraphics[scale=0.4]{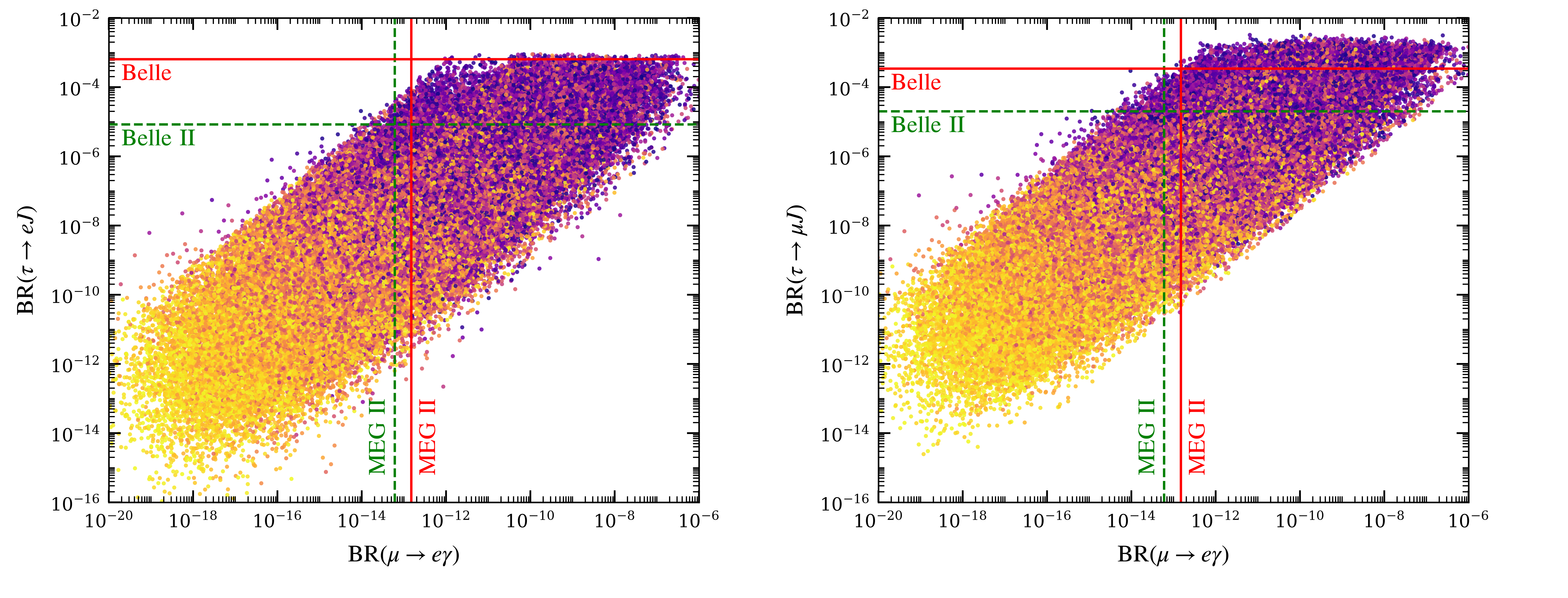} \\
    \includegraphics[scale=0.4]{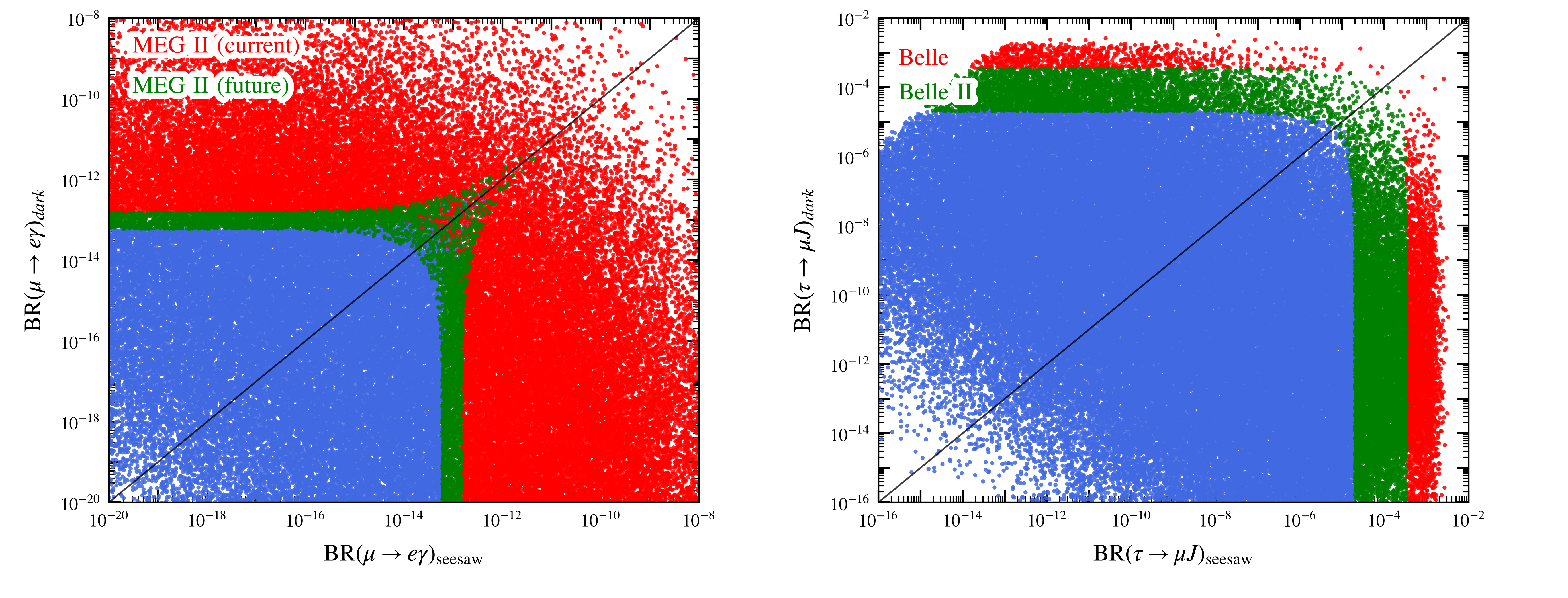} 
    \caption{Top: $\BR(\mu \rightarrow e \gamma)$ versus $\BR(\tau \rightarrow e \gamma)$ (left) and $\BR(\tau \rightarrow \mu \gamma)$ (right). Current cLFV bounds and projected sensitivities, given in Table~\ref{tab:cLFV}, are marked by red-solid and green-dashed lines, respectively. The color bar indicates~$\kappa$ values. Middle: $\BR(\tau \rightarrow e J)$ versus $\BR(\mu \rightarrow e \gamma)$ (left) and $\BR(\tau \rightarrow \mu J)$ versus $\BR(\mu \rightarrow e \gamma)$ (right). Bottom: $\BR(\mu \rightarrow e \gamma)_{\emph{dark}}$ versus $\BR(\mu \rightarrow e \gamma)_{\mathrm{seesaw}}$ (left) and $\BR(\tau \rightarrow \mu J)_{\emph{dark}}$ versus $\BR(\tau \rightarrow \mu J)_{\mathrm{seesaw}}$ (right). Along the black oblique line the \emph{dark} and seesaw-sector contributions are equal. The red points are excluded by current cLFV bounds, the green ones can be probed in the future, and blue points lie beyond projected sensitivities.}
  \label{fig:cLFV}
\end{figure}

Our numerical results presented in Fig.~\ref{fig:cLFV}, deserve several comments:
\begin{itemize}

    \item \textbf{Muon versus tau:} The top panels show $\BR(\mu \to e \gamma)$ versus $\BR(\tau \to e \gamma)$ (left) and $\BR(\tau \to \mu \gamma)$ (right), with the current bounds and projected sensitivities of Table~\ref{tab:cLFV} marked by red-solid and green-dashed lines, respectively. Once the current MEG~II bound on $\BR(\mu \to e \gamma)$ is imposed (horizontal red-solid line), the BaBar and Belle limits on the radiative $\tau$ decays (vertical red-solid lines) are automatically satisfied, and the surviving points predict $\tau \to \ell \gamma$ rates below the projected Belle~II sensitivity (vertical green-dashed lines). In the top and middle panels, the colour bar shows the cubic coupling $\kappa$~\eqref{eq:Vpotentialfull}. As mentioned before, the smallness of $\mathbf{m}_S$ is controlled by the naturally small parameter $\kappa$ [see Eq.~\eqref{eq:Vpotential}]. Reproducing the neutrino data at fixed masses then requires $|\mathbf{Y}_{\nu,\chi}|^2 \propto 1/\kappa$, so that the cLFV BRs scale as $\kappa^{-2}$. In particular, MEG~II constrains $\kappa\lesssim 5\times10^{-5}$~GeV.

    \item \textbf{Enhanced Majoron flavour violation:} The middle panels show $\BR(\tau\to eJ)$ (left) and $\BR(\tau\to \mu J)$ (right) versus $\BR(\mu\to e\gamma)$ for $m_J=1$~GeV, with the current bounds and projected sensitivities of Table~\ref{tab:cLFV}. The Majoron-emitting decays have BRs typically five to six orders of magnitude larger than those of the radiative decays, $\BR(\tau\to\ell J)/\BR(\tau\to\ell\gamma)\sim10^{5}-10^{6}$, and they exceed $\BR(\mu\to e\gamma)$ by a similar factor. This is a distinctive feature of our model, in which both the seesaw and \emph{dark} sectors generate a Majoron coupling to charged leptons that is not suppressed by light neutrino masses, in contrast with the canonical seesaw Majoron models~\cite{Heeck:2019guh,Escribano:2021ymx,Herrero-Brocal:2023czw,CentellesChulia:2024uzv}. As a consequence, and contrary to the radiative decays $\tau\to\ell\gamma$, the searches for $\tau\to\ell J$ at Belle (solid-red line) and Belle~II (dashed-green line) probe regions of the parameter space that MEG~II cannot reach. Taken at face value, the Belle limit on $\tau\to\mu J$ already excludes a few points that satisfy MEG~II, and Belle~II may probe a small region of parameter space beyond the reach of MEG~II in $\tau\to\mu J$ and $\tau\to eJ$. These correspond to regions where $\kappa\lesssim5\times10^{-5}$~GeV (see colour bar). Note, however, that these searches assume an invisible boson. In $\tau\to\ell J$ at Belle the Majoron has $p_J\simeq3.5$~GeV on average~\cite{Belle:2025bpu}, so Eq.~\eqref{eq:LJ} gives $L_J\simeq1.3$~m, comparable to the outer radius of the calorimeter $R\simeq1.6$~m~\cite{Belle:2000cnh}. As a simple estimate, only a fraction $e^{-R/L_J}\simeq0.3$, of the Majorons decays beyond it and escapes detection. This weakens the limits shown by a factor of about three. Furthermore, the Majorons that decay inside the detector open the visible channel $\tau\to\ell J$ with a displaced $J\to\gamma\gamma$ decay, giving $\tau\to\ell(\gamma\gamma)$ with $\ell=e,\mu$~\cite{Heeck:2017xmg}. A displaced diphoton search and a proper lifetime-dependent recast of the missing-energy limits are left for future work.
    
    \item \textbf{\emph{Dark} and seesaw-sector interplay:} The bottom panels show $\BR(\mu \to e \gamma)$ (left) and $\BR(\tau \to \mu J)$ (right), with the vertical axis including only the \emph{dark}-sector contribution and the horizontal axis only the seesaw one. Red points are excluded by current cLFV bounds, while green ones can be probed in the future -- see Table~\ref{tab:cLFV}. Cancellations between the \emph{dark} and seesaw amplitudes, which occur along the black-oblique line, allow for smaller total BRs even when both individual contributions are large. For both $\mu \to e \gamma$ (left) and $\tau \to \mu J$ (right), the \emph{dark} and seesaw sectors contribute on an equal footing to the total BR. Both BRs are quartic in the corresponding Yukawa couplings, so their ratio scales as $|\mathbf{Y}_\chi|^4/|\mathbf{Y}_\nu|^4$. The relative size of the two contributions is controlled by the Casas-Ibarra angle (see Appendix~\ref{sec:CasasIbarra}). Additionally, for $\mu \to e \gamma$ (left), both contributions also depend on the masses entering the loop functions of Eqs.~\eqref{eq:cLFV} and~\eqref{eq:loopfunctionmeg}. A similar interplay of contributions was present in our previous work~\cite{Batra:2023bqj} for $\mu \to e \gamma$ . Here, owing to the spontaneous breaking of lepton number, it also shows up in channels such as $\tau \to \mu J$. Overall, this \emph{dark} and seesaw sector interplay in both the typical cLFV decays and Majoron-emitting ones is a distinctive and, to the best of our knowledge, unique feature of the present model.
    
\end{itemize}

\newpage

\clearpage

\subsection{Dark matter}
\label{sec:DM}

%
\begin{figure}[!t]
    \centering
    \includegraphics[scale=0.4]{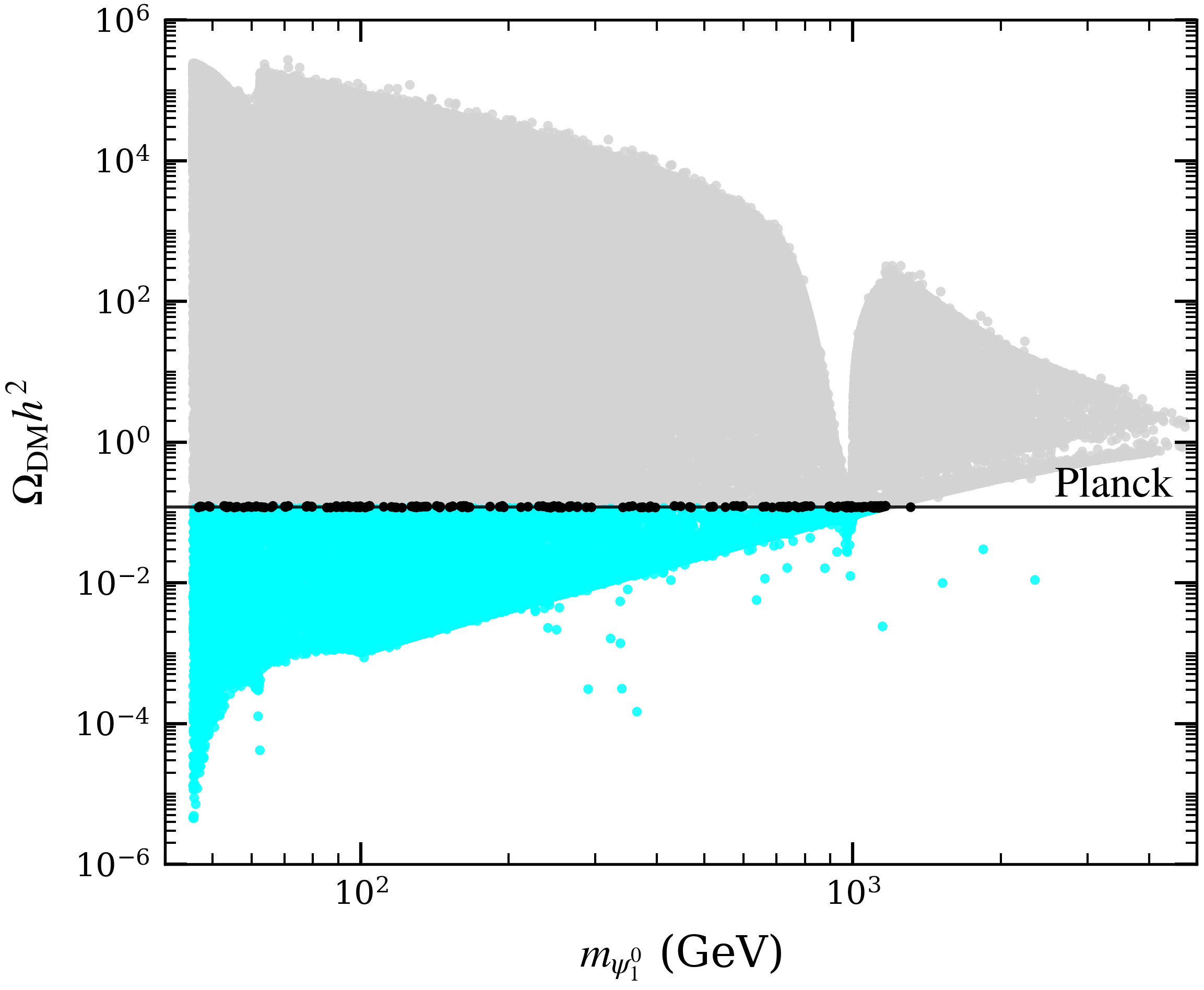} \\ \vspace{+0.2cm} \includegraphics[scale=0.4]{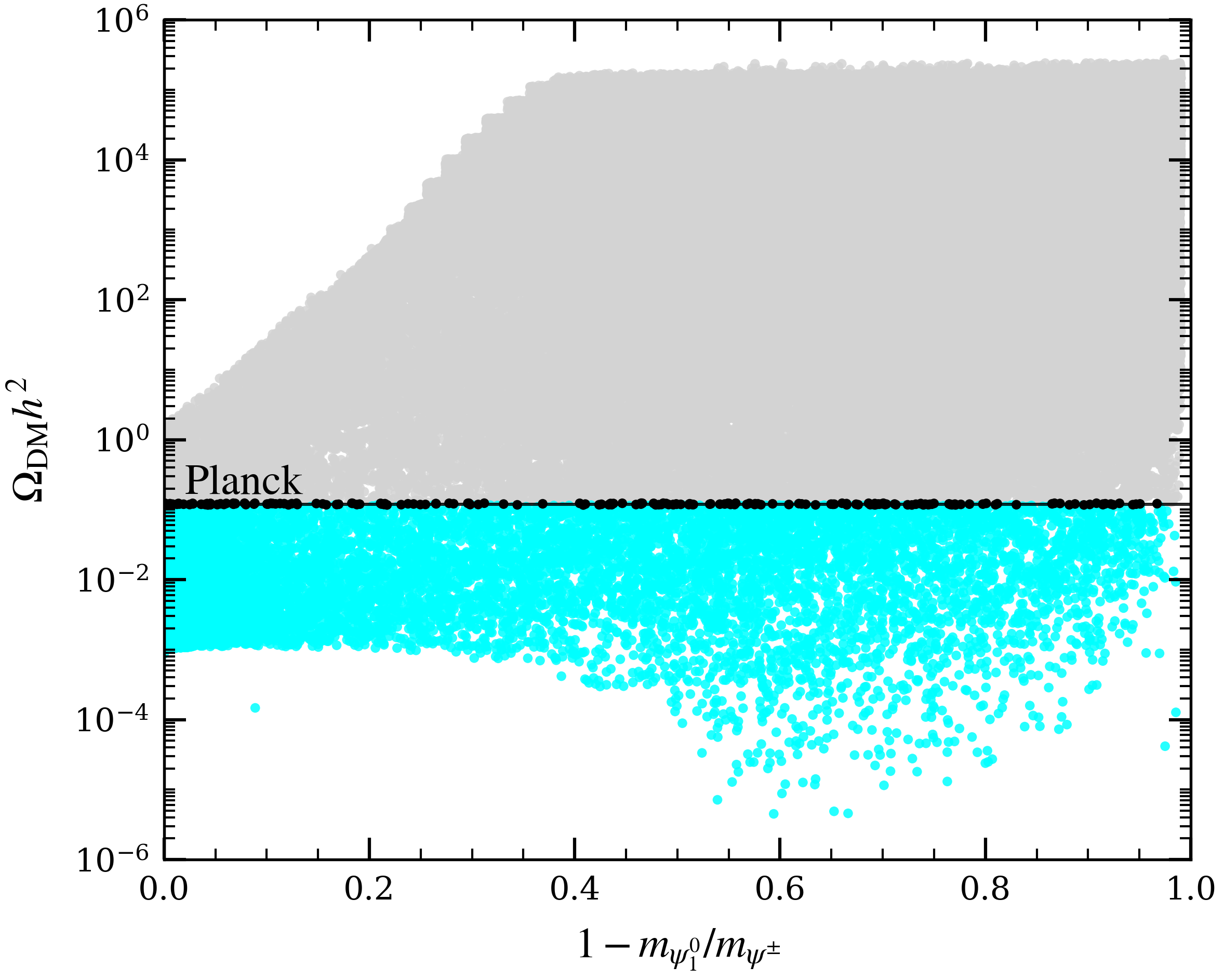} \hspace{+0.1cm} \includegraphics[scale=0.4]{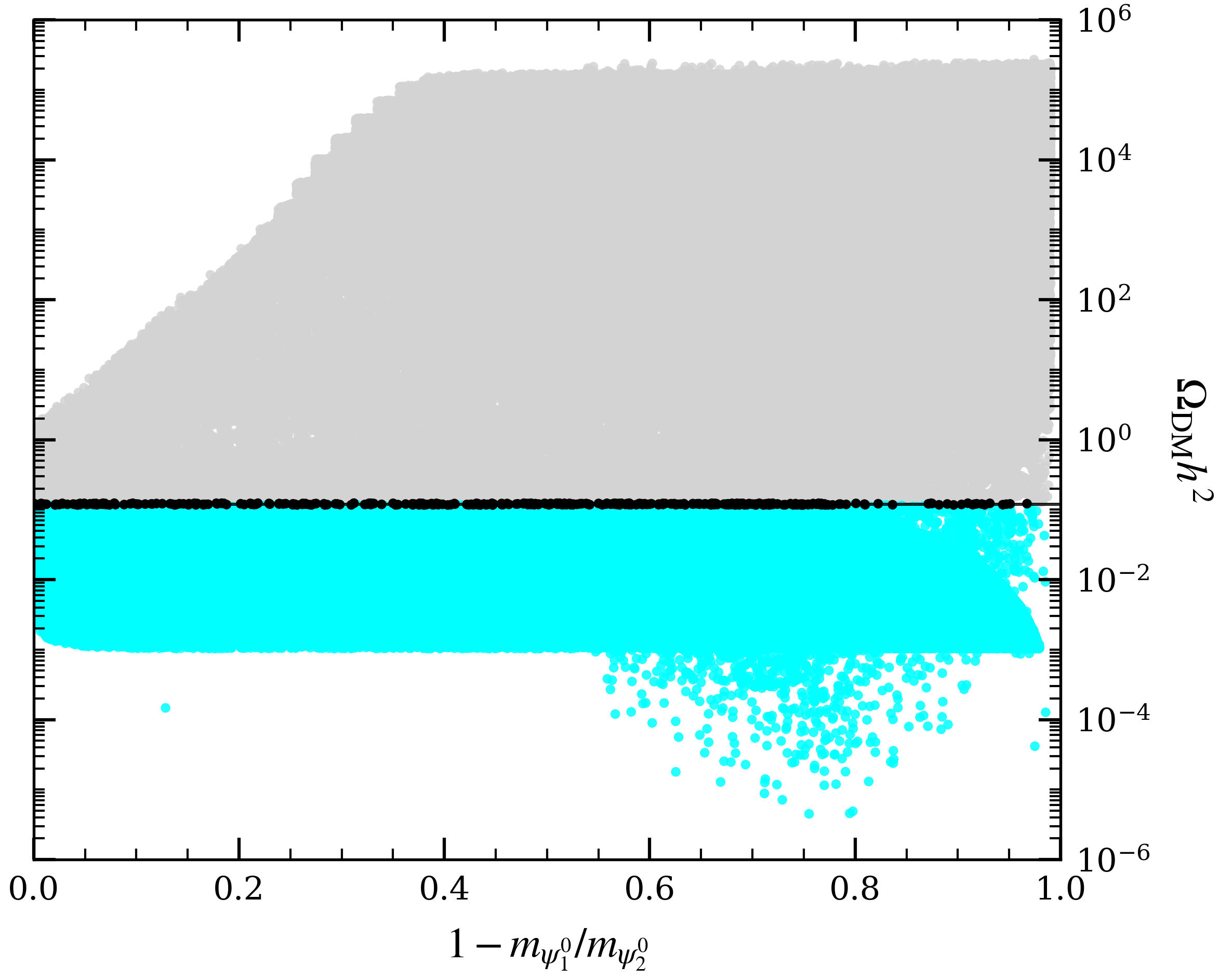}
    \caption{Relic density $\Omega h^2$ as a function of the fermion DM mass $m_{\psi^0_1}$ (top), the relative mass difference $1-m_{\psi^0_1}/m_{\psi^\pm}$ (bottom left), and $1-m_{\psi^0_1}/m_{\psi^0_2}$ (bottom right). Black points lie within the $3 \sigma$ range of the observed relic abundance obtained by Planck~\cite{Planck:2018vyg} [see Eq.~\eqref{eq:Oh2Planck}]. Grey (cyan) points are overabundant (underabundant).}
    \label{fig:DMrelic}
\end{figure}
\begin{figure}[!t]
    \centering
    \includegraphics[scale=0.4]{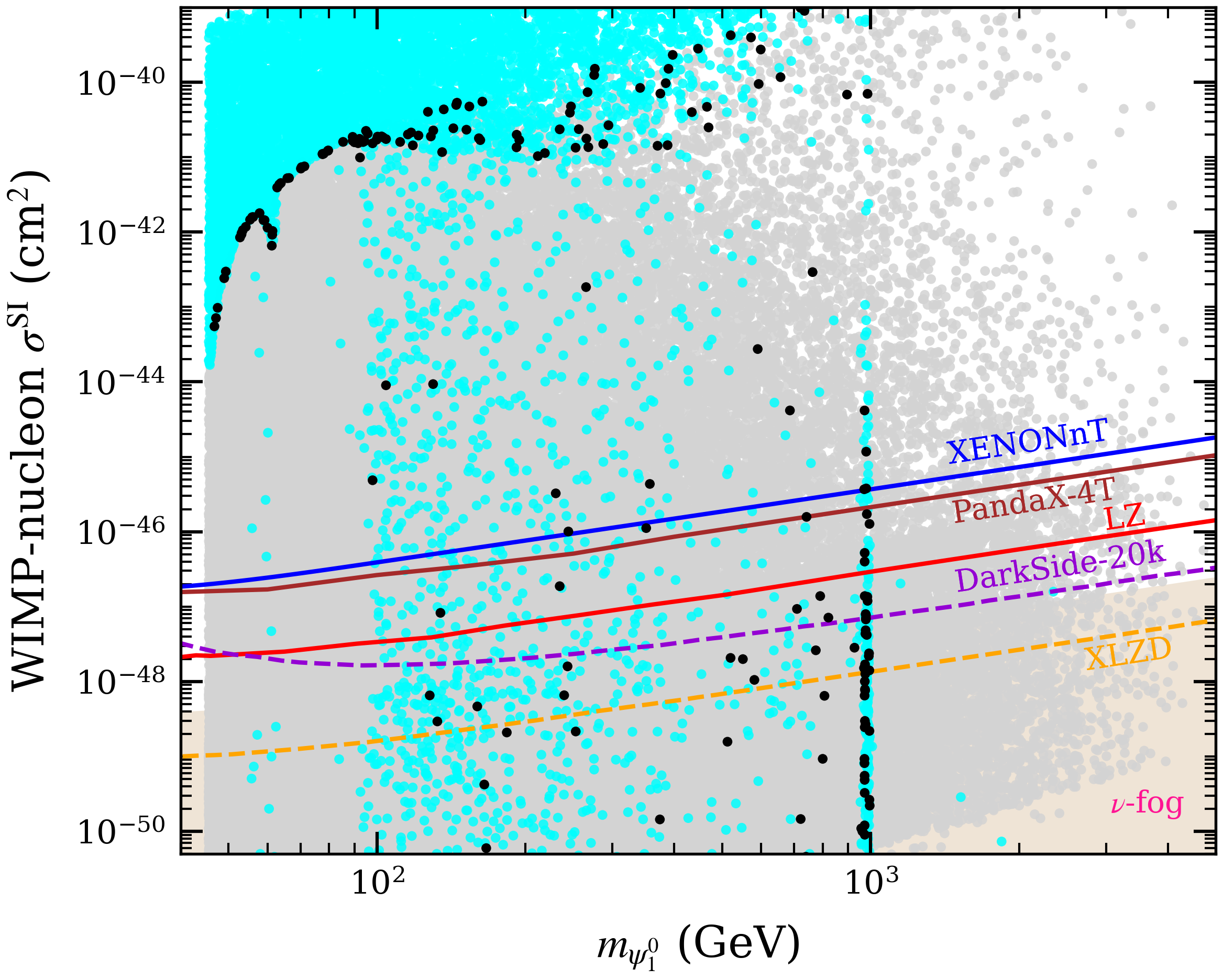} \hspace{+0.2cm}
    \includegraphics[scale=0.4]{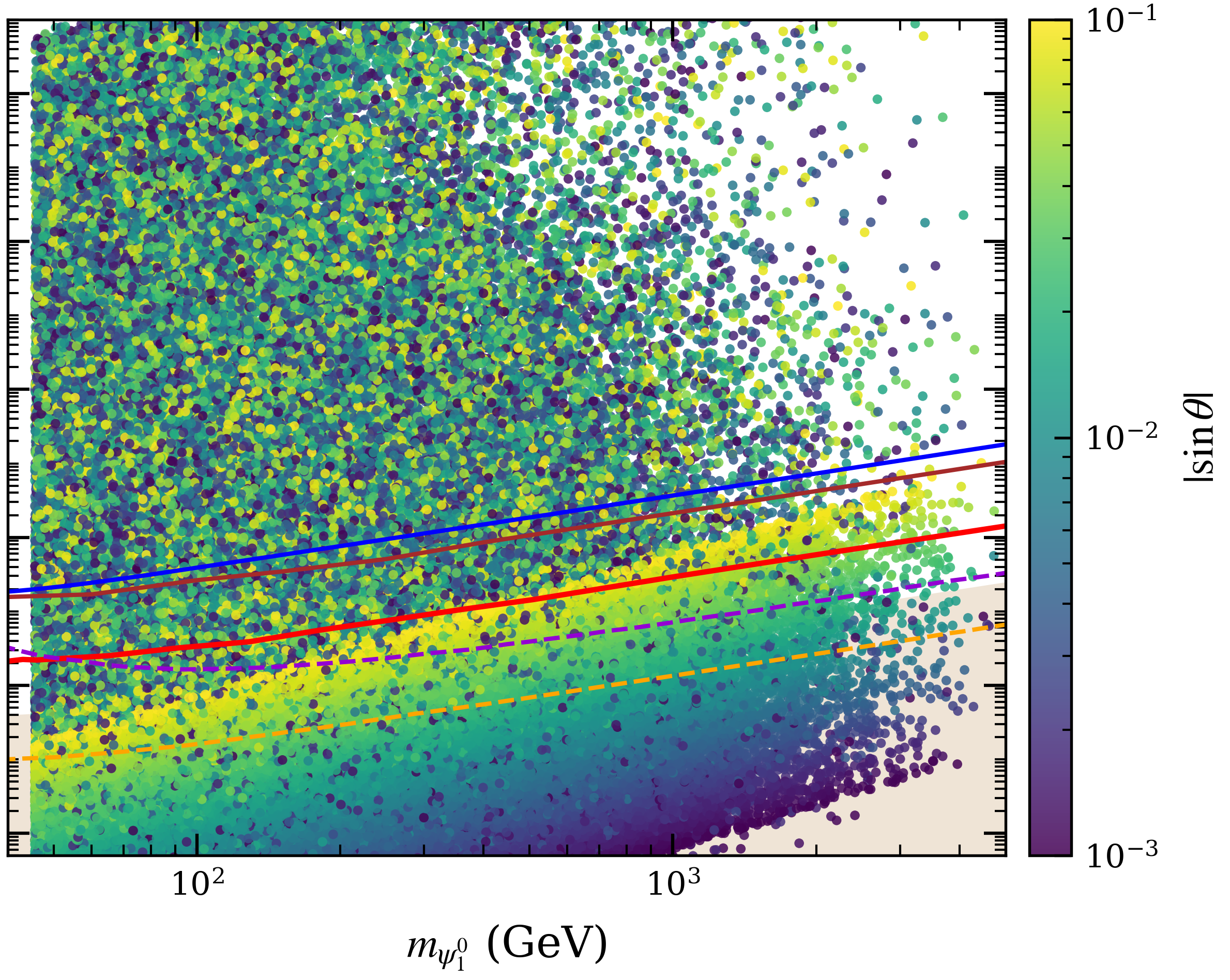}
    \caption{WIMP-nucleon spin-independent elastic scattering cross-section $\sigma^{\text{SI}}$ as a function of the fermion DM mass~$m_{\psi_1^0}$. The solid blue, brown and red lines indicate current constraints from the DD experiments XENONnT~\cite{XENON:2025vwd}, PandaX-4T~\cite{PandaX:2024qfu} and LZ~\cite{LZ:2024zvo}, respectively. The purple and orange-dashed contours indicate the projected sensitivities for DarkSide-20k~\cite{DarkSide-20k:2017zyg}, and XLZD~\cite{XLZD:2024nsu}, respectively. The pink-shaded region is the ``neutrino fog"~\cite{OHare:2021utq}. The points on the left panel follow the relic density colour code of Fig.~\ref{fig:DMrelic}. In the right panel the points are coloured according to $|\sin\theta|$ values as indicated by the colour bar.}
    \label{fig:DMDD}
\end{figure}
In our previous work~\cite{Batra:2023bqj}, where we introduced the \emph{dark} linear seesaw, the \emph{dark} sector was odd under an imposed $\mathcal{Z}_2$ symmetry. This symmetry stabilizes the lightest \emph{dark} particle and thus provides a viable scalar or fermionic WIMP DM candidate. In that work we studied the scalar case, in which the DM is the neutral scalar arising from the mixing between a doublet and a singlet scalar, as in the scotogenic model~\cite{Tao:1996vb,Ma:2006km}. In the present model, by contrast, the $\mathcal{Z}_2$ is not imposed by hand. It is the residual symmetry left by the spontaneous breaking of the global lepton number. The \emph{dark} sector is odd under it, so its lightest particle is stable. The DM candidate can then be either the scalar $\zeta_1$ or the fermion $\psi_1^0$, which stems from the vector-like doublet $D_{L,R}$ neutral component mixing with the singlet $F_{L,R}$. Here we focus on fermionic DM, i.e.\ on the case in which $\psi_1^0$ is the lightest \emph{dark} particle. The thermally averaged cross section, and hence the relic density, then receives contributions from $\psi_1^0-\psi_1^0$ annihilation and from the $\psi_1^0-\psi_2^0$, $\psi_1^0-\psi^\pm$ and $\psi_1^0-\zeta_k$ ($k=1,\ldots,4$) coannihilations into $\mathcal{Z}_2$-even final states, including the Majoron $J$. The current observed value for DM relic abundance obtained by the Planck satellite data, at the $3 \sigma$ level, is~\cite{Planck:2018vyg},
\begin{equation}
0.1164 \leq \Omega_{\text{DM}} h^2 \leq 0.1236 \; ,
\label{eq:Oh2Planck}
\end{equation}
where $h$ is the reduced Hubble constant.

To test whether $\psi_1^0$ is a viable DM candidate, we implement our model in \texttt{SARAH}~\cite{Staub:2013tta} and use \texttt{SPheno}~\cite{Porod:2003um} to obtain mass matrices, vertices and tadpole equations. The relic density $\Omega_{\text{DM}} h^2$ and the WIMP-nucleon spin-independent elastic scattering cross-section $\sigma^{\text{SI}}$ are computed at tree level with \texttt{micrOMEGAs}~\cite{Belanger:2001fz}. In our numerical scan (see Table~\ref{tab:Scan}), we impose the collider constraints discussed in Sec.~\ref{sec:collider} and the cLFV bounds discussed in Sec.~\ref{sec:cLFV}. The results of our numerical analysis are presented in Figs.~\ref{fig:DMrelic} and~\ref{fig:DMDD}:
\begin{itemize}

    \item \textbf{Relic density:} In Fig.~\ref{fig:DMrelic}, we show $\Omega h^2$ as a function of $m_{\psi^0_1}$ (top panel) and of the relative mass splitting between $\psi^0_1$ and the charged fermion $\psi^\pm$ (bottom left) or the heavier neutral fermion~$\psi^0_2$ (bottom right). Black points lie within the $3\sigma$ range of the relic abundance measured by Planck~\cite{Planck:2018vyg} [see Eq.~\eqref{eq:Oh2Planck}], while grey (cyan) points correspond to overabundant (underabundant) DM. Hence, the grey points are excluded, whereas the cyan ones remain viable with $\psi^0_1$ accounting for only a fraction of DM. From the top plot, it is clear that the observed relic abundance (black points) is obtained for DM masses in the range $\sim[45~\text{GeV},1~\text{TeV}]$, the lower limit corresponding to the bound of Eq.~\eqref{eq:mpsi0_LEPbound}. Two dips are clearly visible in the top panel. The first, at $m_{\psi^0_1}\simeq m_h/2\simeq62.6$~GeV, is due to $s$-channel annihilation through the SM Higgs, which becomes very efficient when it is on-shell. The second, at $m_{\psi^0_1}\simeq m_H/2\simeq1$~TeV, for our benchmark $m_H=2$~TeV, stems from the $s$-channel exchange of the heavy CP-even scalar $H$. In fact, since the \emph{dark}-fermion masses are generated by the $\sigma$ VEV, $\psi^0_1$ couples to $H$ and $J$ with strength $\sim m_{\psi^0_1}/v_\sigma$. Near this resonance, the final states follow the $H$ decays, which depend on $|\sin\theta|$ and include a sizeable fraction of Majoron pairs, dominant for small $|\sin\theta|$. Away from resonances, a mostly-singlet~$\psi^0_1$ also annihilates mainly into Majorons, through $\psi^0_1\bar{\psi}^0_1\to JJ$ and, for $2m_{\psi^0_1}>m_H$, $\psi^0_1\bar{\psi}^0_1\to JH$, as in the dynamical scotogenic model~\cite{Chun:2023vbh}. The opening of the latter $s$-wave channel is visible as the sharp drop of the overabundant region above $m_{\psi^0_1}\simeq1$~TeV. However, for $m_{\psi^0_1}\ll v_\sigma$, these channels are too weak, and most points are overabundant (grey points). Additionally, the dense band of cyan points with a sharp lower edge corresponds to a mostly-doublet $\psi^0_1$, which annihilates efficiently through $Z$ and $W$ exchange. Along this edge $\Omega h^2$ grows as $m_{\psi^0_1}^2$ and reaches the observed value at $m_{\psi^0_1}\simeq1.1$~TeV, close to the value expected for a pure Dirac doublet~\cite{Cirelli:2005uq}. Finally, the bottom panels show that coannihilations with $\psi^\pm$ and $\psi^0_2$ play an important role. Namely, they reduce $\Omega h^2$ for relative mass splittings below $\sim 40\%$, the overabundant region shrinking as the splitting decreases, yielding viable relic-density points over the mass range $\sim[45~\text{GeV},1~\text{TeV}]$. Since $y_{DF} \in [10^{-5},1]$ and is thus small for most points, $\psi^0_1$ is mostly singlet, with $\psi^\pm$ and $\psi^0_2$ being quasi-degenerate, which is why the two bottom panels look similar.
    
    \item \textbf{Direct detection (DD):} In Fig.~\ref{fig:DMDD}, we show the WIMP-nucleon $\sigma^{\text{SI}}$ as a function of $m_{\psi_1^0}$, together with the current bounds from XENONnT~\cite{XENON:2025vwd}, PandaX-4T~\cite{PandaX:2024qfu} and LZ~\cite{LZ:2024zvo} (solid blue, brown and red lines) and the projected sensitivities of DarkSide-20k~\cite{DarkSide-20k:2017zyg} and XLZD~\cite{XLZD:2024nsu} (purple and orange dashed lines). The pink-shaded region is the ``neutrino fog", where coherent elastic neutrino-nucleus scattering becomes an irreducible background~\cite{OHare:2021utq}. The points in the left panel follow the relic-density colour code of Fig.~\ref{fig:DMrelic}, and, in the right panel, the points are coloured according to $|\sin\theta|$ values. At tree level, $\psi_1^0$ scatters off nuclei through two contributions. The first is $Z$ exchange, induced by the doublet component of $\psi_1^0$, which is controlled by $y_{DF}$. Since $\psi_1^0$ is a Dirac fermion, this component couples vectorially to the $Z$-boson, yielding $\sigma_n\simeq G_F^2\mu_n^2(\cos^2\theta_L+\cos^2\theta_R)^2/(8\pi)$, with~$\mu_n$ being the DM-neutron reduced mass and the angles $\theta_{R,L}$ control the mixing between the singlet fermion and the neutral component of the doublet -- see Appendix~\ref{sec:darkfermion}. The second is Higgs exchange induced by the $\Phi-\sigma$ mixing. Since $\psi_1^0$ couples to $\sigma$ with a strength $\sim m_{\psi_1^0}/v_\sigma$, it gives $\sigma^{\text{SI}}\propto\sin^2\theta\,m_{\psi_1^0}^2/v_\sigma^2$~\cite{Barreiros:2022aqu}, the only contribution left for vanishing singlet-doublet mixing. The right panel clearly separates the two regimes: the $Z$-dominated points in the upper part show no correlation with $|\sin\theta|$, whereas the Higgs-dominated ones below display a colour gradient, with $\sigma^{\text{SI}}$ growing with $|\sin\theta|$ and $m_{\psi_1^0}$. The LZ bound (solid-red line), currently the strongest, requires $\cos^2\theta_L+\cos^2\theta_R\lesssim10^{-4}$, i.e.\ an almost pure singlet, so that relic-compatible points with a sizeable doublet component are excluded by several orders of magnitude, as in the minimal singlet-doublet Dirac DM scenario~\cite{Yaguna:2015mva}. For a mostly-singlet $\psi_1^0$, LZ requires $|\sin\theta|\lesssim0.1$ at $m_{\psi_1^0}\simeq500$~GeV and $|\sin\theta|\lesssim0.05$ at $1$~TeV (see the color bar). Nevertheless, as shown in the left panel, several relic-compatible points with $45~\text{GeV} \lesssim m_{\psi_1^0} \lesssim 1$~TeV satisfy this bound, most of them around the $h$ and $H$ resonances, $m_{\psi_1^0}\simeq62.6$~GeV and $1$~TeV. Future experiments such as DarkSide-20k (purple-dashed contour) and XLZD (orange-dashed contour) will probe a large part of this remaining parameter space, down to $|\sin\theta|\simeq0.01$ at $1$~TeV, up to the ``neutrino fog'' (pink-shaded region).

\end{itemize}
Lastly, we wish to comment on the alternative scalar DM possibility, i.e.\ the case where the lightest \emph{dark} particle is~$\zeta_1$. Note that, as the real and imaginary parts of the \emph{dark} singlets $\chi_{1,2}$ share the same mass matrix up to the sign of $\kappa$, the scalars $\zeta_1$ and $\zeta_3$ are degenerate [see Eqs.~\eqref{eq:darkscalarmass} and~\eqref{eq:neutraldarkscalarmass}], leading to complex scalar singlet DM. Unlike the scalar DM candidate of our previous work~\cite{Batra:2023bqj}, $\zeta_1$ has no doublet component and hence no gauge interactions. It thus resembles the scalar-singlet Higgs-portal DM, whose parameter space is strongly constrained by DD and, for DM masses below $m_h/2$, by the Higgs invisible width (see Sec.~\ref{sec:collider})~\cite{Cline:2013gha,Feng:2014vea,GAMBIT:2017gge,Casas:2017jjg,Athron:2018ipf,Arcadi:2019lka,Lebedev:2021xey,Arcadi:2024ukq,Yu:2024xsy}. After the latest LZ results~\cite{LZ:2024zvo}, only a narrow region around the Higgs resonance, at DM masses around $m_h/2\simeq62.6$~GeV, remains viable~\cite{Yu:2024xsy}. Our scenario, however, differs from this minimal case through the scalar $H$ and Majoron $J$ arising from~$\sigma$, to which $\zeta_1$ couples via $\kappa$ and $\lambda_{\sigma\chi_{1,2}}$ [see Eq.~\eqref{eq:Vpotentialfull}]. Besides the Higgs dip, controlled by $\lambda_{\Phi\chi_{1,2}}$, these states open new annihilation channels that can reproduce the observed relic abundance. Namely, a second resonance at $m_{\zeta_1}\simeq m_H/2$~\cite{Camara:2023hhn}, and the annihilation into Majoron pairs $\zeta_1\zeta_1\to JJ$, which is enhanced near the $H$ resonance. Furthermore, the WIMP-nucleon $\sigma^{\text{SI}}$, relevant for DD, receives contributions from both $h$ and $H$ exchange. Thus, the relic-compatible points around the $H$ resonance, where annihilations into Majorons are important, can evade the stringent LZ bound for a sufficiently suppressed Higgs-portal or small $|\sin \theta|$. A detailed analysis is left for future work.

\section{Concluding remarks}
\label{sec:concl}

We have proposed a minimal fermionic realisation of the \emph{dark} linear seesaw in which, in contrast to our previous work~\cite{Batra:2023bqj}, lepton number is broken spontaneously. The SM is extended by a single $\nu_R-S_R$ heavy-neutrino pair, a complex scalar singlet $\sigma$ carrying lepton number, and a \emph{dark} sector consisting of a fermion doublet, a fermion singlet and two complex scalar singlets. The small LNV parameter responsible for the linear-seesaw neutrino masses vanishes at tree level and is generated at one loop by the \emph{dark} sector. Since U(1)$_L$ forbids bare fermion mass terms, the heavy-fermion masses arise from the VEV of $\sigma$. Likewise, the cubic scalar coupling that closes the \emph{dark} loop, which broke lepton number explicitly in Ref.~\cite{Batra:2023bqj}, is now generated dynamically. The spontaneous breaking U(1)$_L \rightarrow \mathcal{Z}_2$, with only the \emph{dark} fields odd under the residual $\mathcal{Z}_2$ symmetry, yields a Majoron $J$ and makes the lightest \emph{dark} particle stable. DM stability is thus a remnant of lepton number breaking. This minimal field content reproduces both neutrino mass-squared differences and predicts that the lightest neutrino is massless.

The spontaneous breaking of lepton number provides further tests of the model. The \emph{dark} fermion doublet plays two roles in our scenario. First, its neutral component enters the \emph{dark} loop that generates the linear seesaw. Second, its charged component $\psi^\pm$ has left- and right-handed parts that carry different lepton numbers. As a result, U(1)$_L$ has an EM anomaly and the Majoron couples to photons. This coupling is fixed by $v_\sigma$ and cannot be removed by redefining lepton number. A massless Majoron would require $v_\sigma \gtrsim 10^{8}$~GeV to evade the bound from the CAST helioscope and HB stars, far above the TeV scale that sets the heavy-neutrino and \emph{dark}-sector masses. Hence, we consider the possibility that the Majoron is massive. We introduce a term linear in $\sigma$ that breaks U(1)$_L$ softly. This term preserves the residual $\mathcal{Z}_2$, so that DM stability and neutrino mass generation are unaffected. For our benchmark $m_J=1$~GeV and $v_\sigma/\sqrt{2}=5$~TeV, the Majoron decays mainly into two photons, with a lifetime of about a nanosecond. Thus, it is not a DM candidate. Such a Majoron is a prime target for future beam-dump experiments such as DarkQuest (Phase~II) and SHiP.

We have imposed the relevant collider constraints. In particular, Higgs signal strengths and invisible Higgs decays into Majoron pairs bound the $\Phi-\sigma$ mixing. As a low-scale seesaw, the model allows TeV-scale heavy neutrinos with sizeable Yukawa couplings, which leads to sizeable cLFV rates. The radiative decay $\mu \to e\gamma$ receives contributions from the seesaw sector, through $W$-boson loops with heavy neutrinos, and from the \emph{dark} sector, through loops of the charged \emph{dark} fermion and the \emph{dark} scalars. Either contribution may dominate, and large regions of the parameter space lie within the reach of future MEG~II searches.

The Majoron couplings to charged leptons arise at one loop, again from both the seesaw and \emph{dark} sectors. They are controlled by the heavy-neutrino and \emph{dark} Yukawa couplings rather than by the light-neutrino masses, and are therefore strongly enhanced relative to the canonical type-I and low-scale seesaw Majoron models. For $m_J=1$~GeV, the decay $\mu \to eJ$ is kinematically forbidden. The off-diagonal couplings of the Majoron to the charged leptons are instead probed by $\tau \to eJ$ and $\tau \to \mu J$. Their BRs typically exceed those of $\tau \to e\gamma$ and $\tau \to \mu\gamma$ by orders of magnitude. Unlike the radiative $\tau$ decays, the Majoron-emitting ones can thus be probed at Belle~II in a small parameter space region beyond the reach of MEG~II. As for $\mu \to e\gamma$, either the seesaw or the \emph{dark} contribution may dominate. To the best of our knowledge, this interplay between the \emph{dark} and seesaw sectors in both the conventional cLFV processes and the Majoron-emitting ones is unique to the present model.

Regarding DM, we focused on the case where the stable particle is the lightest \emph{dark}-neutral fermion WIMP $\psi_1^0$, stemming from the mixing between the singlet and the neutral component of the doublet. The observed DM relic abundance is obtained over a DM mass range $\sim[45~\text{GeV},1~\text{TeV}]$, mainly through resonant annihilation via $h$ and $H$, with a large fraction of Majoron final states in the latter case, and through coannihilations with $\psi^\pm$ and $\psi_2^0$. Since the current most stringent DD constraint from LZ excludes a sizeable doublet component, which induces $Z$-mediated scattering, viable DM must be mostly singlet. In such a case, DM scatters off nuclei through Higgs exchange. The resulting spin-independent cross sections lie below the current XENONnT, PandaX-4T and LZ bounds. Future DD experiments such as DarkSide-20k and XLZD will probe a large part of the remaining parameter space up to the ``neutrino fog''.

Overall, our setup is a minimal framework where a \emph{dark} sector seeds the linear seesaw through spontaneous lepton-number breaking. This sector induces the LNV parameter radiatively and contains a viable WIMP DM candidate, stabilised by the residual $\mathcal{Z}_2$ of lepton number. A single scale, the VEV of $\sigma$, sets the heavy-neutrino and \emph{dark}-fermion masses. Through the charged \emph{dark} fermion, the same scale also fixes the anomalous Majoron-to-photon coupling. The interplay between Majoron, cLFV and DM signatures makes this scenario testable at future experiments.

\begin{acknowledgments}
This work is supported by Fundação para a Ciência e a Tecnologia (FCT), Portugal, through the projects \href{https://doi.org/10.54499/UID/00777/2025}{UID/00777/2025}, \href{https://doi.org/10.54499/UID/PRR/00777/2025}{UID/PRR/00777/2025}, \href{https://doi.org/10.54499/UID/PRR2/00777/2025}{UID/PRR2/00777/2025}, which are partially funded through POCTI (FEDER), COMPETE, QREN and the EU. The work of A.B. is supported by the PhD FCT grant UI/BD/154391/2023 (https://doi.org/10.54499/UI/BD/154391/2023). H.B.C. thanks the CERN Department for Theoretical Physics for hospitality and financial support during the final stage of this work. \\

The authors used AI-assisted tools, namely ChatGPT (OpenAI) and Claude (Anthropic), to support literature searches and the verification of numerical results. All AI-assisted outputs were critically reviewed and validated by the authors, who take full responsibility for the accuracy, integrity, and final content of the manuscript.
\end{acknowledgments}

\appendix

\section{Scalar sector}
\label{sec:scalarsector}

As shown in Table~\ref{tab:model}, the scalar content of our model contains, besides the usual Higgs doublet $\Phi$, three scalar singlets $\sigma$ and $\chi_{1,2}$ which we define as
\begin{align}
\Phi&=\begin{pmatrix}
\phi^{+} \\
\phi^0
\end{pmatrix}= \frac{1}{\sqrt{2}}  \begin{pmatrix}
 \sqrt{2} \phi^{+} \\
 v_\Phi + \phi_{\text{R}} + i \phi_{\text{I}}
\end{pmatrix} \; ; \; \sigma = \frac{v_\sigma + \sigma_{\text{R}} + i J }{\sqrt{2}} \; ; \; \chi_{1,2} = \frac{\chi_{1, 2 \text{R}} + i \chi_{1, 2 \text{I}}}{\sqrt{2}} \; .
\label{eq:scalars}
\end{align}
The scalar content of our model can be divided into two classes, depending on the charge carried by the fields under the residual $\mathcal{Z}_2$ symmetry, i.e. we can have even scalars $\Phi$ and $\sigma$ or odd \textit{dark} scalars $\chi_{1,2}$. The vacuum configuration is such that the Higgs doublet and scalar $\sigma$ acquire non-zero VEV, where $\left< \phi^0 \right> = v_\Phi/\sqrt{2} \simeq 174$ GeV and $\left< \sigma \right> = v_\sigma/\sqrt{2}$, with the \emph{dark} singlets $\chi_{1,2}$ being VEVless, i.e. $\left< \chi_{1,2} \right> = 0$. The full scalar potential $V(\Phi,\sigma,\chi_{1,2})$ allowed by the symmetries of our model is,
\begin{align}
V &=  m_{\Phi}^2 \left(\Phi^\dagger \Phi\right) + m_{\sigma}^2 |\sigma|^2 + m_{\chi_1}^2 |\chi_1|^2 + m_{\chi_2}^2 |\chi_2|^2 + \frac{\lambda_{\Phi}}{2} \left(\Phi^\dagger \Phi\right)^2 + \frac{\lambda_{\sigma}}{2} |\sigma|^4  + \frac{\lambda_{\chi_1}}{2} |\chi_1|^4 + \frac{\lambda_{\chi_2}}{2} |\chi_2|^4  \nonumber \\
& + \lambda_{\Phi \sigma} \left(\Phi^\dagger \Phi\right) |\sigma|^2 + \lambda_{\Phi \chi_1} \left(\Phi^\dagger \Phi\right) |\chi_1|^2 + \lambda_{\Phi \chi_2} \left(\Phi^\dagger \Phi\right) |\chi_2|^2 + \lambda_{\sigma \chi_1} |\sigma|^2 |\chi_1|^2 + \lambda_{\sigma \chi_2} |\sigma|^2 |\chi_2|^2 + \lambda_{\chi_1 \chi_2} |\chi_1|^2 |\chi_2|^2 \nonumber \\
& + \left[\kappa \ \chi_1 \chi_2 \sigma^\ast + \text{H.c.} \right]\; ,
\label{eq:Vpotentialfull}
\end{align}
where $\kappa$ can be made real. Additionally, we assume that U(1)$_L$ is broken explicitly via the linear soft-breaking term
\begin{align}
V_{\text{soft}} = -\frac{\mu_J^3}{\sqrt{2}}\left(\sigma+\sigma^\ast\right) \; ,
\label{eq:linearsoft}
\end{align}
with $\mu_J>0$, being naturally small in the 't~Hooft sense~\cite{tHooft:1979rat} since lepton number is restored in its absence. Since $V_{\text{soft}}$ is linear in $\sigma_{\text{R}}$, it enters the scalar masses only through the minimisation conditions
\begin{align}
m_\Phi^2 = -\frac{\lambda_\Phi}{2}\, v_\Phi^2 - \frac{\lambda_{\Phi\sigma}}{2}\, v_\sigma^2 \; , \;
m_\sigma^2 = -\frac{\lambda_\sigma}{2}\, v_\sigma^2 - \frac{\lambda_{\Phi\sigma}}{2}\, v_\Phi^2 + \frac{\mu_J^3}{v_\sigma} \; .
\label{eq:tadpoles}
\end{align}

Besides the usual GB $\phi_{\text{I}}$ and $\phi^{\pm}$ of the SM Higgs doublet $\Phi$, due to the spontaneous breaking of the global lepton number symmetry U(1)$_L$ there will be an associated GB commonly referred to as the Majoron $J = \sqrt{2} \text{Im} (\sigma)$. The soft term turns $J$ into a pseudo-GB with mass
\begin{align}
m_J^2=\frac{\mu_J^3}{v_\sigma} \; .
\end{align}
Furthermore, the CP-even scalars $\phi_{\text{R}}$ and $\sigma_{\text{R}}$ will mix resulting in two massive neutral scalars. Namely, their masses are obtained by computing the eigenvalues of the following mass matrix, written in the $\left(\phi_{\text{R}}, \sigma_{\text{R}} \right)$ basis,
\begin{align}
    \boldsymbol{\mathcal{M}}^2_{\phi\sigma}=\begin{pmatrix}
    v_\Phi^2\,\lambda_\Phi & v_\sigma\,v_\Phi\,\lambda_{\Phi \sigma} \\
   \cdot & v_\sigma^2\lambda_\sigma + m_J^2 \\
    \end{pmatrix} \; ,
    \label{eq:ENSMassMatrixlim}
\end{align}
where '$\cdot$' reflects the symmetric nature of the matrix. The $m_J^2$ in the $\sigma_{\text{R}}$ entry follows from the shift of $m_\sigma^2$ in Eq.~\eqref{eq:tadpoles}. The above is diagonalised via the orthogonal matrix $\mathbf{O}$:
\begin{equation}
\begin{pmatrix}
\phi_{\text{R}}\\
\sigma_{\text{R}}
\end{pmatrix}= \mathbf{O} \begin{pmatrix}
h \\
H
\end{pmatrix} \; , \; \mathbf{O} = \begin{pmatrix}
        \cos \theta & \sin \theta \\
        - \sin \theta & \cos \theta \end{pmatrix} \; , \; \tan 2 \theta = \frac{2 v_\sigma v_\Phi \lambda_{\Phi \sigma}}{v_\sigma^2 \lambda_{\sigma} + m_J^2 - v_\Phi^2 \lambda_{\Phi}} \; ,
\label{eq:mixingnondark}
\end{equation}
with the masses being,
\begin{equation}
   m_{h,H}^2 = \frac{1}{2} \left[v_\Phi^2 \lambda_{\Phi} + v_\sigma^2 \lambda_{\sigma} + m_J^2 \mp \sqrt{\left(v_\sigma^2 \lambda_{\sigma} + m_J^2 - v_\Phi^2 \lambda_{\Phi} \right)^2 + 4 v_\sigma^2 v_\Phi^2 \lambda_{\Phi \sigma}^2} \right] \; ,
   \label{eq:nondarkscalarmasses}
\end{equation}
where the lighter state $h$ corresponds to the SM Higgs boson with $m_h \simeq 125$ GeV~\cite{ParticleDataGroup:2026mpi}. The quartic couplings $\lambda_\Phi$, $\lambda_\sigma$ and $\lambda_{\Phi\sigma}$ can be written in terms of the masses $m_{h,H}$ and $m_J$ and the mixing angle $\theta$ as
\begin{align}
\lambda_\Phi =\frac{m_{h}^2\cos^2\theta+m_{H}^2\sin^2\theta}{v_\Phi^2} \; , \;
\lambda_\sigma =\frac{m_{h}^2\sin^2\theta+m_{H}^2\cos^2\theta-m_J^2}{v_\sigma^2} \; , \;
\lambda_{\Phi\sigma} =\frac{(m_{H}^2-m_{h}^2)}{2 v_\Phi v_\sigma} \sin 2\theta \; .
\label{eq:quarticreconstruction}
\end{align}

The \textit{dark}-scalar sector contains four \textit{dark}-neutral scalars, with their mass matrix in the $\left(\chi_{1 \text{R}}, \chi_{2 \text{R}}, \chi_{1 \text{I}}, \chi_{2 \text{I}} \right)$ basis written as,
\begin{align}
  \boldsymbol{\mathcal{M}}^2_{\chi} = \begin{pmatrix}
   \boldsymbol{\mathcal{M}}^2_{\chi_{\text{R}}} 
    & 0 \\
    0 & \boldsymbol{\mathcal{M}}^2_{\chi_{\text{I}}}
    \end{pmatrix} \; , \;  \boldsymbol{\mathcal{M}}^2_{\chi_{\text{R,I}}}= \begin{pmatrix}
   m_{\chi_1}^2 + \frac{1}{2} \left(v_\Phi^2 \lambda_{\Phi \chi_1}+ v_\sigma^2 \lambda_{\sigma \chi_1} \right) 
    & \pm \frac{v_\sigma \kappa}{\sqrt{2}} \\
    \cdot&m_{\chi_2}^2 + \frac{1}{2} \left(v_\Phi^2 \lambda_{\Phi \chi_2}+ v_\sigma^2 \lambda_{\sigma \chi_2} \right)
    \end{pmatrix} \; ,
    \label{eq:darkscalarmass}
\end{align}
where its block diagonal form shows that the CP-even and CP-odd components do not mix. The components of the $\chi_{1,2}$ scalars are related to the mass-eigenstates~$\zeta_{i}$ through the $4\times 4$ orthogonal matrix $\mathbf{V}$:
\begin{equation}
(\chi_{1 \text{R}},
\chi_{2 \text{R}},
\chi_{1 \text{I}}, \chi_{2 \text{I}})^T
 = \mathbf{V} 
(\zeta_1, \zeta_2, \zeta_3, \zeta_4)^T \; , \; \sqrt{2} \chi_1 = \sum_{k=1}^{4} \left(\mathbf{V}_{1 k} + i \mathbf{V}_{3 k} \right) \zeta_k \; , \; \sqrt{2} \chi_2 = \sum_{k=1}^{4} \left(\mathbf{V}_{2 k} + i \mathbf{V}_{4 k} \right) \zeta_k \; ,
\label{eq:mixneutralscotoDM}
\end{equation}
given by,
\begin{equation}
\mathbf{V} = \begin{pmatrix}
        \cos \xi & \sin \xi & 0 & 0 \\
        - \sin \xi & \cos \xi & 0 & 0 \\
        0 & 0 & \cos \xi & - \sin \xi \\
        0 & 0 & \sin \xi & \cos \xi \end{pmatrix} \; , \; \tan 2 \xi = \frac{- 2 \sqrt{2} v_\sigma \kappa}{2 \left(m_{\chi_1}^2 - m_{\chi_2}^2 \right) + v_\Phi^2 \left(\lambda_{\Phi \chi_1} - \lambda_{\Phi \chi_2} \right) + v_\sigma^2 \left(\lambda_{\sigma \chi_1} - \lambda_{\sigma \chi_2} \right)} \; .
\label{eq:mixingdark}
\end{equation}
The mass spectrum is degenerate,
\begin{align}
4 m_{\zeta_{1,2}}^2 = 4 m_{\zeta_{3,4}}^2 &= 2 \left(m_{\chi_1}^2 + m_{\chi_2}^2 \right) + v_\Phi^2 \left(\lambda_{\Phi \chi_1} + \lambda_{\Phi \chi_2} \right) + v_\sigma^2 \left(\lambda_{\sigma \chi_1} + \lambda_{\sigma \chi_2} \right) \nonumber \\ 
&\mp \sqrt{\left[2 \left(m_{\chi_1}^2 - m_{\chi_2}^2 \right) + v_\Phi^2 \left(\lambda_{\Phi \chi_1} - \lambda_{\Phi \chi_2} \right) + v_\sigma^2 \left(\lambda_{\sigma \chi_1} - \lambda_{\sigma \chi_2} \right) \right]^2 + 8 v_\sigma^2 \kappa^2} \; .
\label{eq:neutraldarkscalarmass}
\end{align}
%

\section{\emph{Dark}-fermion masses and mixing}
\label{sec:darkfermion}

The \emph{dark}-fermion sector, as shown in Table~\ref{tab:model}, is comprised of a vector-like doublet $D_{L,R}$, defined as
\begin{align}
D_{L,R} &=\begin{pmatrix}
f \\
\psi^{-} 
\end{pmatrix}_{L,R} \; ,
\label{eq:doubletfermion}
\end{align}
as well as, a vector-like singlet $F_{L,R}$. The doublet's neutral component $f$ mixes with singlet $F$ via
\begin{align}
- \mathcal{L} \supset \left(\overline{f_L} \; \overline{F_L} \right) \mathbf{M}_{DF} \begin{pmatrix}
 f_R \\
 F_R
\end{pmatrix} + \text{H. c.} \; , \; \mathbf{M}_{DF} = \begin{pmatrix}
 m_D & m_{DF} \\
 0 & m_F
\end{pmatrix} \; , \; m_{D,F} = \frac{v_\sigma \; y_{D,F}}{\sqrt{2}} \; , \; m_{DF} = \frac{v_\Phi \; y_{DF}}{\sqrt{2}} \; .
\label{eq:MassMatrixMixingdarkfermions}
\end{align}
The above mass matrix is bidiagonalised through the unitary transformations $\left(f, F\right)_{L,R}^T \rightarrow \mathbf{U}_{L,R} \left(\psi^0_1, \psi^0_2 \right)_{L,R}^T$:
\begin{equation}
    \mathbf{U}_{L}^\dagger \mathbf{M}_{DF} \mathbf{U}_{R} = \diag(m_{\psi^0_1}, m_{\psi^0_2}) \; ,
    \label{eq:neutraldarkfermiondiagonalise}
\end{equation}
with $m_{\psi^0_{1,2}}$ denoting the physical \emph{dark}-neutral fermion masses. The unitary matrices $\mathbf{U}_{L}$ and $\mathbf{U}_{R}$ diagonalize the Hermitian matrices $\mathbf{H}_{DF} = \mathbf{M}_{DF} \mathbf{M}_{DF}^\dagger$ and $\mathbf{H}_{DF}^\prime = \mathbf{M}_{DF}^\dagger \mathbf{M}_{DF}$, respectively, given by
\begin{equation}
    \mathbf{H}_{DF} = \begin{pmatrix}
        m_D^2 + m_{DF}^2 &  m_F m_{DF} \\
        \cdot & m_F^2
    \end{pmatrix} \; , \; \mathbf{H}_{DF}^\prime = \begin{pmatrix}
        m_D^2 & m_D m_{DF} \\
        \cdot & m_F^2 +  m_{DF}^2
    \end{pmatrix} \; , 
\end{equation}
so that,
\begin{equation}
    \mathbf{U}_{L}^\dagger \mathbf{H}_{DF} \mathbf{U}_{L} = \mathbf{U}_{R}^\dagger \mathbf{H}_{DF}^\prime \mathbf{U}_{R} = \diag(m_{\psi^0_1}^2, m_{\psi^0_2}^2) \; ,
\end{equation}
where,
\begin{equation}
  \mathbf{U}_{L,R} = \begin{pmatrix}
        \cos \theta_{L,R} & \sin \theta_{L,R} \\
        - \sin \theta_{L,R} & \cos \theta_{L,R}
    \end{pmatrix} \; , \; \tan 2 \theta_{L} = - \frac{2 m_F m_{DF}}{m_D^2 - m_F^2 + m_{DF}^2} \; , \; \tan 2 \theta_{R} = - \frac{2 m_D m_{DF} }{m_D^2 - m_F^2 - m_{DF}^2} \; .
    \label{eq:neutraldarkfermionmixing}
\end{equation}
The resulting \emph{dark}-neutral fermion masses are,
\begin{equation}
   m_{\psi^0_{1,2}}^2 = \frac{1}{2} \left[m_{D}^2 + m_{DF}^2 + m_F^2 \mp \sqrt{\left(m_{D}^2 + m_{DF}^2 + m_F^2 \right)^2 - 4 m_D^2 m_F^2} \right] \; ,
   \label{eq:neutraldarkfermionmass}
\end{equation}
while the mass of the \emph{dark}-charged fermion $\psi^{\pm}$ is simply given by,
\begin{equation}
    m_{\psi^{\pm}} = m_D \; .
    \label{eq:chargedfemionmass}
\end{equation}
%

\section{\emph{Dark}-loop functions}
\label{sec:darkloop}

The dimensionless \emph{dark} one-loop factors relevant for neutrino mass generation are written as,
\begin{align}
\mathcal{F}_S\left(m_{\psi^0_a} , m_{\zeta_k}\right) &  = \frac{1}{m_{\psi^0_2}} \sum_{a=1}^{2} \sum_{k=1}^{4} \mathcal{F}\left(m_{\psi^0_a} , m_{\zeta_k}\right) m_{\psi^0_a} \; \left(\mathbf{V}_{1 k} + i \mathbf{V}_{3 k} \right) \left(\mathbf{V}_{2 k} + i \mathbf{V}_{4 k} \right)  \; \left(\mathbf{U}_{L}\right)_{a 2}^\ast \left(\mathbf{U}_{R}\right)_{1 a} \;, \nonumber \\ 
\mathcal{F}_R\left(m_{\psi^0_a} , m_{\zeta_k}\right)  & = \frac{1}{m_{\psi^0_2}} \sum_{a=1}^{2} \sum_{k=1}^{4} \mathcal{F}\left(m_{\psi^0_a} , m_{\zeta_k}\right) m_{\psi^0_a} \; \left|\mathbf{V}_{2 k} + i \mathbf{V}_{4 k} \right|^2 \; \left(\mathbf{U}_{L}\right)_{a 2}^\ast \left(\mathbf{U}_{R}\right)_{2 a} \; ,
\label{eq:loopfactors}
\end{align}
which depend on the \emph{dark} neutral scalar and fermion masses, denoted $m_{\zeta_k}$ ($k=1,\cdots,4$) and $m_{\psi^0_a}$ ($a=1,2$), respectively, given by Eq.~\eqref{eq:neutraldarkscalarmass} and~\eqref{eq:neutraldarkfermionmass}, as well as, the \emph{dark} neutral scalar and fermion mixing, encoded, respectively, by the orthogonal matrix $\mathbf{V}$ of Eq.~\eqref{eq:mixingdark} and the unitary matrices $\mathbf{U}_{L,R}$ of Eq.~\eqref{eq:neutraldarkfermionmixing} -- see Appendix~\ref{sec:scalarsector} and~\ref{sec:darkfermion}. In the above, the loop function $\mathcal{F}$ is given by,
\begin{align}
 \mathcal{F}\left(m_{\psi^0_a} , m_{\zeta_k}\right) = & \; \frac{1}{32 \pi^2}  \frac{1}{m_{\psi^0_a}^2 - m_{\zeta_k}^2} \left(m_{\psi^0_a}^2 \ln m_{\psi^0_a}^2 - m_{\zeta_k}^2 \ln m_{\zeta_k}^2 \right) \; .
\label{eq:loopfunction}
\end{align}
%

\section{Generalised Casas-Ibarra parameterisation}
\label{sec:CasasIbarra}

To obtain the Yukawa coupling vectors $\mathbf{Y}_{\nu, \chi}$ of Eq.~\eqref{eq:LYuk}, that are compatible with neutrino data, we make use of a generalised version of the Casas-Ibarra parameterisation~\cite{Casas:2001sr}. Namely, we start by writing $\mathbf{M}_\nu$ of Eq.~\eqref{eq:Mnu} as~\cite{Batra:2023bqj},
\begin{equation}
  \mathbf{M}_{\nu} = - \frac{v_\Phi^2}{2} \mathbf{Y} \; \mathbf{M}^{-1} \mathbf{Y}^T \; ,\; \mathbf{Y} = \left(\mathbf{Y}_{\nu} \; \mathbf{Y}_{\chi}\right) \; , \; \mathbf{M} = \begin{pmatrix}
      0 & |\tilde{M}| e^{i \phi}\\
      |\tilde{M}| e^{i \phi} & 0
  \end{pmatrix} \; , \; \tilde{M} = \frac{v_\Phi}{\sqrt{2}} \frac{M_R}{\mathcal{F}_S~m_{\psi^0_2}~y_S} \; , \; \phi = \arg \tilde{M} \; ,
\end{equation}
where we define the $3\times 2$ Yukawa matrix $\mathbf{Y}$ and $2\times 2$ mass matrix $\mathbf{M}$. The latter is written in terms of $M_R$ given by Eqs.~\eqref{eq:treemassmatrix}, \eqref{eq:loopmassmatrix} and~\eqref{eq:mSandDeltaMR}, and the Yukawa parameter $y_S$ and \emph{dark}-loop function $\mathcal{F}_S$ of Eqs.~\eqref{eq:LYuk} and~\eqref{eq:loopfactors}, respectively. The matrix $\mathbf{M}$ can be diagonalised as follows
\begin{align}
\U_M^{\dagger}\, \M\, \U_M^\ast &= \mathbf{d}_{M} = \text{diag}\left(|\tilde{M}| , |\tilde{M}| \right) \; , \;
\U_M = \frac{e^{i \phi}}{\sqrt{2}} \begin{pmatrix}
    i & 1 \\
    -i & 1
\end{pmatrix} \; .
\label{eq:Mheavy}
\end{align}
From the above we obtain the following parameterisation for $\mathbf{Y}$,
\begin{equation} \label{eq:CI}
    \mathbf{Y} = i \frac{\sqrt{2}}{v_\Phi} \mathbf{U}_{\nu}^\ast\, \mathbf{d}_\nu^{1/2}\, \mathbf{R}\, \mathbf{d}_M^{1/2} \mathbf{U}_M^T \; , \; \mathbf{R}_\text{NO} = \begin{pmatrix}
        0 & 0 \\
        \cos (z) & - \sin (z) \\
        \pm \sin (z) & \pm \cos (z)
    \end{pmatrix} \; , \; 
   \mathbf{R}_\text{IO} = \begin{pmatrix}
        \cos (z) & - \sin (z) \\
        \pm \sin (z) & \pm \cos (z) \\
         0 & 0
    \end{pmatrix} \,,
\end{equation}
where $\mathbf{d}_\nu$, $\mathbf{U}_{\nu}$, $\mathbf{d}_M$ and $\mathbf{U}_M$ are given by Eqs.~\eqref{eq:Unudef} and~\eqref{eq:Mheavy}, respectively. $\mathbf{R}$ is an orthogonal $3\times 2$ matrix
with $z$ being a complex parameter.




\begin{thebibliography}{100}

\bibitem{Kajita:2016cak}
T.~Kajita, ``{Nobel Lecture: Discovery of atmospheric neutrino oscillations},'' \href{http://dx.doi.org/10.1103/RevModPhys.88.030501}{{\em Rev. Mod. Phys.} {\bfseries 88} no.~3, (2016) 030501}.

\bibitem{McDonald:2016ixn}
A.~B. McDonald, ``{Nobel Lecture: The Sudbury Neutrino Observatory: Observation of flavor change for solar neutrinos},'' \href{http://dx.doi.org/10.1103/RevModPhys.88.030502}{{\em Rev. Mod. Phys.} {\bfseries 88} no.~3, (2016) 030502}.

\bibitem{Planck:2018vyg}
{\bfseries Planck} Collaboration, N.~Aghanim {\em et~al.}, ``{Planck 2018 results. VI. Cosmological parameters},'' \href{http://dx.doi.org/10.1051/0004-6361/201833910}{{\em Astron. Astrophys.} {\bfseries 641} (2020) A6}, \href{http://arxiv.org/abs/1807.06209}{{\ttfamily arXiv:1807.06209 [astro-ph.CO]}}. [Erratum: Astron.Astrophys. 652, C4 (2021)].

\bibitem{Cirelli:2024ssz}
M.~Cirelli, A.~Strumia, and J.~Zupan, ``{Dark Matter},'' \href{http://arxiv.org/abs/2406.01705}{{\ttfamily arXiv:2406.01705 [hep-ph]}}.

\bibitem{Minkowski:1977sc}
P.~Minkowski, ``{$\mu \to e\gamma$ at a Rate of One Out of $10^{9}$ Muon Decays?},'' \href{http://dx.doi.org/10.1016/0370-2693(77)90435-X}{{\em Phys. Lett. B} {\bfseries 67} (1977) 421--428}.

\bibitem{Gell-Mann:1979vob}
M.~Gell-Mann, P.~Ramond, and R.~Slansky, ``{Complex Spinors and Unified Theories},'' {\em Conf. Proc. C} {\bfseries 790927} (1979) 315--321, \href{http://arxiv.org/abs/1306.4669}{{\ttfamily arXiv:1306.4669 [hep-th]}}.

\bibitem{Yanagida:1979as}
T.~Yanagida, ``{Horizontal gauge symmetry and masses of neutrinos},'' {\em Conf. Proc. C} {\bfseries 7902131} (1979) 95--99.

\bibitem{Schechter:1980gr}
J.~Schechter and J.~W.~F. Valle, ``{Neutrino Masses in SU(2) x U(1) Theories},'' \href{http://dx.doi.org/10.1103/PhysRevD.22.2227}{{\em Phys. Rev. D} {\bfseries 22} (1980) 2227}.

\bibitem{Glashow:1979nm}
S.~L. Glashow, ``{The Future of Elementary Particle Physics},'' \href{http://dx.doi.org/10.1007/978-1-4684-7197-7_15}{{\em NATO Sci. Ser. B} {\bfseries 61} (1980) 687}.

\bibitem{Mohapatra:1979ia}
R.~N. Mohapatra and G.~Senjanovic, ``{Neutrino Mass and Spontaneous Parity Nonconservation},'' \href{http://dx.doi.org/10.1103/PhysRevLett.44.912}{{\em Phys. Rev. Lett.} {\bfseries 44} (1980) 912}.

\bibitem{Zee:1980ai}
A.~Zee, ``{A Theory of Lepton Number Violation, Neutrino Majorana Mass, and Oscillation},'' \href{http://dx.doi.org/10.1016/0370-2693(80)90349-4}{{\em Phys. Lett. B} {\bfseries 93} (1980) 389}. [Erratum: Phys.Lett.B 95, 461 (1980)].

\bibitem{Ma:1998dn}
E.~Ma, ``{Pathways to naturally small neutrino masses},'' \href{http://dx.doi.org/10.1103/PhysRevLett.81.1171}{{\em Phys. Rev. Lett.} {\bfseries 81} (1998) 1171--1174}, \href{http://arxiv.org/abs/hep-ph/9805219}{{\ttfamily arXiv:hep-ph/9805219}}.

\bibitem{Cai:2017jrq}
Y.~Cai, J.~Herrero-Garc{\'\i}a, M.~A. Schmidt, A.~Vicente, and R.~R. Volkas, ``{From the trees to the forest: a review of radiative neutrino mass models},'' \href{http://dx.doi.org/10.3389/fphy.2017.00063}{{\em Front. in Phys.} {\bfseries 5} (2017) 63}, \href{http://arxiv.org/abs/1706.08524}{{\ttfamily arXiv:1706.08524 [hep-ph]}}.

\bibitem{Mohapatra:1986bd}
R.~N. Mohapatra and J.~W.~F. Valle, ``{Neutrino Mass and Baryon Number Nonconservation in Superstring Models},'' \href{http://dx.doi.org/10.1103/PhysRevD.34.1642}{{\em Phys. Rev. D} {\bfseries 34} (1986) 1642}.

\bibitem{Akhmedov:1995ip}
E.~K. Akhmedov, M.~Lindner, E.~Schnapka, and J.~W.~F. Valle, ``{Left-right symmetry breaking in NJL approach},'' \href{http://dx.doi.org/10.1016/0370-2693(95)01504-3}{{\em Phys. Lett. B} {\bfseries 368} (1996) 270--280}, \href{http://arxiv.org/abs/hep-ph/9507275}{{\ttfamily arXiv:hep-ph/9507275}}.

\bibitem{Akhmedov:1995vm}
E.~K. Akhmedov, M.~Lindner, E.~Schnapka, and J.~W.~F. Valle, ``{Dynamical left-right symmetry breaking},'' \href{http://dx.doi.org/10.1103/PhysRevD.53.2752}{{\em Phys. Rev. D} {\bfseries 53} (1996) 2752--2780}, \href{http://arxiv.org/abs/hep-ph/9509255}{{\ttfamily arXiv:hep-ph/9509255}}.

\bibitem{Malinsky:2005bi}
M.~Malinsky, J.~C. Romao, and J.~W.~F. Valle, ``{Novel supersymmetric SO(10) seesaw mechanism},'' \href{http://dx.doi.org/10.1103/PhysRevLett.95.161801}{{\em Phys. Rev. Lett.} {\bfseries 95} (2005) 161801}, \href{http://arxiv.org/abs/hep-ph/0506296}{{\ttfamily arXiv:hep-ph/0506296}}.

\bibitem{tHooft:1979rat}
G.~'t~Hooft, ``{Naturalness, chiral symmetry, and spontaneous chiral symmetry breaking},'' \href{http://dx.doi.org/10.1007/978-1-4684-7571-5_9}{{\em NATO Sci. Ser. B} {\bfseries 59} (1980) 135--157}.

\bibitem{Alonso:2012ji}
R.~Alonso, M.~Dhen, M.~B. Gavela, and T.~Hambye, ``{Muon conversion to electron in nuclei in type-I seesaw models},'' \href{http://dx.doi.org/10.1007/JHEP01(2013)118}{{\em JHEP} {\bfseries 01} (2013) 118}, \href{http://arxiv.org/abs/1209.2679}{{\ttfamily arXiv:1209.2679 [hep-ph]}}.

\bibitem{Abada:2015oba}
A.~Abada, V.~De~Romeri, and A.~M. Teixeira, ``{Impact of sterile neutrinos on nuclear-assisted cLFV processes},'' \href{http://dx.doi.org/10.1007/JHEP02(2016)083}{{\em JHEP} {\bfseries 02} (2016) 083}, \href{http://arxiv.org/abs/1510.06657}{{\ttfamily arXiv:1510.06657 [hep-ph]}}.

\bibitem{Abada:2018nio}
A.~Abada and A.~M. Teixeira, ``{Heavy neutral leptons and high-intensity observables},'' \href{http://dx.doi.org/10.3389/fphy.2018.00142}{{\em Front. in Phys.} {\bfseries 6} (2018) 142}, \href{http://arxiv.org/abs/1812.08062}{{\ttfamily arXiv:1812.08062 [hep-ph]}}.

\bibitem{Hernandez-Tome:2019lkb}
G.~Hern{\'a}ndez-Tom{\'e}, J.~I. Illana, M.~Masip, G.~L{\'o}pez~Castro, and P.~Roig, ``{Effects of heavy Majorana neutrinos on lepton flavor violating processes},'' \href{http://dx.doi.org/10.1103/PhysRevD.101.075020}{{\em Phys. Rev. D} {\bfseries 101} no.~7, (2020) 075020}, \href{http://arxiv.org/abs/1912.13327}{{\ttfamily arXiv:1912.13327 [hep-ph]}}.

\bibitem{Camara:2020efq}
H.~B. Camara, R.~G. Felipe, and F.~R. Joaquim, ``{Minimal inverse-seesaw mechanism with Abelian flavour symmetries},'' \href{http://dx.doi.org/10.1007/JHEP05(2021)021}{{\em JHEP} {\bfseries 05} (2021) 021}, \href{http://arxiv.org/abs/2012.04557}{{\ttfamily arXiv:2012.04557 [hep-ph]}}.

\bibitem{Batra:2022arl}
A.~Batra, P.~Bharadwaj, S.~Mandal, R.~Srivastava, and J.~W.~F. Valle, ``{W-mass anomaly in the simplest linear seesaw mechanism},'' \href{http://dx.doi.org/10.1016/j.physletb.2022.137408}{{\em Phys. Lett. B} {\bfseries 834} (2022) 137408}, \href{http://arxiv.org/abs/2208.04983}{{\ttfamily arXiv:2208.04983 [hep-ph]}}.

\bibitem{Batra:2023ssq}
A.~Batra, P.~Bharadwaj, S.~Mandal, R.~Srivastava, and J.~W.~F. Valle, ``{Large lepton number violation at colliders: Predictions from the minimal linear seesaw mechanism},'' \href{http://dx.doi.org/10.1016/j.physletb.2024.139204}{{\em Phys. Lett. B} {\bfseries 860} (2025) 139204}, \href{http://arxiv.org/abs/2304.06080}{{\ttfamily arXiv:2304.06080 [hep-ph]}}.

\bibitem{Batra:2023mds}
A.~Batra, P.~Bharadwaj, S.~Mandal, R.~Srivastava, and J.~W.~F. Valle, ``{Phenomenology of the simplest linear seesaw mechanism},'' \href{http://dx.doi.org/10.1007/JHEP07(2023)221}{{\em JHEP} {\bfseries 07} (2023) 221}, \href{http://arxiv.org/abs/2305.00994}{{\ttfamily arXiv:2305.00994 [hep-ph]}}.

\bibitem{Mandal:2026wbz}
S.~Mandal, R.~Roshan, and J.~W.~F. Valle, ``{Particle and gravitational wave probes of minimal seesaw neutrinos},'' \href{http://dx.doi.org/10.1007/JHEP09(2026)173}{{\em JHEP} {\bfseries 09} (2026) 173}, \href{http://arxiv.org/abs/2606.15316}{{\ttfamily arXiv:2606.15316 [hep-ph]}}.

\bibitem{Tao:1996vb}
Z.-j. Tao, ``{Radiative seesaw mechanism at weak scale},'' \href{http://dx.doi.org/10.1103/PhysRevD.54.5693}{{\em Phys. Rev. D} {\bfseries 54} (1996) 5693--5697}, \href{http://arxiv.org/abs/hep-ph/9603309}{{\ttfamily arXiv:hep-ph/9603309}}.

\bibitem{Ma:2006km}
E.~Ma, ``{Verifiable radiative seesaw mechanism of neutrino mass and dark matter},'' \href{http://dx.doi.org/10.1103/PhysRevD.73.077301}{{\em Phys. Rev. D} {\bfseries 73} (2006) 077301}, \href{http://arxiv.org/abs/hep-ph/0601225}{{\ttfamily arXiv:hep-ph/0601225}}.

\bibitem{Avila:2025qsc}
I.~M. {\'A}vila, A.~Karan, S.~Mandal, S.~Sadhukhan, and J.~W.~F. Valle, ``{Dark matter as the source of neutrino mass: Theory overview and experimental prospects},'' \href{http://dx.doi.org/10.1016/j.physrep.2026.02.003}{{\em Phys. Rept.} {\bfseries 1173} (2026) 1--81}, \href{http://arxiv.org/abs/2506.24027}{{\ttfamily arXiv:2506.24027 [hep-ph]}}.

\bibitem{Batra:2023bqj}
A.~Batra, H.~B. C{\^a}mara, and F.~R. Joaquim, ``{Dark linear seesaw mechanism},'' \href{http://dx.doi.org/10.1016/j.physletb.2023.138012}{{\em Phys. Lett. B} {\bfseries 843} (2023) 138012}, \href{http://arxiv.org/abs/2305.01687}{{\ttfamily arXiv:2305.01687 [hep-ph]}}.

\bibitem{CarcamoHernandez:2023atk}
A.~E. C{\'a}rcamo~Hern{\'a}ndez, V.~K.~N., and J.~W.~F. Valle, ``{Linear seesaw mechanism from dark sector},'' \href{http://dx.doi.org/10.1007/JHEP09(2023)046}{{\em JHEP} {\bfseries 09} (2023) 046}, \href{http://arxiv.org/abs/2305.02273}{{\ttfamily arXiv:2305.02273 [hep-ph]}}.

\bibitem{Wang:2015saa}
W.~Wang and Z.-L. Han, ``{Radiative linear seesaw model, dark matter, and $U(1)_{B-L}$},'' \href{http://dx.doi.org/10.1103/PhysRevD.92.095001}{{\em Phys. Rev. D} {\bfseries 92} (2015) 095001}, \href{http://arxiv.org/abs/1508.00706}{{\ttfamily arXiv:1508.00706 [hep-ph]}}.

\bibitem{Das:2017ski}
A.~Das, T.~Nomura, H.~Okada, and S.~Roy, ``{Generation of a radiative neutrino mass in the linear seesaw framework, charged lepton flavor violation, and dark matter},'' \href{http://dx.doi.org/10.1103/PhysRevD.96.075001}{{\em Phys. Rev. D} {\bfseries 96} no.~7, (2017) 075001}, \href{http://arxiv.org/abs/1704.02078}{{\ttfamily arXiv:1704.02078 [hep-ph]}}.

\bibitem{CarcamoHernandez:2021tlv}
A.~E. C{\'a}rcamo~Hern{\'a}ndez, S.~Kovalenko, F.~S. Queiroz, and Y.~S. Villamizar, ``{An extended 3-3-1 model with radiative linear seesaw mechanism},'' \href{http://dx.doi.org/10.1016/j.physletb.2022.137082}{{\em Phys. Lett. B} {\bfseries 829} (2022) 137082}, \href{http://arxiv.org/abs/2105.01731}{{\ttfamily arXiv:2105.01731 [hep-ph]}}.

\bibitem{CarcamoHernandez:2024edi}
A.~E. C{\'a}rcamo~Hern{\'a}ndez, Y.~H. Vel{\'a}squez, S.~Kovalenko, N.~A. P{\'e}rez-Julve, and I.~Schmidt, ``{Models of Radiative Linear Seesaw with Electrically Charged Mediators},'' \href{http://dx.doi.org/10.1093/ptep/ptae142}{{\em PTEP} {\bfseries 2024} no.~10, (2024) 103B02}, \href{http://arxiv.org/abs/2403.05637}{{\ttfamily arXiv:2403.05637 [hep-ph]}}.

\bibitem{Abada:2025ozu}
A.~Abada, A.~E. C{\'a}rcamo~Hern{\'a}ndez, and S.~Urrea, ``{Dynamical scotogenic generation of the linear and inverse seesaws},'' \href{http://dx.doi.org/10.1007/JHEP05(2026)086}{{\em JHEP} {\bfseries 05} (2026) 086}, \href{http://arxiv.org/abs/2512.12029}{{\ttfamily arXiv:2512.12029 [hep-ph]}}.

\bibitem{Ma:2009gu}
E.~Ma, ``{Radiative inverse seesaw mechanism for nonzero neutrino mass},'' \href{http://dx.doi.org/10.1103/PhysRevD.80.013013}{{\em Phys. Rev. D} {\bfseries 80} (2009) 013013}, \href{http://arxiv.org/abs/0904.4450}{{\ttfamily arXiv:0904.4450 [hep-ph]}}.

\bibitem{Bazzocchi:2009kc}
F.~Bazzocchi, D.~G. Cerdeno, C.~Munoz, and J.~W.~F. Valle, ``{Calculable inverse-seesaw neutrino masses in supersymmetry},'' \href{http://dx.doi.org/10.1103/PhysRevD.81.051701}{{\em Phys. Rev. D} {\bfseries 81} (2010) 051701}, \href{http://arxiv.org/abs/0907.1262}{{\ttfamily arXiv:0907.1262 [hep-ph]}}.

\bibitem{Baldes:2013eva}
I.~Baldes, N.~F. Bell, K.~Petraki, and R.~R. Volkas, ``{Two radiative inverse seesaw models, dark matter, and baryogenesis},'' \href{http://dx.doi.org/10.1088/1475-7516/2013/07/029}{{\em JCAP} {\bfseries 07} (2013) 029}, \href{http://arxiv.org/abs/1304.6162}{{\ttfamily arXiv:1304.6162 [hep-ph]}}.

\bibitem{Mandal:2019oth}
S.~Mandal, N.~Rojas, R.~Srivastava, and J.~W.~F. Valle, ``{Dark matter as the origin of neutrino mass in the inverse seesaw mechanism},'' \href{http://dx.doi.org/10.1016/j.physletb.2021.136609}{{\em Phys. Lett. B} {\bfseries 821} (2021) 136609}, \href{http://arxiv.org/abs/1907.07728}{{\ttfamily arXiv:1907.07728 [hep-ph]}}.

\bibitem{Abada:2021yot}
A.~Abada, N.~Bernal, A.~E.~C. Hern{\'a}ndez, X.~Marcano, and G.~Piazza, ``{Gauged inverse seesaw from dark matter},'' \href{http://dx.doi.org/10.1140/epjc/s10052-021-09535-5}{{\em Eur. Phys. J. C} {\bfseries 81} no.~8, (2021) 758}, \href{http://arxiv.org/abs/2107.02803}{{\ttfamily arXiv:2107.02803 [hep-ph]}}.

\bibitem{MEGII:2018kmf}
{\bfseries MEG II} Collaboration, A.~M. Baldini {\em et~al.}, ``{The design of the MEG II experiment},'' \href{http://dx.doi.org/10.1140/epjc/s10052-018-5845-6}{{\em Eur. Phys. J. C} {\bfseries 78} no.~5, (2018) 380}, \href{http://arxiv.org/abs/1801.04688}{{\ttfamily arXiv:1801.04688 [physics.ins-det]}}.

\bibitem{Cohen:2011ec}
T.~Cohen, J.~Kearney, A.~Pierce, and D.~Tucker-Smith, ``{Singlet-Doublet Dark Matter},'' \href{http://dx.doi.org/10.1103/PhysRevD.85.075003}{{\em Phys. Rev. D} {\bfseries 85} (2012) 075003}, \href{http://arxiv.org/abs/1109.2604}{{\ttfamily arXiv:1109.2604 [hep-ph]}}.

\bibitem{Cheung:2013dua}
C.~Cheung and D.~Sanford, ``{Simplified Models of Mixed Dark Matter},'' \href{http://dx.doi.org/10.1088/1475-7516/2014/02/011}{{\em JCAP} {\bfseries 02} (2014) 011}, \href{http://arxiv.org/abs/1311.5896}{{\ttfamily arXiv:1311.5896 [hep-ph]}}.

\bibitem{Calibbi:2015nha}
L.~Calibbi, A.~Mariotti, and P.~Tziveloglou, ``{Singlet-Doublet Model: Dark matter searches and LHC constraints},'' \href{http://dx.doi.org/10.1007/JHEP10(2015)116}{{\em JHEP} {\bfseries 10} (2015) 116}, \href{http://arxiv.org/abs/1505.03867}{{\ttfamily arXiv:1505.03867 [hep-ph]}}.

\bibitem{Yaguna:2015mva}
C.~E. Yaguna, ``{Singlet-Doublet Dirac Dark Matter},'' \href{http://dx.doi.org/10.1103/PhysRevD.92.115002}{{\em Phys. Rev. D} {\bfseries 92} no.~11, (2015) 115002}, \href{http://arxiv.org/abs/1510.06151}{{\ttfamily arXiv:1510.06151 [hep-ph]}}.

\bibitem{Banerjee:2016hsk}
S.~Banerjee, S.~Matsumoto, K.~Mukaida, and Y.-L.~S. Tsai, ``{WIMP Dark Matter in a Well-Tempered Regime: A case study on Singlet-Doublets Fermionic WIMP},'' \href{http://dx.doi.org/10.1007/JHEP11(2016)070}{{\em JHEP} {\bfseries 11} (2016) 070}, \href{http://arxiv.org/abs/1603.07387}{{\ttfamily arXiv:1603.07387 [hep-ph]}}.

\bibitem{Paul:2024prs}
P.~K. Paul, S.~K. Sahoo, and N.~Sahu, ``{Anatomy of singlet-doublet dark matter relic: annihilation, co-annihilation, co-scattering, and freeze-in},'' \href{http://dx.doi.org/10.1088/1475-7516/2025/10/053}{{\em JCAP} {\bfseries 10} (2025) 053}, \href{http://arxiv.org/abs/2412.02607}{{\ttfamily arXiv:2412.02607 [hep-ph]}}. [Erratum: JCAP 01, E02 (2026)].

\bibitem{Restrepo:2015ura}
D.~Restrepo, A.~Rivera, M.~S{\'a}nchez-Pel{\'a}ez, O.~Zapata, and W.~Tangarife, ``{Radiative Neutrino Masses in the Singlet-Doublet Fermion Dark Matter Model with Scalar Singlets},'' \href{http://dx.doi.org/10.1103/PhysRevD.92.013005}{{\em Phys. Rev. D} {\bfseries 92} no.~1, (2015) 013005}, \href{http://arxiv.org/abs/1504.07892}{{\ttfamily arXiv:1504.07892 [hep-ph]}}.

\bibitem{Bhattacharya:2018fus}
S.~Bhattacharya, P.~Ghosh, N.~Sahoo, and N.~Sahu, ``{Mini Review on Vector-Like Leptonic Dark Matter, Neutrino Mass, and Collider Signatures},'' \href{http://dx.doi.org/10.3389/fphy.2019.00080}{{\em Front. in Phys.} {\bfseries 7} (2019) 80}, \href{http://arxiv.org/abs/1812.06505}{{\ttfamily arXiv:1812.06505 [hep-ph]}}.

\bibitem{Konar:2020wvl}
P.~Konar, A.~Mukherjee, A.~K. Saha, and S.~Show, ``{Linking pseudo-Dirac dark matter to radiative neutrino masses in a singlet-doublet scenario},'' \href{http://dx.doi.org/10.1103/PhysRevD.102.015024}{{\em Phys. Rev. D} {\bfseries 102} no.~1, (2020) 015024}, \href{http://arxiv.org/abs/2001.11325}{{\ttfamily arXiv:2001.11325 [hep-ph]}}.

\bibitem{Paul:2026snc}
P.~K. Paul, N.~Sahu, and S.~Sharma, ``{Singlet-doublet dark matter induced radiative neutrino mass and TeV scale leptogenesis},'' \href{http://dx.doi.org/10.1088/1475-7516/2026/10/092}{{\em JCAP} {\bfseries 10} (2026) 092}, \href{http://arxiv.org/abs/2605.04824}{{\ttfamily arXiv:2605.04824 [hep-ph]}}.

\bibitem{Esch:2018ccs}
S.~Esch, M.~Klasen, and C.~E. Yaguna, ``{A singlet doublet dark matter model with radiative neutrino masses},'' \href{http://dx.doi.org/10.1007/JHEP10(2018)055}{{\em JHEP} {\bfseries 10} (2018) 055}, \href{http://arxiv.org/abs/1804.03384}{{\ttfamily arXiv:1804.03384 [hep-ph]}}.

\bibitem{Sarazin:2021nwo}
M.~Sarazin, J.~Bernigaud, and B.~Herrmann, ``{Dark matter and lepton flavour phenomenology in a singlet-doublet scotogenic model},'' \href{http://dx.doi.org/10.1007/JHEP12(2021)116}{{\em JHEP} {\bfseries 12} (2021) 116}, \href{http://arxiv.org/abs/2107.04613}{{\ttfamily arXiv:2107.04613 [hep-ph]}}.

\bibitem{Alvarez:2023dzz}
A.~Alvarez, A.~Banik, R.~Cepedello, B.~Herrmann, W.~Porod, M.~Sarazin, and M.~Schnelke, ``{Accommodating muon (g {\ensuremath{-}} 2) and leptogenesis in a scotogenic model},'' \href{http://dx.doi.org/10.1007/JHEP06(2023)163}{{\em JHEP} {\bfseries 06} (2023) 163}, \href{http://arxiv.org/abs/2301.08485}{{\ttfamily arXiv:2301.08485 [hep-ph]}}.

\bibitem{Darricau:2025vcs}
A.~Darricau, H.~Lee, J.~Orloff, and A.~M. Teixeira, ``{Flavour and precision probes of a class of scotogenic models},'' \href{http://dx.doi.org/10.1140/epjc/s10052-025-14946-9}{{\em Eur. Phys. J. C} {\bfseries 85} no.~10, (2025) 1234}, \href{http://arxiv.org/abs/2506.23383}{{\ttfamily arXiv:2506.23383 [hep-ph]}}.

\bibitem{Darricau:2026iwg}
A.~Darricau and A.~M. Teixeira, ``{Revisiting cLFV in ''T1-2-A'' scotogenic models: asymmetries in three-body lepton decays},'' \href{http://arxiv.org/abs/2607.07484}{{\ttfamily arXiv:2607.07484 [hep-ph]}}.

\bibitem{Chikashige:1980ui}
Y.~Chikashige, R.~N. Mohapatra, and R.~D. Peccei, ``{Are There Real Goldstone Bosons Associated with Broken Lepton Number?},'' \href{http://dx.doi.org/10.1016/0370-2693(81)90011-3}{{\em Phys. Lett. B} {\bfseries 98} (1981) 265--268}.

\bibitem{Schechter:1981cv}
J.~Schechter and J.~W.~F. Valle, ``{Neutrino Decay and Spontaneous Violation of Lepton Number},'' \href{http://dx.doi.org/10.1103/PhysRevD.25.774}{{\em Phys. Rev. D} {\bfseries 25} (1982) 774}.

\bibitem{Ma:2015xla}
E.~Ma, ``{Derivation of Dark Matter Parity from Lepton Parity},'' \href{http://dx.doi.org/10.1103/PhysRevLett.115.011801}{{\em Phys. Rev. Lett.} {\bfseries 115} no.~1, (2015) 011801}, \href{http://arxiv.org/abs/1502.02200}{{\ttfamily arXiv:1502.02200 [hep-ph]}}.

\bibitem{Escribano:2020wua}
P.~Escribano and A.~Vicente, ``{Ultralight scalars in leptonic observables},'' \href{http://dx.doi.org/10.1007/JHEP03(2021)240}{{\em JHEP} {\bfseries 03} (2021) 240}, \href{http://arxiv.org/abs/2008.01099}{{\ttfamily arXiv:2008.01099 [hep-ph]}}.

\bibitem{Herrero-Brocal:2023czw}
A.~Herrero-Brocal and A.~Vicente, ``{The majoron coupling to charged leptons},'' \href{http://dx.doi.org/10.1007/JHEP01(2024)078}{{\em JHEP} {\bfseries 01} (2024) 078}, \href{http://arxiv.org/abs/2311.10145}{{\ttfamily arXiv:2311.10145 [hep-ph]}}.

\bibitem{Heeck:2019guh}
J.~Heeck and H.~H. Patel, ``{Majoron at two loops},'' \href{http://dx.doi.org/10.1103/PhysRevD.100.095015}{{\em Phys. Rev. D} {\bfseries 100} no.~9, (2019) 095015}, \href{http://arxiv.org/abs/1909.02029}{{\ttfamily arXiv:1909.02029 [hep-ph]}}.

\bibitem{Escribano:2021uhf}
P.~Escribano, M.~Hirsch, J.~Nava, and A.~Vicente, ``{Observable flavor violation from spontaneous lepton number breaking},'' \href{http://dx.doi.org/10.1007/JHEP01(2022)098}{{\em JHEP} {\bfseries 01} (2022) 098}, \href{http://arxiv.org/abs/2108.01101}{{\ttfamily arXiv:2108.01101 [hep-ph]}}.

\bibitem{CentellesChulia:2024uzv}
S.~Centelles~Chuli{\'a}, A.~Herrero-Brocal, and A.~Vicente, ``{The Type-I Seesaw family},'' \href{http://dx.doi.org/10.1007/JHEP07(2024)060}{{\em JHEP} {\bfseries 07} (2024) 060}, \href{http://arxiv.org/abs/2404.15415}{{\ttfamily arXiv:2404.15415 [hep-ph]}}.

\bibitem{CentellesChulia:2025eck}
S.~Centelles~Chuli{\'a}, T.~Herbermann, A.~Herrero-Brocal, and A.~Vicente, ``{Flavour and cosmological probes of Diracon models},'' \href{http://dx.doi.org/10.1007/JHEP09(2025)110}{{\em JHEP} {\bfseries 09} (2025) 110}, \href{http://arxiv.org/abs/2506.06449}{{\ttfamily arXiv:2506.06449 [hep-ph]}}.

\bibitem{Bonilla:2019ipe}
C.~Bonilla, L.~M.~G. de~la Vega, J.~M. Lamprea, R.~A. Lineros, and E.~Peinado, ``{Fermion Dark Matter and Radiative Neutrino Masses from Spontaneous Lepton Number Breaking},'' \href{http://dx.doi.org/10.1088/1367-2630/ab7254}{{\em New J. Phys.} {\bfseries 22} no.~3, (2020) 033009}, \href{http://arxiv.org/abs/1908.04276}{{\ttfamily arXiv:1908.04276 [hep-ph]}}.

\bibitem{Escribano:2021ymx}
P.~Escribano and A.~Vicente, ``{An ultraviolet completion for the Scotogenic model},'' \href{http://dx.doi.org/10.1016/j.physletb.2021.136717}{{\em Phys. Lett. B} {\bfseries 823} (2021) 136717}, \href{http://arxiv.org/abs/2107.10265}{{\ttfamily arXiv:2107.10265 [hep-ph]}}.

\bibitem{DeRomeri:2022cem}
V.~De~Romeri, J.~Nava, M.~Puerta, and A.~Vicente, ``{Dark matter in the scotogenic model with spontaneous lepton number violation},'' \href{http://dx.doi.org/10.1103/PhysRevD.107.095019}{{\em Phys. Rev. D} {\bfseries 107} no.~9, (2023) 095019}, \href{http://arxiv.org/abs/2210.07706}{{\ttfamily arXiv:2210.07706 [hep-ph]}}.

\bibitem{Portillo-Sanchez:2023kbz}
D.~Portillo-S{\'a}nchez, P.~Escribano, and A.~Vicente, ``{Ultraviolet extensions of the Scotogenic model},'' \href{http://dx.doi.org/10.1007/JHEP08(2023)023}{{\em JHEP} {\bfseries 08} (2023) 023}, \href{http://arxiv.org/abs/2301.05249}{{\ttfamily arXiv:2301.05249 [hep-ph]}}.

\bibitem{Chun:2023vbh}
E.~J. Chun, A.~Roy, S.~Mandal, and M.~Mitra, ``{Fermionic dark matter in Dynamical Scotogenic Model},'' \href{http://dx.doi.org/10.1007/JHEP08(2023)130}{{\em JHEP} {\bfseries 08} (2023) 130}, \href{http://arxiv.org/abs/2303.02681}{{\ttfamily arXiv:2303.02681 [hep-ph]}}.

\bibitem{Bonilla:2023egs}
C.~Bonilla, A.~E. Carcamo~Hernandez, B.~Saez~D{\i}az, S.~Kovalenko, and J.~Marchant~Gonzalez, ``{Dark matter from a radiative inverse seesaw majoron model},'' \href{http://dx.doi.org/10.1016/j.physletb.2023.138282}{{\em Phys. Lett. B} {\bfseries 847} (2023) 138282}, \href{http://arxiv.org/abs/2306.08453}{{\ttfamily arXiv:2306.08453 [hep-ph]}}.

\bibitem{Ardila-Tafurth:2025qdf}
G.~Ardila-Tafurth, A.~Fl{\'o}rez, C.~Rodr{\'\i}guez, M.~Sarazin, and {\'O}.~Zapata, ``{Feasibility to probe the dynamical scotogenic model at the LHC},'' \href{http://dx.doi.org/10.1103/f3fq-gbct}{{\em Phys. Rev. D} {\bfseries 113} no.~11, (2026) 115013}, \href{http://arxiv.org/abs/2512.17903}{{\ttfamily arXiv:2512.17903 [hep-ph]}}.

\bibitem{Datta:2026aks}
A.~Datta, H.~M. Lee, and J.-H. Song, ``{Spontaneous Scoto-leptogenesis},'' \href{http://arxiv.org/abs/2608.12497}{{\ttfamily arXiv:2608.12497 [hep-ph]}}.

\bibitem{CentellesChulia:2026ats}
S.~Centelles~Chuli{\'a}, A.~Chattaraj, R.~Srivastava, and S.~Srivastava, ``{Escaping Solar Capture: Majoron mediated Elastic Dark Matter for the LZ 248{\textasciitilde}keV Event},'' \href{http://arxiv.org/abs/2609.32435}{{\ttfamily arXiv:2609.32435 [hep-ph]}}.

\bibitem{Hieu:2026pfh}
T.~T. Hieu, P.~N. Thu, N.~T. Huong, and D.~T. Huong, ``{Majoron phenomenology in a radiative inverse seesaw embedded in an extended inert doublet model},'' \href{http://dx.doi.org/10.1142/s0217751x2650137x}{{\em Int. J. Mod. Phys. A} {\bfseries 41} no.~23, (2026) 2650137}.

\bibitem{Liang:2024vnd}
Q.~Liang, X.~Ponce~D{\'\i}az, and T.~T. Yanagida, ``{Axion Detection Experiments Can Probe Majoron Models},'' \href{http://dx.doi.org/10.1103/PhysRevLett.134.151803}{{\em Phys. Rev. Lett.} {\bfseries 134} no.~15, (2025) 151803}, \href{http://arxiv.org/abs/2406.19083}{{\ttfamily arXiv:2406.19083 [hep-ph]}}.

\bibitem{Herrero-Brocal:2026nmc}
A.~Herrero-Brocal, ``{Non-anomalous axions: Lessons from the Majoron},'' \href{http://dx.doi.org/10.1016/j.physletb.2026.141045}{{\em Phys. Lett. B} {\bfseries 882} (2026) 141045}, \href{http://arxiv.org/abs/2602.22311}{{\ttfamily arXiv:2602.22311 [hep-ph]}}.

\bibitem{ParticleDataGroup:2026mpi}
{\bfseries Particle Data Group} Collaboration, F.~Takahashi {\em et~al.}, ``{Review of Particle Physics},'' \href{http://dx.doi.org/10.1142/s0217751x26300115}{{\em Int. J. Mod. Phys. A} {\bfseries 41} no.~22, (2026) 2630011}.

\bibitem{Schechter:1981bd}
J.~Schechter and J.~W.~F. Valle, ``{Neutrinoless Double beta Decay in SU(2) x U(1) Theories},'' \href{http://dx.doi.org/10.1103/PhysRevD.25.2951}{{\em Phys. Rev. D} {\bfseries 25} (1982) 2951}.

\bibitem{Barreiros:2018bju}
D.~M. Barreiros, R.~G. Felipe, and F.~R. Joaquim, ``{Combining texture zeros with a remnant CP symmetry in the minimal type-I seesaw},'' \href{http://dx.doi.org/10.1007/JHEP01(2019)223}{{\em JHEP} {\bfseries 01} (2019) 223}, \href{http://arxiv.org/abs/1810.05454}{{\ttfamily arXiv:1810.05454 [hep-ph]}}.

\bibitem{SNO:2015wyx}
{\bfseries SNO+} Collaboration, S.~Andringa {\em et~al.}, ``{Current Status and Future Prospects of the SNO+ Experiment},'' \href{http://dx.doi.org/10.1155/2016/6194250}{{\em Adv. High Energy Phys.} {\bfseries 2016} (2016) 6194250}, \href{http://arxiv.org/abs/1508.05759}{{\ttfamily arXiv:1508.05759 [physics.ins-det]}}.

\bibitem{LEGEND:2017cdu}
{\bfseries LEGEND} Collaboration, N.~Abgrall {\em et~al.}, ``{The Large Enriched Germanium Experiment for Neutrinoless Double Beta Decay (LEGEND)},'' \href{http://dx.doi.org/10.1063/1.5007652}{{\em AIP Conf. Proc.} {\bfseries 1894} no.~1, (2017) 020027}, \href{http://arxiv.org/abs/1709.01980}{{\ttfamily arXiv:1709.01980 [physics.ins-det]}}.

\bibitem{nEXO:2017nam}
{\bfseries nEXO} Collaboration, J.~B. Albert {\em et~al.}, ``{Sensitivity and Discovery Potential of nEXO to Neutrinoless Double Beta Decay},'' \href{http://dx.doi.org/10.1103/PhysRevC.97.065503}{{\em Phys. Rev. C} {\bfseries 97} no.~6, (2018) 065503}, \href{http://arxiv.org/abs/1710.05075}{{\ttfamily arXiv:1710.05075 [nucl-ex]}}.

\bibitem{Esteban:2026phq}
I.~Esteban, M.~C. Gonzalez-Garcia, M.~Maltoni, I.~Martinez-Soler, J.~P. Pinheiro, and T.~Schwetz, ``{Lessons from the first JUNO results},'' \href{http://dx.doi.org/10.1007/JHEP04(2026)089}{{\em JHEP} {\bfseries 04} (2026) 089}, \href{http://arxiv.org/abs/2601.09791}{{\ttfamily arXiv:2601.09791 [hep-ph]}}.

\bibitem{deSalas:2020pgw}
P.~F. de~Salas, D.~V. Forero, S.~Gariazzo, P.~Mart{\'\i}nez-Mirav{\'e}, O.~Mena, C.~A. Ternes, M.~T{\'o}rtola, and J.~W.~F. Valle, ``{2020 global reassessment of the neutrino oscillation picture},'' \href{http://dx.doi.org/10.1007/JHEP02(2021)071}{{\em JHEP} {\bfseries 02} (2021) 071}, \href{http://arxiv.org/abs/2006.11237}{{\ttfamily arXiv:2006.11237 [hep-ph]}}.

\bibitem{Esteban:2024eli}
I.~Esteban, M.~C. Gonzalez-Garcia, M.~Maltoni, I.~Martinez-Soler, J.~P. Pinheiro, and T.~Schwetz, ``{NuFit-6.0: updated global analysis of three-flavor neutrino oscillations},'' \href{http://dx.doi.org/10.1007/JHEP12(2024)216}{{\em JHEP} {\bfseries 12} (2024) 216}, \href{http://arxiv.org/abs/2410.05380}{{\ttfamily arXiv:2410.05380 [hep-ph]}}.

\bibitem{Capozzi:2025wyn}
F.~Capozzi, W.~Giar{\`e}, E.~Lisi, A.~Marrone, A.~Melchiorri, and A.~Palazzo, ``{Neutrino masses and mixing: Entering the era of subpercent precision},'' \href{http://dx.doi.org/10.1103/PhysRevD.111.093006}{{\em Phys. Rev. D} {\bfseries 111} no.~9, (2025) 093006}, \href{http://arxiv.org/abs/2503.07752}{{\ttfamily arXiv:2503.07752 [hep-ph]}}.

\bibitem{Capozzi:2025ovi}
F.~Capozzi, E.~Lisi, F.~Marcone, A.~Marrone, and A.~Palazzo, ``{Updated bounds on the (1, 2) neutrino oscillation parameters after first JUNO results},'' \href{http://dx.doi.org/10.1103/cxqw-1bty}{{\em Phys. Rev. D} {\bfseries 114} no.~1, (2026) 016026}, \href{http://arxiv.org/abs/2511.21650}{{\ttfamily arXiv:2511.21650 [hep-ph]}}.

\bibitem{Izaguirre:2016dfi}
E.~Izaguirre, T.~Lin, and B.~Shuve, ``{Searching for Axionlike Particles in Flavor-Changing Neutral Current Processes},'' \href{http://dx.doi.org/10.1103/PhysRevLett.118.111802}{{\em Phys. Rev. Lett.} {\bfseries 118} no.~11, (2017) 111802}, \href{http://arxiv.org/abs/1611.09355}{{\ttfamily arXiv:1611.09355 [hep-ph]}}.

\bibitem{Bauer:2017ris}
M.~Bauer, M.~Neubert, and A.~Thamm, ``{Collider Probes of Axion-Like Particles},'' \href{http://dx.doi.org/10.1007/JHEP12(2017)044}{{\em JHEP} {\bfseries 12} (2017) 044}, \href{http://arxiv.org/abs/1708.00443}{{\ttfamily arXiv:1708.00443 [hep-ph]}}.

\bibitem{Bauer:2021mvw}
M.~Bauer, M.~Neubert, S.~Renner, M.~Schnubel, and A.~Thamm, ``{Flavor probes of axion-like particles},'' \href{http://dx.doi.org/10.1007/JHEP09(2022)056}{{\em JHEP} {\bfseries 09} (2022) 056}, \href{http://arxiv.org/abs/2110.10698}{{\ttfamily arXiv:2110.10698 [hep-ph]}}.

\bibitem{BaBar:2013npw}
{\bfseries BaBar} Collaboration, J.~P. Lees {\em et~al.}, ``{Search for $B \to K^{(*)} \nu \overline \nu$ and invisible quarkonium decays},'' \href{http://dx.doi.org/10.1103/PhysRevD.87.112005}{{\em Phys. Rev. D} {\bfseries 87} no.~11, (2013) 112005}, \href{http://arxiv.org/abs/1303.7465}{{\ttfamily arXiv:1303.7465 [hep-ex]}}.

\bibitem{Belle-II:2023esi}
{\bfseries Belle-II} Collaboration, I.~Adachi {\em et~al.}, ``{Evidence for B+{\textrightarrow}K+{\ensuremath{\nu}}{\ensuremath{\nu}}{\textasciimacron} decays},'' \href{http://dx.doi.org/10.1103/PhysRevD.109.112006}{{\em Phys. Rev. D} {\bfseries 109} no.~11, (2024) 112006}, \href{http://arxiv.org/abs/2311.14647}{{\ttfamily arXiv:2311.14647 [hep-ex]}}.

\bibitem{CAST:2024eil}
{\bfseries CAST} Collaboration, K.~Altenm{\"u}ller {\em et~al.}, ``{New Upper Limit on the Axion-Photon Coupling with an Extended CAST Run with a Xe-Based Micromegas Detector},'' \href{http://dx.doi.org/10.1103/PhysRevLett.133.221005}{{\em Phys. Rev. Lett.} {\bfseries 133} no.~22, (2024) 221005}, \href{http://arxiv.org/abs/2406.16840}{{\ttfamily arXiv:2406.16840 [hep-ex]}}.

\bibitem{Ayala:2014pea}
A.~Ayala, I.~Dom{\'\i}nguez, M.~Giannotti, A.~Mirizzi, and O.~Straniero, ``{Revisiting the bound on axion-photon coupling from Globular Clusters},'' \href{http://dx.doi.org/10.1103/PhysRevLett.113.191302}{{\em Phys. Rev. Lett.} {\bfseries 113} no.~19, (2014) 191302}, \href{http://arxiv.org/abs/1406.6053}{{\ttfamily arXiv:1406.6053 [astro-ph.SR]}}.

\bibitem{Dolan:2022kul}
M.~J. Dolan, F.~J. Hiskens, and R.~R. Volkas, ``{Advancing globular cluster constraints on the axion-photon coupling},'' \href{http://dx.doi.org/10.1088/1475-7516/2022/10/096}{{\em JCAP} {\bfseries 10} (2022) 096}, \href{http://arxiv.org/abs/2207.03102}{{\ttfamily arXiv:2207.03102 [hep-ph]}}.

\bibitem{Payez:2014xsa}
A.~Payez, C.~Evoli, T.~Fischer, M.~Giannotti, A.~Mirizzi, and A.~Ringwald, ``{Revisiting the SN1987A gamma-ray limit on ultralight axion-like particles},'' \href{http://dx.doi.org/10.1088/1475-7516/2015/02/006}{{\em JCAP} {\bfseries 02} (2015) 006}, \href{http://arxiv.org/abs/1410.3747}{{\ttfamily arXiv:1410.3747 [astro-ph.HE]}}.

\bibitem{Hoof:2022xbe}
S.~Hoof and L.~Schulz, ``{Updated constraints on axion-like particles from temporal information in supernova SN1987A gamma-ray data},'' \href{http://dx.doi.org/10.1088/1475-7516/2023/03/054}{{\em JCAP} {\bfseries 03} (2023) 054}, \href{http://arxiv.org/abs/2212.09764}{{\ttfamily arXiv:2212.09764 [hep-ph]}}.

\bibitem{Barger:2008jx}
V.~Barger, P.~Langacker, M.~McCaskey, M.~Ramsey-Musolf, and G.~Shaughnessy, ``{Complex Singlet Extension of the Standard Model},'' \href{http://dx.doi.org/10.1103/PhysRevD.79.015018}{{\em Phys. Rev. D} {\bfseries 79} (2009) 015018}, \href{http://arxiv.org/abs/0811.0393}{{\ttfamily arXiv:0811.0393 [hep-ph]}}.

\bibitem{Gonderinger:2012rd}
M.~Gonderinger, H.~Lim, and M.~J. Ramsey-Musolf, ``{Complex Scalar Singlet Dark Matter: Vacuum Stability and Phenomenology},'' \href{http://dx.doi.org/10.1103/PhysRevD.86.043511}{{\em Phys. Rev. D} {\bfseries 86} (2012) 043511}, \href{http://arxiv.org/abs/1202.1316}{{\ttfamily arXiv:1202.1316 [hep-ph]}}.

\bibitem{Chiang:2017nmu}
C.-W. Chiang, M.~J. Ramsey-Musolf, and E.~Senaha, ``{Standard Model with a Complex Scalar Singlet: Cosmological Implications and Theoretical Considerations},'' \href{http://dx.doi.org/10.1103/PhysRevD.97.015005}{{\em Phys. Rev. D} {\bfseries 97} no.~1, (2018) 015005}, \href{http://arxiv.org/abs/1707.09960}{{\ttfamily arXiv:1707.09960 [hep-ph]}}.

\bibitem{Alanne:2020jwx}
T.~Alanne, N.~Benincasa, M.~Heikinheimo, K.~Kannike, V.~Keus, N.~Koivunen, and K.~Tuominen, ``{Pseudo-Goldstone dark matter: gravitational waves and direct-detection blind spots},'' \href{http://dx.doi.org/10.1007/JHEP10(2020)080}{{\em JHEP} {\bfseries 10} (2020) 080}, \href{http://arxiv.org/abs/2008.09605}{{\ttfamily arXiv:2008.09605 [hep-ph]}}.

\bibitem{Bodas:2020yho}
A.~Bodas, S.~Kumar, and R.~Sundrum, ``{The Scalar Chemical Potential in Cosmological Collider Physics},'' \href{http://dx.doi.org/10.1007/JHEP02(2021)079}{{\em JHEP} {\bfseries 02} (2021) 079}, \href{http://arxiv.org/abs/2010.04727}{{\ttfamily arXiv:2010.04727 [hep-ph]}}.

\bibitem{Coito:2021fgo}
L.~Coito, C.~Faubel, J.~Herrero-Garcia, and A.~Santamaria, ``{Dark matter from a complex scalar singlet: the role of dark CP and other discrete symmetries},'' \href{http://dx.doi.org/10.1007/JHEP11(2021)202}{{\em JHEP} {\bfseries 11} (2021) 202}, \href{http://arxiv.org/abs/2106.05289}{{\ttfamily arXiv:2106.05289 [hep-ph]}}.

\bibitem{Kim:2024tnn}
J.~Kim and P.~Ko, ``{Pseudo Nambu-Goldstone Boson DM with Linear Symmetry Breaking: Revisited},'' \href{http://arxiv.org/abs/2401.02243}{{\ttfamily arXiv:2401.02243 [hep-ph]}}.

\bibitem{Chiang:2026ayw}
C.-W. Chiang, S.-Y. Ho, and V.~Q. Tran, ``{TeV-scale origin of light dark matter and neutrino mass},'' \href{http://dx.doi.org/10.1103/s986-nbtm}{{\em Phys. Rev. D} {\bfseries 114} no.~5, (2026) 055017}, \href{http://arxiv.org/abs/2603.05200}{{\ttfamily arXiv:2603.05200 [hep-ph]}}.

\bibitem{Giddings:1987cg}
S.~B. Giddings and A.~Strominger, ``{Axion Induced Topology Change in Quantum Gravity and String Theory},'' \href{http://dx.doi.org/10.1016/0550-3213(88)90446-4}{{\em Nucl. Phys. B} {\bfseries 306} (1988) 890--907}.

\bibitem{Kallosh:1995hi}
R.~Kallosh, A.~D. Linde, D.~A. Linde, and L.~Susskind, ``{Gravity and global symmetries},'' \href{http://dx.doi.org/10.1103/PhysRevD.52.912}{{\em Phys. Rev. D} {\bfseries 52} (1995) 912--935}, \href{http://arxiv.org/abs/hep-th/9502069}{{\ttfamily arXiv:hep-th/9502069}}.

\bibitem{Banks:2010zn}
T.~Banks and N.~Seiberg, ``{Symmetries and Strings in Field Theory and Gravity},'' \href{http://dx.doi.org/10.1103/PhysRevD.83.084019}{{\em Phys. Rev. D} {\bfseries 83} (2011) 084019}, \href{http://arxiv.org/abs/1011.5120}{{\ttfamily arXiv:1011.5120 [hep-th]}}.

\bibitem{Kamionkowski:1992mf}
M.~Kamionkowski and J.~March-Russell, ``{Planck scale physics and the Peccei-Quinn mechanism},'' \href{http://dx.doi.org/10.1016/0370-2693(92)90492-M}{{\em Phys. Lett. B} {\bfseries 282} (1992) 137--141}, \href{http://arxiv.org/abs/hep-th/9202003}{{\ttfamily arXiv:hep-th/9202003}}.

\bibitem{Holman:1992us}
R.~Holman, S.~D.~H. Hsu, T.~W. Kephart, E.~W. Kolb, R.~Watkins, and L.~M. Widrow, ``{Solutions to the strong CP problem in a world with gravity},'' \href{http://dx.doi.org/10.1016/0370-2693(92)90491-L}{{\em Phys. Lett. B} {\bfseries 282} (1992) 132--136}, \href{http://arxiv.org/abs/hep-ph/9203206}{{\ttfamily arXiv:hep-ph/9203206}}.

\bibitem{Barr:1992qq}
S.~M. Barr and D.~Seckel, ``{Planck scale corrections to axion models},'' \href{http://dx.doi.org/10.1103/PhysRevD.46.539}{{\em Phys. Rev. D} {\bfseries 46} (1992) 539--549}.

\bibitem{Akhmedov:1992hi}
E.~K. Akhmedov, Z.~G. Berezhiani, R.~N. Mohapatra, and G.~Senjanovic, ``{Planck scale effects on the majoron},'' \href{http://dx.doi.org/10.1016/0370-2693(93)90887-N}{{\em Phys. Lett. B} {\bfseries 299} (1993) 90--93}, \href{http://arxiv.org/abs/hep-ph/9209285}{{\ttfamily arXiv:hep-ph/9209285}}.

\bibitem{Rothstein:1992rh}
I.~Z. Rothstein, K.~S. Babu, and D.~Seckel, ``{Planck scale symmetry breaking and majoron physics},'' \href{http://dx.doi.org/10.1016/0550-3213(93)90368-Y}{{\em Nucl. Phys. B} {\bfseries 403} (1993) 725--748}, \href{http://arxiv.org/abs/hep-ph/9301213}{{\ttfamily arXiv:hep-ph/9301213}}.

\bibitem{Zeldovich:1974uw}
Y.~B. Zeldovich, I.~Y. Kobzarev, and L.~B. Okun, ``{Cosmological Consequences of the Spontaneous Breakdown of Discrete Symmetry},'' {\em Zh. Eksp. Teor. Fiz.} {\bfseries 67} (1974) 3--11.

\bibitem{Sikivie:1982qv}
P.~Sikivie, ``{Of Axions, Domain Walls and the Early Universe},'' \href{http://dx.doi.org/10.1103/PhysRevLett.48.1156}{{\em Phys. Rev. Lett.} {\bfseries 48} (1982) 1156--1159}.

\bibitem{Vilenkin:1984ib}
A.~Vilenkin, ``{Cosmic Strings and Domain Walls},'' \href{http://dx.doi.org/10.1016/0370-1573(85)90033-X}{{\em Phys. Rept.} {\bfseries 121} (1985) 263--315}.

\bibitem{Khan:2026eqx}
M.~Khan, W.~Chao, and Y.~Gao, ``{Soft Breaking Naturalness in Majorana Fermion Loop induced Scalar and Majoron Amplitudes},'' \href{http://arxiv.org/abs/2610.28198}{{\ttfamily arXiv:2610.28198 [hep-ph]}}.

\bibitem{Saikawa:2018rcs}
K.~Saikawa and S.~Shirai, ``{Primordial gravitational waves, precisely: The role of thermodynamics in the Standard Model},'' \href{http://dx.doi.org/10.1088/1475-7516/2018/05/035}{{\em JCAP} {\bfseries 05} (2018) 035}, \href{http://arxiv.org/abs/1803.01038}{{\ttfamily arXiv:1803.01038 [hep-ph]}}.

\bibitem{Fields:2019pfx}
B.~D. Fields, K.~A. Olive, T.-H. Yeh, and C.~Young, ``{Big-Bang Nucleosynthesis after Planck},'' \href{http://dx.doi.org/10.1088/1475-7516/2020/03/010}{{\em JCAP} {\bfseries 03} (2020) 010}, \href{http://arxiv.org/abs/1912.01132}{{\ttfamily arXiv:1912.01132 [astro-ph.CO]}}. [Erratum: JCAP 11, E02 (2020)].

\bibitem{Depta:2020wmr}
P.~F. Depta, M.~Hufnagel, and K.~Schmidt-Hoberg, ``{Robust cosmological constraints on axion-like particles},'' \href{http://dx.doi.org/10.1088/1475-7516/2020/05/009}{{\em JCAP} {\bfseries 05} (2020) 009}, \href{http://arxiv.org/abs/2002.08370}{{\ttfamily arXiv:2002.08370 [hep-ph]}}.

\bibitem{Foster:2021ngm}
J.~W. Foster, M.~Kongsore, C.~Dessert, Y.~Park, N.~L. Rodd, K.~Cranmer, and B.~R. Safdi, ``{Deep Search for Decaying Dark Matter with XMM-Newton Blank-Sky Observations},'' \href{http://dx.doi.org/10.1103/PhysRevLett.127.051101}{{\em Phys. Rev. Lett.} {\bfseries 127} no.~5, (2021) 051101}, \href{http://arxiv.org/abs/2102.02207}{{\ttfamily arXiv:2102.02207 [astro-ph.CO]}}.

\bibitem{Roach:2022lgo}
B.~M. Roach, S.~Rossland, K.~C.~Y. Ng, K.~Perez, J.~F. Beacom, B.~W. Grefenstette, S.~Horiuchi, R.~Krivonos, and D.~R. Wik, ``{Long-exposure NuSTAR constraints on decaying dark matter in the Galactic halo},'' \href{http://dx.doi.org/10.1103/PhysRevD.107.023009}{{\em Phys. Rev. D} {\bfseries 107} no.~2, (2023) 023009}, \href{http://arxiv.org/abs/2207.04572}{{\ttfamily arXiv:2207.04572 [astro-ph.HE]}}.

\bibitem{Calore:2022pks}
F.~Calore, A.~Dekker, P.~D. Serpico, and T.~Siegert, ``{Constraints on light decaying dark matter candidates from 16~yr of INTEGRAL/SPI observations},'' \href{http://dx.doi.org/10.1093/mnras/stad457}{{\em Mon. Not. Roy. Astron. Soc.} {\bfseries 520} no.~3, (2023) 4167--4172}, \href{http://arxiv.org/abs/2209.06299}{{\ttfamily arXiv:2209.06299 [hep-ph]}}. [Erratum: Mon.Not.Roy.Astron.Soc. 538, (2025), Erratum: Mon.Not.Roy.Astron.Soc. 547, stag045 (2026)].

\bibitem{Cadamuro:2011fd}
D.~Cadamuro and J.~Redondo, ``{Cosmological bounds on pseudo Nambu-Goldstone bosons},'' \href{http://dx.doi.org/10.1088/1475-7516/2012/02/032}{{\em JCAP} {\bfseries 02} (2012) 032}, \href{http://arxiv.org/abs/1110.2895}{{\ttfamily arXiv:1110.2895 [hep-ph]}}.

\bibitem{AxionLimits}
C.~O'Hare, ``cajohare/axionlimits: Axionlimits.'' \url{https://cajohare.github.io/AxionLimits/}, July, 2020.

\bibitem{Jerhot:2022chi}
J.~Jerhot, B.~D{\"o}brich, F.~Ertas, F.~Kahlhoefer, and T.~Spadaro, ``{ALPINIST: Axion-Like Particles In Numerous Interactions Simulated and Tabulated},'' \href{http://dx.doi.org/10.1007/JHEP07(2022)094}{{\em JHEP} {\bfseries 07} (2022) 094}, \href{http://arxiv.org/abs/2201.05170}{{\ttfamily arXiv:2201.05170 [hep-ph]}}.

\bibitem{Jaeckel:2017tud}
J.~Jaeckel, P.~C. Malta, and J.~Redondo, ``{Decay photons from the axionlike particles burst of type II supernovae},'' \href{http://dx.doi.org/10.1103/PhysRevD.98.055032}{{\em Phys. Rev. D} {\bfseries 98} no.~5, (2018) 055032}, \href{http://arxiv.org/abs/1702.02964}{{\ttfamily arXiv:1702.02964 [hep-ph]}}.

\bibitem{Lucente:2020whw}
G.~Lucente, P.~Carenza, T.~Fischer, M.~Giannotti, and A.~Mirizzi, ``{Heavy axion-like particles and core-collapse supernovae: constraints and impact on the explosion mechanism},'' \href{http://dx.doi.org/10.1088/1475-7516/2020/12/008}{{\em JCAP} {\bfseries 12} (2020) 008}, \href{http://arxiv.org/abs/2008.04918}{{\ttfamily arXiv:2008.04918 [hep-ph]}}.

\bibitem{Caputo:2022mah}
A.~Caputo, H.-T. Janka, G.~Raffelt, and E.~Vitagliano, ``{Low-Energy Supernovae Severely Constrain Radiative Particle Decays},'' \href{http://dx.doi.org/10.1103/PhysRevLett.128.221103}{{\em Phys. Rev. Lett.} {\bfseries 128} no.~22, (2022) 221103}, \href{http://arxiv.org/abs/2201.09890}{{\ttfamily arXiv:2201.09890 [astro-ph.HE]}}.

\bibitem{Muller:2023vjm}
E.~M{\"u}ller, F.~Calore, P.~Carenza, C.~Eckner, and M.~C.~D. Marsh, ``{Investigating the gamma-ray burst from decaying MeV-scale axion-like particles produced in supernova explosions},'' \href{http://dx.doi.org/10.1088/1475-7516/2023/07/056}{{\em JCAP} {\bfseries 07} (2023) 056}, \href{http://arxiv.org/abs/2304.01060}{{\ttfamily arXiv:2304.01060 [astro-ph.HE]}}.

\bibitem{CHARM:1985anb}
{\bfseries CHARM} Collaboration, F.~Bergsma {\em et~al.}, ``{Search for Axion Like Particle Production in 400-{GeV} Proton - Copper Interactions},'' \href{http://dx.doi.org/10.1016/0370-2693(85)90400-9}{{\em Phys. Lett. B} {\bfseries 157} (1985) 458--462}.

\bibitem{Riordan:1987aw}
E.~M. Riordan {\em et~al.}, ``{A Search for Short Lived Axions in an Electron Beam Dump Experiment},'' \href{http://dx.doi.org/10.1103/PhysRevLett.59.755}{{\em Phys. Rev. Lett.} {\bfseries 59} (1987) 755}.

\bibitem{Bjorken:1988as}
J.~D. Bjorken, S.~Ecklund, W.~R. Nelson, A.~Abashian, C.~Church, B.~Lu, L.~W. Mo, T.~A. Nunamaker, and P.~Rassmann, ``{Search for Neutral Metastable Penetrating Particles Produced in the SLAC Beam Dump},'' \href{http://dx.doi.org/10.1103/PhysRevD.38.3375}{{\em Phys. Rev. D} {\bfseries 38} (1988) 3375}.

\bibitem{Blumlein:1990ay}
J.~Blumlein {\em et~al.}, ``{Limits on neutral light scalar and pseudoscalar particles in a proton beam dump experiment},'' \href{http://dx.doi.org/10.1007/BF01548556}{{\em Z. Phys. C} {\bfseries 51} (1991) 341--350}.

\bibitem{Dolan:2017osp}
M.~J. Dolan, T.~Ferber, C.~Hearty, F.~Kahlhoefer, and K.~Schmidt-Hoberg, ``{Revised constraints and Belle II sensitivity for visible and invisible axion-like particles},'' \href{http://dx.doi.org/10.1007/JHEP12(2017)094}{{\em JHEP} {\bfseries 12} (2017) 094}, \href{http://arxiv.org/abs/1709.00009}{{\ttfamily arXiv:1709.00009 [hep-ph]}}. [Erratum: JHEP 03, 190 (2021)].

\bibitem{NA64:2020qwq}
{\bfseries NA64} Collaboration, D.~Banerjee {\em et~al.}, ``{Search for Axionlike and Scalar Particles with the NA64 Experiment},'' \href{http://dx.doi.org/10.1103/PhysRevLett.125.081801}{{\em Phys. Rev. Lett.} {\bfseries 125} no.~8, (2020) 081801}, \href{http://arxiv.org/abs/2005.02710}{{\ttfamily arXiv:2005.02710 [hep-ex]}}.

\bibitem{Jaeckel:2015jla}
J.~Jaeckel and M.~Spannowsky, ``{Probing MeV to 90 GeV axion-like particles with LEP and LHC},'' \href{http://dx.doi.org/10.1016/j.physletb.2015.12.037}{{\em Phys. Lett. B} {\bfseries 753} (2016) 482--487}, \href{http://arxiv.org/abs/1509.00476}{{\ttfamily arXiv:1509.00476 [hep-ph]}}.

\bibitem{Mariotti:2017vtv}
A.~Mariotti, D.~Redigolo, F.~Sala, and K.~Tobioka, ``{New LHC bound on low-mass diphoton resonances},'' \href{http://dx.doi.org/10.1016/j.physletb.2018.06.039}{{\em Phys. Lett. B} {\bfseries 783} (2018) 13--18}, \href{http://arxiv.org/abs/1710.01743}{{\ttfamily arXiv:1710.01743 [hep-ph]}}.

\bibitem{Knapen:2016moh}
S.~Knapen, T.~Lin, H.~K. Lou, and T.~Melia, ``{Searching for Axionlike Particles with Ultraperipheral Heavy-Ion Collisions},'' \href{http://dx.doi.org/10.1103/PhysRevLett.118.171801}{{\em Phys. Rev. Lett.} {\bfseries 118} no.~17, (2017) 171801}, \href{http://arxiv.org/abs/1607.06083}{{\ttfamily arXiv:1607.06083 [hep-ph]}}.

\bibitem{CMS:2018erd}
{\bfseries CMS} Collaboration, A.~M. Sirunyan {\em et~al.}, ``{Evidence for light-by-light scattering and searches for axion-like particles in ultraperipheral PbPb collisions at $\sqrt{s_\mathrm{NN}} =$ 5.02 TeV},'' \href{http://dx.doi.org/10.1016/j.physletb.2019.134826}{{\em Phys. Lett. B} {\bfseries 797} (2019) 134826}, \href{http://arxiv.org/abs/1810.04602}{{\ttfamily arXiv:1810.04602 [hep-ex]}}.

\bibitem{Aloni:2019ruo}
D.~Aloni, C.~Fanelli, Y.~Soreq, and M.~Williams, ``{Photoproduction of Axionlike Particles},'' \href{http://dx.doi.org/10.1103/PhysRevLett.123.071801}{{\em Phys. Rev. Lett.} {\bfseries 123} no.~7, (2019) 071801}, \href{http://arxiv.org/abs/1903.03586}{{\ttfamily arXiv:1903.03586 [hep-ph]}}.

\bibitem{ATLAS:2020hii}
{\bfseries ATLAS} Collaboration, G.~Aad {\em et~al.}, ``{Measurement of light-by-light scattering and search for axion-like particles with 2.2 nb$^{-1}$ of Pb+Pb data with the ATLAS detector},'' \href{http://dx.doi.org/10.1007/JHEP03(2021)243}{{\em JHEP} {\bfseries 03} (2021) 243}, \href{http://arxiv.org/abs/2008.05355}{{\ttfamily arXiv:2008.05355 [hep-ex]}}. [Erratum: JHEP 11, 050 (2021)].

\bibitem{Belle-II:2020jti}
{\bfseries Belle-II} Collaboration, F.~Abudin{\'e}n {\em et~al.}, ``{Search for Axion-Like Particles produced in $e^+e^-$ collisions at Belle II},'' \href{http://dx.doi.org/10.1103/PhysRevLett.125.161806}{{\em Phys. Rev. Lett.} {\bfseries 125} no.~16, (2020) 161806}, \href{http://arxiv.org/abs/2007.13071}{{\ttfamily arXiv:2007.13071 [hep-ex]}}.

\bibitem{BESIII:2022rzz}
{\bfseries BESIII} Collaboration, M.~Ablikim {\em et~al.}, ``{Search for an axion-like particle in radiative J/{\ensuremath{\psi}} decays},'' \href{http://dx.doi.org/10.1016/j.physletb.2023.137698}{{\em Phys. Lett. B} {\bfseries 838} (2023) 137698}, \href{http://arxiv.org/abs/2211.12699}{{\ttfamily arXiv:2211.12699 [hep-ex]}}.

\bibitem{BESIII:2024hdv}
{\bfseries BESIII} Collaboration, M.~Ablikim {\em et~al.}, ``{Search for diphoton decays of an axionlike particle in radiative J/{\ensuremath{\psi}} decays},'' \href{http://dx.doi.org/10.1103/PhysRevD.110.L031101}{{\em Phys. Rev. D} {\bfseries 110} no.~3, (2024) L031101}, \href{http://arxiv.org/abs/2404.04640}{{\ttfamily arXiv:2404.04640 [hep-ex]}}.

\bibitem{Dobrich:2015jyk}
B.~D{\"o}brich, J.~Jaeckel, F.~Kahlhoefer, A.~Ringwald, and K.~Schmidt-Hoberg, ``{ALPtraum: ALP production in proton beam dump experiments},'' \href{http://dx.doi.org/10.1007/JHEP02(2016)018}{{\em JHEP} {\bfseries 02} (2016) 018}, \href{http://arxiv.org/abs/1512.03069}{{\ttfamily arXiv:1512.03069 [hep-ph]}}.

\bibitem{Berlin:2018pwi}
A.~Berlin, S.~Gori, P.~Schuster, and N.~Toro, ``{Dark Sectors at the Fermilab SeaQuest Experiment},'' \href{http://dx.doi.org/10.1103/PhysRevD.98.035011}{{\em Phys. Rev. D} {\bfseries 98} no.~3, (2018) 035011}, \href{http://arxiv.org/abs/1804.00661}{{\ttfamily arXiv:1804.00661 [hep-ph]}}.

\bibitem{Blinov:2021say}
N.~Blinov, E.~Kowalczyk, and M.~Wynne, ``{Axion-like particle searches at DarkQuest},'' \href{http://dx.doi.org/10.1007/JHEP02(2022)036}{{\em JHEP} {\bfseries 02} (2022) 036}, \href{http://arxiv.org/abs/2112.09814}{{\ttfamily arXiv:2112.09814 [hep-ph]}}.

\bibitem{Alekhin:2015byh}
S.~Alekhin {\em et~al.}, ``{A facility to Search for Hidden Particles at the CERN SPS: the SHiP physics case},'' \href{http://dx.doi.org/10.1088/0034-4885/79/12/124201}{{\em Rept. Prog. Phys.} {\bfseries 79} no.~12, (2016) 124201}, \href{http://arxiv.org/abs/1504.04855}{{\ttfamily arXiv:1504.04855 [hep-ph]}}.

\bibitem{Climescu:2026omx}
M.~Climescu, ``{The SHiP/NA67 experiment at the ECN3 high-intensity beam facility at the CERN SPS},''
\newblock 6, 2026.
\newblock \href{http://arxiv.org/abs/2606.11039}{{\ttfamily arXiv:2606.11039 [hep-ex]}}.

\bibitem{Abe:2021nih}
S.~Abe, G.-C. Cho, and K.~Mawatari, ``{Probing a degenerate-scalar scenario in a pseudoscalar dark-matter model},'' \href{http://dx.doi.org/10.1103/PhysRevD.104.035023}{{\em Phys. Rev. D} {\bfseries 104} no.~3, (2021) 035023}, \href{http://arxiv.org/abs/2101.04887}{{\ttfamily arXiv:2101.04887 [hep-ph]}}.

\bibitem{Joshipura:1992hp}
A.~S. Joshipura and J.~W.~F. Valle, ``{Invisible Higgs decays and neutrino physics},'' \href{http://dx.doi.org/10.1016/0550-3213(93)90337-O}{{\em Nucl. Phys. B} {\bfseries 397} (1993) 105--122}.

\bibitem{Joshipura:1992ua}
A.~S. Joshipura and S.~D. Rindani, ``{Majoron models and the Higgs search},'' \href{http://dx.doi.org/10.1103/PhysRevLett.69.3269}{{\em Phys. Rev. Lett.} {\bfseries 69} (1992) 3269--3273}.

\bibitem{ATLAS:2022vkf}
{\bfseries ATLAS} Collaboration, G.~Aad {\em et~al.}, ``{A detailed map of Higgs boson interactions by the ATLAS experiment ten years after the discovery},'' \href{http://dx.doi.org/10.1038/s41586-022-04893-w}{{\em Nature} {\bfseries 607} no.~7917, (2022) 52--59}, \href{http://arxiv.org/abs/2207.00092}{{\ttfamily arXiv:2207.00092 [hep-ex]}}. [Erratum: Nature 612, E24 (2022)].

\bibitem{CMS:2022dwd}
{\bfseries CMS} Collaboration, A.~Tumasyan {\em et~al.}, ``{A portrait of the Higgs boson by the CMS experiment ten years after the discovery.},'' \href{http://dx.doi.org/10.1038/s41586-022-04892-x}{{\em Nature} {\bfseries 607} no.~7917, (2022) 60--68}, \href{http://arxiv.org/abs/2207.00043}{{\ttfamily arXiv:2207.00043 [hep-ex]}}. [Erratum: Nature 623, (2023)].

\bibitem{ATLAS:2025qxq}
{\bfseries ATLAS} Collaboration, ``{Combined measurements of Higgs boson production and decay at $\sqrt{s} =$ 13 TeV using up to 140 fb$^{-1}$ of data collected by the ATLAS Experiment},''.

\bibitem{CMS:2026zxu}
{\bfseries CMS} Collaboration, A.~Hayrapetyan {\em et~al.}, ``{Combined measurements and interpretations of Higgs boson production and decay in proton{\textendash}proton collisions at $\sqrt{s} = 13\,\textrm{TeV}$},'' \href{http://dx.doi.org/10.1088/1361-6633/ae72c4}{{\em Rept. Prog. Phys.} {\bfseries 89} no.~8, (2026) 087801}, \href{http://arxiv.org/abs/2602.18611}{{\ttfamily arXiv:2602.18611 [hep-ex]}}.

\bibitem{Djouadi:2005gj}
A.~Djouadi, ``{The Anatomy of electro-weak symmetry breaking. II. The Higgs bosons in the minimal supersymmetric model},'' \href{http://dx.doi.org/10.1016/j.physrep.2007.10.005}{{\em Phys. Rept.} {\bfseries 459} (2008) 1--241}, \href{http://arxiv.org/abs/hep-ph/0503173}{{\ttfamily arXiv:hep-ph/0503173}}.

\bibitem{Posch:2010hx}
P.~Posch, ``{Enhancement of h ---{\ensuremath{>}} gamma gamma in the Two Higgs Doublet Model Type I},'' \href{http://dx.doi.org/10.1016/j.physletb.2011.01.003}{{\em Phys. Lett. B} {\bfseries 696} (2011) 447--453}, \href{http://arxiv.org/abs/1001.1759}{{\ttfamily arXiv:1001.1759 [hep-ph]}}.

\bibitem{Fontes:2014xva}
D.~Fontes, J.~C. Rom{\~a}o, and J.~P. Silva, ``{$h \rightarrow Z \gamma$ in the complex two Higgs doublet model},'' \href{http://dx.doi.org/10.1007/JHEP12(2014)043}{{\em JHEP} {\bfseries 12} (2014) 043}, \href{http://arxiv.org/abs/1408.2534}{{\ttfamily arXiv:1408.2534 [hep-ph]}}.

\bibitem{Aguilar-Saavedra:2023tql}
J.~A. Aguilar-Saavedra, H.~B. C{\^a}mara, F.~R. Joaquim, and J.~F. Seabra, ``{Confronting the 95 GeV excesses within the U(1)'-extended next-to-minimal 2HDM},'' \href{http://dx.doi.org/10.1103/PhysRevD.108.075020}{{\em Phys. Rev. D} {\bfseries 108} no.~7, (2023) 075020}, \href{http://arxiv.org/abs/2307.03768}{{\ttfamily arXiv:2307.03768 [hep-ph]}}.

\bibitem{Belyaev:2016lok}
A.~Belyaev, G.~Cacciapaglia, I.~P. Ivanov, F.~Rojas-Abatte, and M.~Thomas, ``{Anatomy of the Inert Two Higgs Doublet Model in the light of the LHC and non-LHC Dark Matter Searches},'' \href{http://dx.doi.org/10.1103/PhysRevD.97.035011}{{\em Phys. Rev. D} {\bfseries 97} no.~3, (2018) 035011}, \href{http://arxiv.org/abs/1612.00511}{{\ttfamily arXiv:1612.00511 [hep-ph]}}.

\bibitem{Barreiros:2022aqu}
D.~M. Barreiros, H.~B. Camara, and F.~R. Joaquim, ``{Flavour and dark matter in a scoto/type-II seesaw model},'' \href{http://dx.doi.org/10.1007/JHEP08(2022)030}{{\em JHEP} {\bfseries 08} (2022) 030}, \href{http://arxiv.org/abs/2204.13605}{{\ttfamily arXiv:2204.13605 [hep-ph]}}.

\bibitem{Gelmini:1980re}
G.~B. Gelmini and M.~Roncadelli, ``{Left-Handed Neutrino Mass Scale and Spontaneously Broken Lepton Number},'' \href{http://dx.doi.org/10.1016/0370-2693(81)90559-1}{{\em Phys. Lett. B} {\bfseries 99} (1981) 411--415}.

\bibitem{L3:2001xsz}
{\bfseries L3} Collaboration, P.~Achard {\em et~al.}, ``{Search for heavy neutral and charged leptons in $e^{+} e^{-}$ annihilation at LEP},'' \href{http://dx.doi.org/10.1016/S0370-2693(01)01005-X}{{\em Phys. Lett. B} {\bfseries 517} (2001) 75--85}, \href{http://arxiv.org/abs/hep-ex/0107015}{{\ttfamily arXiv:hep-ex/0107015}}.

\bibitem{Marciano:1977wx}
W.~J. Marciano and A.~I. Sanda, ``{Exotic Decays of the Muon and Heavy Leptons in Gauge Theories},'' \href{http://dx.doi.org/10.1016/0370-2693(77)90377-X}{{\em Phys. Lett. B} {\bfseries 67} (1977) 303--305}.

\bibitem{Cheng:1980tp}
T.~P. Cheng and L.-F. Li, ``{$\mu \to e \gamma$ in Theories With Dirac and Majorana Neutrino Mass Terms},'' \href{http://dx.doi.org/10.1103/PhysRevLett.45.1908}{{\em Phys. Rev. Lett.} {\bfseries 45} (1980) 1908}.

\bibitem{Lim:1981kv}
C.~S. Lim and T.~Inami, ``{Lepton Flavor Nonconservation and the Mass Generation Mechanism for Neutrinos},'' \href{http://dx.doi.org/10.1143/PTP.67.1569}{{\em Prog. Theor. Phys.} {\bfseries 67} (1982) 1569}.

\bibitem{Langacker:1988up}
P.~Langacker and D.~London, ``{Lepton Number Violation and Massless Nonorthogonal Neutrinos},'' \href{http://dx.doi.org/10.1103/PhysRevD.38.907}{{\em Phys. Rev. D} {\bfseries 38} (1988) 907}.

\bibitem{Ilakovac:1994kj}
A.~Ilakovac and A.~Pilaftsis, ``{Flavor violating charged lepton decays in seesaw-type models},'' \href{http://dx.doi.org/10.1016/0550-3213(94)00567-X}{{\em Nucl. Phys. B} {\bfseries 437} (1995) 491}, \href{http://arxiv.org/abs/hep-ph/9403398}{{\ttfamily arXiv:hep-ph/9403398}}.

\bibitem{BaBar:2009hkt}
{\bfseries BaBar} Collaboration, B.~Aubert {\em et~al.}, ``{Searches for Lepton Flavor Violation in the Decays $\tau^\pm \to e^\pm \gamma$ and $\tau^\pm \to \mu^\pm \gamma$},'' \href{http://dx.doi.org/10.1103/PhysRevLett.104.021802}{{\em Phys. Rev. Lett.} {\bfseries 104} (2010) 021802}, \href{http://arxiv.org/abs/0908.2381}{{\ttfamily arXiv:0908.2381 [hep-ex]}}.

\bibitem{Belle:2021ysv}
{\bfseries Belle} Collaboration, A.~Abdesselam {\em et~al.}, ``{Search for lepton-flavor-violating tau-lepton decays to $\ell\gamma$ at Belle},'' \href{http://dx.doi.org/10.1007/JHEP10(2021)019}{{\em JHEP} {\bfseries 10} (2021) 19}, \href{http://arxiv.org/abs/2103.12994}{{\ttfamily arXiv:2103.12994 [hep-ex]}}.

\bibitem{Belle-II:2018jsg}
{\bfseries Belle-II} Collaboration, W.~Altmannshofer {\em et~al.}, ``{The Belle II Physics Book},'' \href{http://dx.doi.org/10.1093/ptep/ptz106}{{\em PTEP} {\bfseries 2019} no.~12, (2019) 123C01}, \href{http://arxiv.org/abs/1808.10567}{{\ttfamily arXiv:1808.10567 [hep-ex]}}. [Erratum: PTEP 2020, 029201 (2020)].

\bibitem{MEGII:2025gzr}
{\bfseries MEG II} Collaboration, K.~Afanaciev {\em et~al.}, ``{New limit on the ${\mu ^+ \rightarrow e^+ \gamma }$ decay with the MEG II experiment},'' \href{http://dx.doi.org/10.1140/epjc/s10052-025-14906-3}{{\em Eur. Phys. J. C} {\bfseries 85} no.~10, (2025) 1177}, \href{http://arxiv.org/abs/2504.15711}{{\ttfamily arXiv:2504.15711 [hep-ex]}}. [Erratum: Eur.Phys.J.C 85, 1317 (2025)].

\bibitem{Jodidio:1986mz}
A.~Jodidio {\em et~al.}, ``{Search for Right-Handed Currents in Muon Decay},'' \href{http://dx.doi.org/10.1103/PhysRevD.34.1967}{{\em Phys. Rev. D} {\bfseries 34} (1986) 1967}. [Erratum: Phys.Rev.D 37, 237 (1988)].

\bibitem{Hirsch:2009ee}
M.~Hirsch, A.~Vicente, J.~Meyer, and W.~Porod, ``{Majoron emission in muon and tau decays revisited},'' \href{http://dx.doi.org/10.1103/PhysRevD.79.055023}{{\em Phys. Rev. D} {\bfseries 79} (2009) 055023}, \href{http://arxiv.org/abs/0902.0525}{{\ttfamily arXiv:0902.0525 [hep-ph]}}. [Erratum: Phys.Rev.D 79, 079901 (2009)].

\bibitem{Calibbi:2020jvd}
L.~Calibbi, D.~Redigolo, R.~Ziegler, and J.~Zupan, ``{Looking forward to lepton-flavor-violating ALPs},'' \href{http://dx.doi.org/10.1007/JHEP09(2021)173}{{\em JHEP} {\bfseries 09} (2021) 173}, \href{http://arxiv.org/abs/2006.04795}{{\ttfamily arXiv:2006.04795 [hep-ph]}}.

\bibitem{COMET:2018auw}
{\bfseries COMET} Collaboration, R.~Abramishvili {\em et~al.}, ``{COMET Phase-I Technical Design Report},'' \href{http://dx.doi.org/10.1093/ptep/ptz125}{{\em PTEP} {\bfseries 2020} no.~3, (2020) 033C01}, \href{http://arxiv.org/abs/1812.09018}{{\ttfamily arXiv:1812.09018 [physics.ins-det]}}.

\bibitem{Xing:2022rob}
T.~Xing, C.~Wu, H.~Miao, H.-B. Li, W.~Li, Y.~Yuan, and Y.~Zhang, ``{Search for Majoron at the COMET experiment*},'' \href{http://dx.doi.org/10.1088/1674-1137/ac9897}{{\em Chin. Phys. C} {\bfseries 47} no.~1, (2023) 013108}, \href{http://arxiv.org/abs/2209.12802}{{\ttfamily arXiv:2209.12802 [hep-ex]}}.

\bibitem{Belle:2025bpu}
{\bfseries Belle} Collaboration, K.~Uno {\em et~al.}, ``{Search for lepton-flavor-violating tau decays to {\ensuremath{\ell}}{\ensuremath{\alpha}} at Belle},'' \href{http://dx.doi.org/10.1007/JHEP08(2025)155}{{\em JHEP} {\bfseries 08} (2025) 155}, \href{http://arxiv.org/abs/2503.22195}{{\ttfamily arXiv:2503.22195 [hep-ex]}}.

\bibitem{Jho:2022snj}
Y.~Jho, S.~Knapen, and D.~Redigolo, ``{Lepton-flavor violating axions at MEG II},'' \href{http://dx.doi.org/10.1007/JHEP10(2022)029}{{\em JHEP} {\bfseries 10} (2022) 029}, \href{http://arxiv.org/abs/2203.11222}{{\ttfamily arXiv:2203.11222 [hep-ph]}}.

\bibitem{Knapen:2023zgi}
S.~Knapen, K.~Langhoff, T.~Opferkuch, and D.~Redigolo, ``{A robust search for lepton flavour violating axions at Mu3e},'' \href{http://dx.doi.org/10.1007/JHEP07(2025)243}{{\em JHEP} {\bfseries 07} (2025) 243}, \href{http://arxiv.org/abs/2311.17915}{{\ttfamily arXiv:2311.17915 [hep-ph]}}.

\bibitem{Raffelt:1994ry}
G.~Raffelt and A.~Weiss, ``{Red giant bound on the axion - electron coupling revisited},'' \href{http://dx.doi.org/10.1103/PhysRevD.51.1495}{{\em Phys. Rev. D} {\bfseries 51} (1995) 1495--1498}, \href{http://arxiv.org/abs/hep-ph/9410205}{{\ttfamily arXiv:hep-ph/9410205}}.

\bibitem{DiLuzio:2020wdo}
L.~Di~Luzio, M.~Giannotti, E.~Nardi, and L.~Visinelli, ``{The landscape of QCD axion models},'' \href{http://dx.doi.org/10.1016/j.physrep.2020.06.002}{{\em Phys. Rept.} {\bfseries 870} (2020) 1--117}, \href{http://arxiv.org/abs/2003.01100}{{\ttfamily arXiv:2003.01100 [hep-ph]}}.

\bibitem{Bollig:2020xdr}
R.~Bollig, W.~DeRocco, P.~W. Graham, and H.-T. Janka, ``{Muons in Supernovae: Implications for the Axion-Muon Coupling},'' \href{http://dx.doi.org/10.1103/PhysRevLett.125.051104}{{\em Phys. Rev. Lett.} {\bfseries 125} no.~5, (2020) 051104}, \href{http://arxiv.org/abs/2005.07141}{{\ttfamily arXiv:2005.07141 [hep-ph]}}. [Erratum: Phys.Rev.Lett. 126, 189901 (2021)].

\bibitem{Croon:2020lrf}
D.~Croon, G.~Elor, R.~K. Leane, and S.~D. McDermott, ``{Supernova Muons: New Constraints on $Z$' Bosons, Axions and ALPs},'' \href{http://dx.doi.org/10.1007/JHEP01(2021)107}{{\em JHEP} {\bfseries 01} (2021) 107}, \href{http://arxiv.org/abs/2006.13942}{{\ttfamily arXiv:2006.13942 [hep-ph]}}.

\bibitem{Caputo:2021rux}
A.~Caputo, G.~Raffelt, and E.~Vitagliano, ``{Muonic boson limits: Supernova redux},'' \href{http://dx.doi.org/10.1103/PhysRevD.105.035022}{{\em Phys. Rev. D} {\bfseries 105} no.~3, (2022) 035022}, \href{http://arxiv.org/abs/2109.03244}{{\ttfamily arXiv:2109.03244 [hep-ph]}}.

\bibitem{Belle:2000cnh}
{\bfseries Belle} Collaboration, A.~Abashian {\em et~al.}, ``{The Belle Detector},'' \href{http://dx.doi.org/10.1016/S0168-9002(01)02013-7}{{\em Nucl. Instrum. Meth. A} {\bfseries 479} (2002) 117--232}.

\bibitem{Heeck:2017xmg}
J.~Heeck and W.~Rodejohann, ``{Lepton flavor violation with displaced vertices},'' \href{http://dx.doi.org/10.1016/j.physletb.2017.11.067}{{\em Phys. Lett. B} {\bfseries 776} (2018) 385--390}, \href{http://arxiv.org/abs/1710.02062}{{\ttfamily arXiv:1710.02062 [hep-ph]}}.

\bibitem{XENON:2025vwd}
{\bfseries XENON} Collaboration, E.~Aprile {\em et~al.}, ``{WIMP Dark Matter Search Using a 3.1 Tonne-Year Exposure of the XENONnT Experiment},'' \href{http://dx.doi.org/10.1103/msw4-t342}{{\em Phys. Rev. Lett.} {\bfseries 135} no.~22, (2025) 221003}, \href{http://arxiv.org/abs/2502.18005}{{\ttfamily arXiv:2502.18005 [hep-ex]}}.

\bibitem{PandaX:2024qfu}
{\bfseries PandaX} Collaboration, Z.~Bo {\em et~al.}, ``{Dark Matter Search Results from 1.54{\,}{\,}Tonne{\textperiodcentered}Year Exposure of PandaX-4T},'' \href{http://dx.doi.org/10.1103/PhysRevLett.134.011805}{{\em Phys. Rev. Lett.} {\bfseries 134} no.~1, (2025) 011805}, \href{http://arxiv.org/abs/2408.00664}{{\ttfamily arXiv:2408.00664 [hep-ex]}}.

\bibitem{LZ:2024zvo}
{\bfseries LZ} Collaboration, J.~Aalbers {\em et~al.}, ``{Dark Matter Search Results from 4.2{\,}{\,}Tonne-Years of Exposure of the LUX-ZEPLIN (LZ) Experiment},'' \href{http://dx.doi.org/10.1103/4dyc-z8zf}{{\em Phys. Rev. Lett.} {\bfseries 135} no.~1, (2025) 011802}, \href{http://arxiv.org/abs/2410.17036}{{\ttfamily arXiv:2410.17036 [hep-ex]}}.

\bibitem{DarkSide-20k:2017zyg}
{\bfseries DarkSide-20k} Collaboration, C.~E. Aalseth {\em et~al.}, ``{DarkSide-20k: A 20 tonne two-phase LAr TPC for direct dark matter detection at LNGS},'' \href{http://dx.doi.org/10.1140/epjp/i2018-11973-4}{{\em Eur. Phys. J. Plus} {\bfseries 133} (2018) 131}, \href{http://arxiv.org/abs/1707.08145}{{\ttfamily arXiv:1707.08145 [physics.ins-det]}}.

\bibitem{XLZD:2024nsu}
{\bfseries XLZD} Collaboration, J.~Aalbers {\em et~al.}, ``{The XLZD Design Book: towards the next-generation liquid xenon observatory for dark matter and neutrino physics},'' \href{http://dx.doi.org/10.1140/epjc/s10052-025-14810-w}{{\em Eur. Phys. J. C} {\bfseries 85} no.~10, (2025) 1192}, \href{http://arxiv.org/abs/2410.17137}{{\ttfamily arXiv:2410.17137 [hep-ex]}}.

\bibitem{OHare:2021utq}
C.~A.~J. O'Hare, ``{New Definition of the Neutrino Floor for Direct Dark Matter Searches},'' \href{http://dx.doi.org/10.1103/PhysRevLett.127.251802}{{\em Phys. Rev. Lett.} {\bfseries 127} no.~25, (2021) 251802}, \href{http://arxiv.org/abs/2109.03116}{{\ttfamily arXiv:2109.03116 [hep-ph]}}.

\bibitem{Staub:2013tta}
F.~Staub, ``{SARAH 4 : A tool for (not only SUSY) model builders},'' \href{http://dx.doi.org/10.1016/j.cpc.2014.02.018}{{\em Comput. Phys. Commun.} {\bfseries 185} (2014) 1773--1790}, \href{http://arxiv.org/abs/1309.7223}{{\ttfamily arXiv:1309.7223 [hep-ph]}}.

\bibitem{Porod:2003um}
W.~Porod, ``{SPheno, a program for calculating supersymmetric spectra, SUSY particle decays and SUSY particle production at e+ e- colliders},'' \href{http://dx.doi.org/10.1016/S0010-4655(03)00222-4}{{\em Comput. Phys. Commun.} {\bfseries 153} (2003) 275--315}, \href{http://arxiv.org/abs/hep-ph/0301101}{{\ttfamily arXiv:hep-ph/0301101}}.

\bibitem{Belanger:2001fz}
G.~Belanger, F.~Boudjema, A.~Pukhov, and A.~Semenov, ``{MicrOMEGAs: A Program for calculating the relic density in the MSSM},'' \href{http://dx.doi.org/10.1016/S0010-4655(02)00596-9}{{\em Comput. Phys. Commun.} {\bfseries 149} (2002) 103--120}, \href{http://arxiv.org/abs/hep-ph/0112278}{{\ttfamily arXiv:hep-ph/0112278}}.

\bibitem{Cirelli:2005uq}
M.~Cirelli, N.~Fornengo, and A.~Strumia, ``{Minimal dark matter},'' \href{http://dx.doi.org/10.1016/j.nuclphysb.2006.07.012}{{\em Nucl. Phys. B} {\bfseries 753} (2006) 178--194}, \href{http://arxiv.org/abs/hep-ph/0512090}{{\ttfamily arXiv:hep-ph/0512090}}.

\bibitem{Cline:2013gha}
J.~M. Cline, K.~Kainulainen, P.~Scott, and C.~Weniger, ``{Update on scalar singlet dark matter},'' \href{http://dx.doi.org/10.1103/PhysRevD.88.055025}{{\em Phys. Rev. D} {\bfseries 88} (2013) 055025}, \href{http://arxiv.org/abs/1306.4710}{{\ttfamily arXiv:1306.4710 [hep-ph]}}. [Erratum: Phys.Rev.D 92, 039906 (2015)].

\bibitem{Feng:2014vea}
L.~Feng, S.~Profumo, and L.~Ubaldi, ``{Closing in on singlet scalar dark matter: LUX, invisible Higgs decays and gamma-ray lines},'' \href{http://dx.doi.org/10.1007/JHEP03(2015)045}{{\em JHEP} {\bfseries 03} (2015) 045}, \href{http://arxiv.org/abs/1412.1105}{{\ttfamily arXiv:1412.1105 [hep-ph]}}.

\bibitem{GAMBIT:2017gge}
{\bfseries GAMBIT} Collaboration, P.~Athron {\em et~al.}, ``{Status of the scalar singlet dark matter model},'' \href{http://dx.doi.org/10.1140/epjc/s10052-017-5113-1}{{\em Eur. Phys. J. C} {\bfseries 77} no.~8, (2017) 568}, \href{http://arxiv.org/abs/1705.07931}{{\ttfamily arXiv:1705.07931 [hep-ph]}}.

\bibitem{Casas:2017jjg}
J.~A. Casas, D.~G. Cerde{\~n}o, J.~M. Moreno, and J.~Quilis, ``{Reopening the Higgs portal for single scalar dark matter},'' \href{http://dx.doi.org/10.1007/JHEP05(2017)036}{{\em JHEP} {\bfseries 05} (2017) 036}, \href{http://arxiv.org/abs/1701.08134}{{\ttfamily arXiv:1701.08134 [hep-ph]}}.

\bibitem{Athron:2018ipf}
P.~Athron, J.~M. Cornell, F.~Kahlhoefer, J.~Mckay, P.~Scott, and S.~Wild, ``{Impact of vacuum stability, perturbativity and XENON1T on global fits of $\mathbb {Z}_2$ and $\mathbb {Z}_3$ scalar singlet dark matter},'' \href{http://dx.doi.org/10.1140/epjc/s10052-018-6314-y}{{\em Eur. Phys. J. C} {\bfseries 78} no.~10, (2018) 830}, \href{http://arxiv.org/abs/1806.11281}{{\ttfamily arXiv:1806.11281 [hep-ph]}}.

\bibitem{Arcadi:2019lka}
G.~Arcadi, A.~Djouadi, and M.~Raidal, ``{Dark Matter through the Higgs portal},'' \href{http://dx.doi.org/10.1016/j.physrep.2019.11.003}{{\em Phys. Rept.} {\bfseries 842} (2020) 1--180}, \href{http://arxiv.org/abs/1903.03616}{{\ttfamily arXiv:1903.03616 [hep-ph]}}.

\bibitem{Lebedev:2021xey}
O.~Lebedev, ``{The Higgs portal to cosmology},'' \href{http://dx.doi.org/10.1016/j.ppnp.2021.103881}{{\em Prog. Part. Nucl. Phys.} {\bfseries 120} (2021) 103881}, \href{http://arxiv.org/abs/2104.03342}{{\ttfamily arXiv:2104.03342 [hep-ph]}}.

\bibitem{Arcadi:2024ukq}
G.~Arcadi, D.~Cabo-Almeida, M.~Dutra, P.~Ghosh, M.~Lindner, Y.~Mambrini, J.~P. Neto, M.~Pierre, S.~Profumo, and F.~S. Queiroz, ``{The Waning of the WIMP: Endgame?},'' \href{http://dx.doi.org/10.1140/epjc/s10052-024-13672-y}{{\em Eur. Phys. J. C} {\bfseries 85} no.~2, (2025) 152}, \href{http://arxiv.org/abs/2403.15860}{{\ttfamily arXiv:2403.15860 [hep-ph]}}.

\bibitem{Yu:2024xsy}
Y.~Yu, T.-P. Tang, and L.~Feng, ``{New constraints on singlet scalar dark matter model with LZ, invisible Higgs decay and gamma-ray line observations},'' \href{http://dx.doi.org/10.1016/j.nuclphysb.2025.116910}{{\em Nucl. Phys. B} {\bfseries 1015} (2025) 116910}, \href{http://arxiv.org/abs/2410.21089}{{\ttfamily arXiv:2410.21089 [hep-ph]}}.

\bibitem{Camara:2023hhn}
H.~B. C{\^a}mara, F.~R. Joaquim, and J.~W.~F. Valle, ``{Dark-sector seeded solution to the strong CP problem},'' \href{http://dx.doi.org/10.1103/PhysRevD.108.095003}{{\em Phys. Rev. D} {\bfseries 108} no.~9, (2023) 095003}, \href{http://arxiv.org/abs/2303.00705}{{\ttfamily arXiv:2303.00705 [hep-ph]}}.

\bibitem{Casas:2001sr}
J.~A. Casas and A.~Ibarra, ``{Oscillating neutrinos and $\mu \to e, \gamma$},'' \href{http://dx.doi.org/10.1016/S0550-3213(01)00475-8}{{\em Nucl. Phys. B} {\bfseries 618} (2001) 171--204}, \href{http://arxiv.org/abs/hep-ph/0103065}{{\ttfamily arXiv:hep-ph/0103065}}.

\end{thebibliography}
\end{document}